\documentclass[aip,graphicx,amsmath,amssymb,reprint,citeautoscript]{revtex4-1}
\draft

\usepackage[pdftex]{graphics}
\usepackage{subcaption}
\usepackage{tikz}
\usepackage{xr}
\usetikzlibrary{calc}
\usetikzlibrary{decorations.pathmorphing, patterns,shapes}

\usepackage{siunitx}

\tikzset{snake it/.style={decorate, decoration={snake, amplitude=.5mm, segment length=2mm}}}
\tikzset{zigzag it/.style={decorate, decoration={zigzag, amplitude=.5mm, segment length=2mm}}}

\newcommand{\ket}[1]{|#1\rangle}
\newcommand{\bra}[1]{\langle#1|}
\newcommand{\dd}{\mathrm{d}}
\newcommand{\bs}[1]{\boldsymbol{#1}}
\newcommand{\pihalf}{\frac{\pi}{2}}

\newcommand{\cm}{\mathrm{cm}^{-1}}

\begin{document}

\title{Two-dimensional electronic spectra from trajectory-based dynamics: pure-state Ehrenfest, spin-mapping, and mean classical path approaches}

\author{Annina Z. Lieberherr}
\thanks{These authors contributed equally to this work.}
\affiliation{Department of Chemistry, University of Oxford, Physical and Theoretical Chemistry Laboratory, South Parks Road, Oxford, OX1 3QZ, United Kingdom}

\author{Joseph Kelly}
\thanks{These authors contributed equally to this work.}
\affiliation{Department of Chemistry, Stanford University, Stanford, California, 94305, USA}

\author{Johan E. Runeson}
\affiliation{Department of Chemistry, University of Oxford, Physical and Theoretical Chemistry Laboratory, South Parks Road, Oxford, OX1 3QZ, United Kingdom}

\author{Thomas E. Markland}
\email{tmarkland@stanford.edu}
\affiliation{Department of Chemistry, Stanford University, Stanford, California, 94305, USA}

\author{David E. Manolopoulos}
\email{david.manolopoulos@chem.ox.ac.uk}
\affiliation{Department of Chemistry, University of Oxford, Physical and Theoretical Chemistry Laboratory, South Parks Road, Oxford, OX1 3QZ, United Kingdom}  

\date{\today}
\definecolor{darkred}{RGB}{139,0,0}
\def\rev#1{\color{black}#1 \color{black}}

\begin{abstract}
Two-dimensional electronic spectroscopy (2DES) provides a detailed picture of electronically nonadiabatic dynamics that can be interpreted with the aid of simulations. Here, we develop and contrast trajectory-based nonadiabatic dynamics approaches for simulating 2DES spectra. First, we argue that an improved pure-state Ehrenfest approach can be constructed by decomposing the initial coherence into a sum of equatorial pure states that contain equal contributions from the states in the coherence. We then use this framework to show how one can obtain a more accurate, but computationally more expensive, approximation to the third-order 2DES response function by replacing Ehrenfest dynamics with spin mapping during the pump-probe delay time. We end by comparing and contrasting the accuracy of these methods \rev{and the simpler mean classical path approximation} in reproducing the exact linear, pump-probe, and 2DES spectra of two Frenkel exciton models: a coupled dimer system and the Fenna--Matthews--Olson complex. 
\end{abstract}


\maketitle
\section{Introduction}
Two-dimensional electronic spectroscopy \cite{Hybl1998, Muka2000, Hybl2001,Brix2005} (2DES) is an important experimental tool for investigating the electronically nonadiabatic dynamics of condensed-phase systems. This technique has motivated the development of theoretical methods that can be used to facilitate the interpretation of the resulting spectra. For example, accurate nonadiabatic quantum dynamics simulations have proved invaluable in disentangling features in 2DES attributable to vibronic effects from those associated with electronic coherence,\cite{Enge2007,Smit2011,Cao2020,Schu2024} thereby resolving questions about the role of quantum coherence in photosynthetic processes.

A variety of mixed quantum-classical (MQC) trajectory approaches to electronically nonadiabatic dynamics have now been developed, both to avoid the expense of a fully quantum calculation and to enable the study of more realistic Hamiltonians. They include Ehrenfest dynamics,\cite{McLa1964, Tull1998, Grun2009} stochastic surface hopping,\cite{Tull1990} mapping variable methods,\cite{Meye1979,Stoc1997,Rune2020} the symmetric quasi-classical method,\cite{Mill2016} and the mapping approach to surface hopping.\cite{Mann2023,Rune2023,Mann2024} These methods have been successful in simulating the population dynamics of nonadiabatic systems ranging from system-bath models to systems with atomistic force fields,\cite{Cres2018} and even \textit{ab initio} potentials.\cite{Ceot2009, Aiet2025, Sche2025, Lawr2024b} Extending them to allow for the calculation of advanced spectroscopies such as 2DES offers the opportunity to create practical methods to simulate and understand the information encoded in these spectra. However, this poses an additional challenge because it requires the calculation of multi-time correlation functions.\cite{Muka1999,Hamm2005}

Methods from various tiers of the MQC hierarchy have recently been extended to calculate 2DES spectra. Higher level approaches that employ partial linearization, such as the partially linearized density matrix (PLDM) formulation,\cite{Prov2018,Prov2018a}  have been shown to provide good accuracy but at a high computational cost.\cite{Rune2020} For example, to compute the 2DES spectrum of a Frenkel exciton model representing the Fenna--Matthews--Olson (FMO) complex, with a ground state, 7 singly excited states, and 21 doubly excited states, a PLDM-based approach incorporating spin mapping variables (spin-PLDM) used $1\times10^9$ trajectories, or $4\times10^7$ trajectories when taking advantage of focused initial conditions.\cite{Mann2022} For a biexciton model consisting of just a ground state, two singly excited states, and a doubly excited state, $2.4\times10^5$ trajectories were used to obtain the 2DES with PLDM\cite{Prov2018a} and, in another study, $10^6$ trajectories were used with spin-PLDM.\cite{Mann2022} 

Using a lower level of the MQC hierarchy, with full linearization of both the electronic and nuclear degrees of freedom, the linearized semiclassical initial value representation\cite{Sun1998} (LSC-IVR) has been used to simulate 2DES spectra\cite{Gao2020} within a non-perturbative framework\cite{Seid1995}. However, while fully linearized methods are less computationally expensive than PLDM, they still require roughly an order of magnitude more trajectories to converge than Ehrenfest dynamics because they involve averaging over an initial distribution of mapping variables. The lower computational cost of Ehrenfest dynamics makes it more suitable for simulations with \textit{ab initio} surfaces. For example, it has recently been used to compute the 2DES spectra of molecules such as pyrene for delay times up to \SI{30}{fs} using time-dependent density functional theory (TDDFT).\cite{Krum2024} 
 
When simulating a 2DES spectrum, the dynamics of the electronic system must be initialized in a coherence, which complicates the implementation of the Ehrenfest method. There are various ways to deal with this, two of which have been suggested in the previous literature. Van der Vegte {\em et al.}\cite{Vegt2013} used a mean classical path approximation\footnote{\rev{Note that this is not the same as the other ``classical path approximation'' that is sometimes used in mixed quantum-classical simulations, which is rather more drastic. In that, the back-action of the electronic system on the nuclear motion is neglected entirely, whereas the mean classical path approximation used in Ref.~34 includes the average  back action from the bra and ket states of an electronic coherence}} in which the force on the nuclei was taken to be the average of the Ehrenfest forces of the two states appearing in the coherence. Atsango {\em et al.}\cite{Atsa2023} developed what we shall refer to here as a `polar' pure state Ehrenfest method, in which the coherence is first decomposed into a sum of four pure states which are then propagated using standard Ehrenfest dynamics. 

Here we shall describe a different decomposition of the coherence into `equatorial' pure states, each of which contains an equal contribution from the two states in the coherence. We will show that this equatorial decomposition allows us to collapse two of the three pure state summations that arise in the third-order response function and that this leads to significant improvements in accuracy at a 32-fold reduced computational cost compared to the polar decomposition.~\cite{Atsa2023} The resulting scheme is equivalent to making the mean classical path approximation\cite{Vegt2013} during the first ($t_1$) and last ($t_3$) stages of the time evolution that is used to calculate the response function, but it differs during the delay time $t_2$ between the pump and probe pulses of a pump-probe experiment. We will end by showing that the lack of detailed balance in Ehrenfest dynamics leads to incorrect peak heights in pump-probe and 2DES spectra, and that spin mapping can be used to alleviate this problem. 

Sec.~II summarizes the response function formulation of 2DES. Sec.~III introduces the equatorial pure state Ehrenfest method and describes how it differs from the previously proposed polar pure state method.\cite{Atsa2023} This section explains in detail why the first and last equatorial pure state sums collapse, how the symmetries of the partial response functions can be exploited to further reduce the cost of the calculation, and how the central $t_2$ Ehrenfest evolution can be replaced with spin mapping. Sec.~IV defines the Frenkel exciton Hamiltonian that we have used in our  calculations and describes in detail how the equatorial pure-state decomposition can be used to calculate its linear and nonlinear response functions. Sec.~V uses a comparison with exact HEOM\cite{Kram2018,Tani2020} benchmark results to assess the accuracy and efficiency of the pure-state Ehrenfest and spin mapping approaches for a Frenkel biexciton model and a 7-site model of the FMO complex. \rev{A particularly revealing comparison is made at the end of this section with the results of the mean classical path approximation.\cite{Vegt2013}} Sec.~VI concludes the paper by assessing how practical it would be to combine these methods with {\em ab initio} forces, and by suggesting some possible directions for further research.

\section{Optical spectroscopy}
Two-dimensional electronic spectroscopy probes the third-order response function\cite{Muka1999,Hamm2005,Cho2008}
\begin{equation}
    R(t_3, t_2, t_1) = \mathrm{Tr}\{\hat{\mu}\, \mathcal{G}(t_3)[ \hat{\mu}, \mathcal{G}(t_2)[ \hat{\mu}, \mathcal{G}(t_1)[\hat{\mu}, \hat{\rho}_0]]]\},
    \label{eq:def-response-tot}
\end{equation}
where $\mathcal{G}(t) [\cdot] = e^{-i{\hat{{H}}}t} [\cdot]e^{i\hat{H}t}$ denotes the time evolution of the operator ` $\cdot$ ' over a time interval $t$. $\hat{{H}}$ is the Hamiltonian, $\hat{\mu}$ is the dipole moment operator, $\hat{\rho}_0$ is the density operator of the initial ground state, and $\hbar=1$ throughout. In what follows, we will split the dipole operator into its excitation and de-excitation components, $\hat{\mu} = \hat{\mu}_+ + \hat{\mu}_-$, which allows us to separate the pathways that contribute to the 2DES spectrum.

\begin{figure}
    \begin{subfigure}{.33\linewidth}
        \begin{tikzpicture}[node distance = .6cm]
    
    \node (L0) at (0,0) {};
    \node (R0) at (1.5,0) {};
    \node (L1) at (0,1) {};
    \node (R1) at (1.5,1) {};
    \node (L2) at (0,2) {};
    \node (R2) at (1.5,2) {};
    \node (L3) at (0,3) {};
    \node (R3) at (1.5,3) {};
    \node [below right of=R0] (arrow0) {};
    \node [below left of=L1] (arrow1) {};
    \node [above right of=R2] (arrow2) {};
    \node [above left of=L3] (arrow3) {};
    
    \draw[dashed] (L0.center) -- (R0.center) ;
    \draw[dashed] (L1.center) -- (R1.center) ;
    \draw[dashed] (L2.center) -- (R2.center) ;
    \draw[dashed] (L3.center) -- (R3.center) ;
    
    \node at (0.75,.5) {$|g \rangle \langle s |$};
    \node at (0.75,1.5) {$| s \rangle \langle s |$};
    \node at (0.75,2.5) {$|s \rangle \langle g |$};

    \draw[black, line width=2pt] (0,-.5) -- (L0.center) ;
    \draw[black, line width=2pt] (1.5,-.5) -- (R0.center) ;
    \draw[black, line width=2pt] (L0.center) -- (L1.center) ;
    \draw[red, line width=2pt, zigzag it] (R0.center) -- (R1.center) ;
    \draw[red, line width=2pt, zigzag it] (L1.center) -- (L2.center) ;
    \draw[red, line width=2pt, zigzag it] (R1.center) -- (R2.center) ;
    \draw[red, line width=2pt, zigzag it] (L2.center) -- (L3.center) ;
    \draw[black, line width=2pt] (R2.center) -- (R3.center) ;

    \draw[<-] (R0) --  (arrow0.center);
    \draw[->] (arrow1.center) -- (L1) ;
    \draw[->] (R2) -- (arrow2.center) ;
    \draw[<-] (arrow3.center) -- (L3) ;
    
\end{tikzpicture}
        \subcaption*{$\Phi_1$}
        \label{subfig:feynman-phi1}
    \end{subfigure}
    \begin{subfigure}{.33\linewidth}
        \begin{tikzpicture}[node distance = .6cm]
    
    \node (L0) at (0,0) {};
    \node (R0) at (1.5,0) {};
    \node (L1) at (0,1) {};
    \node (R1) at (1.5,1) {};
    \node (L2) at (0,2) {};
    \node (R2) at (1.5,2) {};
    \node (L3) at (0,3) {};
    \node (R3) at (1.5,3) {};
    \node [below right of=R0] (arrow0) {};
    \node [above right of=R1] (arrow1) {};
    \node [below left of=L2] (arrow2) {};
    \node [above left of=L3] (arrow3) {};
    
    \draw[dashed] (L0.center) -- (R0.center) ;
    \draw[dashed] (L1.center) -- (R1.center) ;
    \draw[dashed] (L2.center) -- (R2.center) ;
    \draw[dashed] (L3.center) -- (R3.center) ;
    
    \node at (0.75,.5) {$|g \rangle \langle s |$};
    \node at (0.75,1.5) {$|g\rangle \langle g |$};
    \node at (0.75,2.5) {$| s \rangle \langle g |$};

    \draw[black, line width=2pt] (0,-.5) -- (L0.center) ;
    \draw[black, line width=2pt] (1.5,-.5) -- (R0.center) ;
    \draw[black, line width=2pt] (L0.center) -- (L1.center) ;
    \draw[red, line width=2pt, zigzag it] (R0.center) -- (R1.center) ;
    \draw[black, line width=2pt] (L1.center) -- (L2.center) ;
    \draw[black, line width=2pt] (R1.center) -- (R2.center) ;
    \draw[red, line width=2pt, zigzag it] (L2.center) -- (L3.center) ;
    \draw[black, line width=2pt] (R2.center) -- (R3.center) ;

    \draw[<-] (R0) --  (arrow0.center);
    \draw[<-] (arrow1.center) -- (R1) ;
    \draw[<-] (L2) -- (arrow2.center) ;
    \draw[<-] (arrow3.center) -- (L3) ;

\end{tikzpicture}
        \subcaption*{$\Phi_2$}
        \label{subfig:feynman-phi2}
    \end{subfigure}
    \begin{subfigure}{.33\linewidth}
        \begin{tikzpicture}[node distance = .6cm]
    
    \node (L0) at (0,0) {};
    \node (R0) at (1.5,0) {};
    \node (L1) at (0,1) {};
    \node (R1) at (1.5,1) {};
    \node (L2) at (0,2) {};
    \node (R2) at (1.5,2) {};
    \node (L3) at (0,3) {};
    \node (R3) at (1.5,3) {};
    \node [below right of=R0] (arrow0) {};
    \node [below left of=L1] (arrow1) {};
    \node [below left of=L2] (arrow2) {};
    \node [above left of=L3] (arrow3) {};
    
    \draw[dashed] (L0.center) -- (R0.center) ;
    \draw[dashed] (L1.center) -- (R1.center) ;
    \draw[dashed] (L2.center) -- (R2.center) ;
    \draw[dashed] (L3.center) -- (R3.center) ;
    
    \node at (0.75,.5) {$|g \rangle \langle s |$};
    \node at (0.75,1.5) {$| s \rangle \langle s |$};
    \node at (0.75,2.5) {$|d \rangle \langle s |$};

    \draw[black, line width=2pt] (0,-.5) -- (L0.center) ;
    \draw[black, line width=2pt] (1.5,-.5) -- (R0.center) ;
    \draw[black, line width=2pt] (L0.center) -- (L1.center) ;
    \draw[red, line width=2pt, zigzag it] (R0.center) -- (R1.center) ;
    \draw[red, line width=2pt, zigzag it] (L1.center) -- (L2.center) ;
    \draw[red, line width=2pt, zigzag it] (R1.center) -- (R2.center) ;
    \draw[blue, line width=2pt, snake it] (L2.center) -- (L3.center) ;
    \draw[red, line width=2pt, zigzag it] (R2.center) -- (R3.center) ;

    \draw[<-] (R0) --  (arrow0.center);
    \draw[->] (arrow1.center) -- (L1) ;
    \draw[<-] (L2) -- (arrow2.center) ;
    \draw[<-] (arrow3.center) -- (L3) ;

\end{tikzpicture}
        \subcaption*{$\Phi_3$}
        \label{subfig:feynman-phi3}
    \end{subfigure}
    \begin{subfigure}{.33\linewidth}
        \begin{tikzpicture}[node distance = .6cm]
    
    \node (L0) at (0,0) {};
    \node (R0) at (1.5,0) {};
    \node (L1) at (0,1) {};
    \node (R1) at (1.5,1) {};
    \node (L2) at (0,2) {};
    \node (R2) at (1.5,2) {};
    \node (L3) at (0,3) {};
    \node (R3) at (1.5,3) {};
    \node [below left of=L0] (arrow0) {};
    \node [below right of=R1] (arrow1) {};
    \node [above right of=R2] (arrow2) {};
    \node [above left of=L3] (arrow3) {};
    
    \draw[dashed] (L0.center) -- (R0.center) ;
    \draw[dashed] (L1.center) -- (R1.center) ;
    \draw[dashed] (L2.center) -- (R2.center) ;
    \draw[dashed] (L3.center) -- (R3.center) ;
    
    \node at (0.75,.5) {$| s \rangle \langle g |$};
    \node at (0.75,1.5) {$| s \rangle \langle s |$};
    \node at (0.75,2.5) {$| s \rangle \langle g |$};

    \draw[black, line width=2pt] (0,-.5) -- (L0.center) ;
    \draw[black, line width=2pt] (1.5,-.5) -- (R0.center) ;
    \draw[red, line width=2pt, zigzag it] (L0.center) -- (L1.center) ;
    \draw[black, line width=2pt] (R0.center) -- (R1.center) ;
    \draw[red, line width=2pt, zigzag it] (L1.center) -- (L2.center) ;
    \draw[red, line width=2pt, zigzag it] (R1.center) -- (R2.center) ;
    \draw[red, line width=2pt, zigzag it] (L2.center) -- (L3.center) ;
    \draw[black, line width=2pt] (R2.center) -- (R3.center) ;

    \draw[<-] (L0) --  (arrow0.center);
    \draw[->] (arrow1.center) -- (R1) ;
    \draw[->] (R2) -- (arrow2.center) ;
    \draw[<-] (arrow3.center) -- (L3) ;

\end{tikzpicture}
        \subcaption*{$\Phi_4$}
        \label{subfig:feynman-phi4}
    \end{subfigure}
    \begin{subfigure}{.33\linewidth}
        \begin{tikzpicture}[node distance = .6cm]
    
    \node (L0) at (0,0) {};
    \node (R0) at (1.5,0) {};
    \node (L1) at (0,1) {};
    \node (R1) at (1.5,1) {};
    \node (L2) at (0,2) {};
    \node (R2) at (1.5,2) {};
    \node (L3) at (0,3) {};
    \node (R3) at (1.5,3) {};
    \node [below left of=L0] (arrow0) {};
    \node [above left of=L1] (arrow1) {};
    \node [below left of=L2] (arrow2) {};
    \node [above left of=L3] (arrow3) {};
    \node [above right of=R2] (ghost) {};
    
    \draw[dashed] (L0.center) -- (R0.center) ;
    \draw[dashed] (L1.center) -- (R1.center) ;
    \draw[dashed] (L2.center) -- (R2.center) ;
    \draw[dashed] (L3.center) -- (R3.center) ;
    
    \node at (0.75,.5) {$| s \rangle \langle g |$};
    \node at (0.75,1.5) {$| g \rangle \langle g |$};
    \node at (0.75,2.5) {$| s \rangle \langle g |$};

    \draw[black, line width=2pt] (0,-.5) -- (L0.center) ;
    \draw[black, line width=2pt] (1.5,-.5) -- (R0.center) ;
    \draw[red, line width=2pt, zigzag it] (L0.center) -- (L1.center) ;
    \draw[black, line width=2pt] (R0.center) -- (R1.center) ;
    \draw[black, line width=2pt] (L1.center) -- (L2.center) ;
    \draw[black, line width=2pt] (R1.center) -- (R2.center) ;
    \draw[red, line width=2pt, zigzag it] (L2.center) -- (L3.center) ;
    \draw[black, line width=2pt] (R2.center) -- (R3.center) ;

    \draw[<-] (L0) --  (arrow0.center);
    \draw[<-] (arrow1.center) -- (L1) ;
    \draw[<-] (L2) -- (arrow2.center) ;
    \draw[<-] (arrow3.center) -- (L3) ;

\end{tikzpicture}
        \subcaption*{$\Phi_5$}
        \label{subfig:feynman-phi5}
    \end{subfigure}
    \begin{subfigure}{.33\linewidth}
        \begin{tikzpicture}[node distance = .6cm]
    
    \node (L0) at (0,0) {};
    \node (R0) at (1.5,0) {};
    \node (L1) at (0,1) {};
    \node (R1) at (1.5,1) {};
    \node (L2) at (0,2) {};
    \node (R2) at (1.5,2) {};
    \node (L3) at (0,3) {};
    \node (R3) at (1.5,3) {};
    \node [below left of=L0] (arrow0) {};
    \node [below right of=R1] (arrow1) {};
    \node [below left of=L2] (arrow2) {};
    \node [above left of=L3] (arrow3) {};
    
    \draw[dashed] (L0.center) -- (R0.center) ;
    \draw[dashed] (L1.center) -- (R1.center) ;
    \draw[dashed] (L2.center) -- (R2.center) ;
    \draw[dashed] (L3.center) -- (R3.center) ;
    
    \node at (0.75,.5) {$|s \rangle \langle g |$};
    \node at (0.75,1.5) {$| s \rangle \langle s |$};
    \node at (0.75,2.5) {$| d \rangle \langle s |$};

    \draw[black, line width=2pt] (0,-.5) -- (L0.center) ;
    \draw[black, line width=2pt] (1.5,-.5) -- (R0.center) ;
    \draw[red, line width=2pt, zigzag it] (L0.center) -- (L1.center) ;
    \draw[black, line width=2pt] (R0.center) -- (R1.center) ;
    \draw[red, line width=2pt, zigzag it] (L1.center) -- (L2.center) ;
    \draw[red, line width=2pt, zigzag it] (R1.center) -- (R2.center) ;
    \draw[blue, line width=2pt, snake it] (L2.center) -- (L3.center) ;
    \draw[red, line width=2pt, zigzag it] (R2.center) -- (R3.center) ;

    \draw[<-] (L0) --  (arrow0.center);
    \draw[->] (arrow1.center) -- (R1) ;
    \draw[<-] (L2) -- (arrow2.center) ;
    \draw[<-] (arrow3.center) -- (L3) ;

\end{tikzpicture}
        \subcaption*{$\Phi_6$}
        \label{subfig:feynman-phi6}
    \end{subfigure}
    
    \caption{Feynman diagrams for the pathways (nonlinear response functions) that survive the rotating wave approximation.
    Right-pointing arrows correspond to the $e^{-i\omega t + i\mathbf{k} \cdot \mathbf{r}}$ component and left-pointing arrows to the $e^{+i\omega t - i\mathbf{k} \cdot \mathbf{r}}$ component of the electric field.
    Arrows pointing towards the center of the diagram symbolize an excitation, while arrows pointing away symbolize a de-excitation.
    The colors denote the ground ($\ket{g}$, black straight), singly excited ($\ket{s}$, red zigzag) and doubly excited ($\ket{d}$, blue wavy) manifolds of the bra and ket states.}
    \label{fig:feynman-diagrams}
\end{figure}
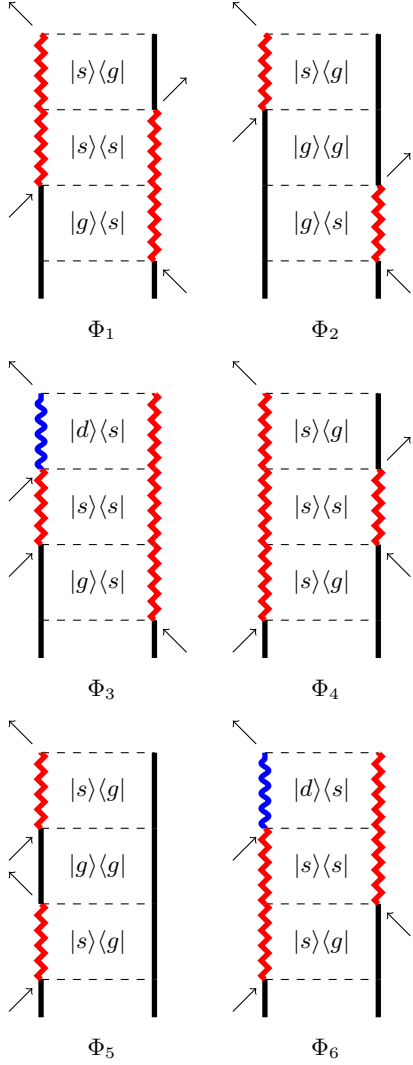

Of the eight terms that arise from the nested commutators in Eq.~\eqref{eq:def-response-tot}, only six survive the rotating wave approximation.~\cite{Muka1999,Hamm2005} These give rise to the following contributions
\begin{subequations}
\label{eq:def-phis}
\begin{align}
    \Phi_1 &= \mathrm{Tr}\{\hat{\mu}_- \mathcal{G}(t_3)[\mathcal{G}(t_2) [\hat{\mu}_+ \mathcal{G}(t_1) [\hat{\rho}_0 \hat{\mu}_-]]\hat{\mu}_+]\}, \label{eq:def_phi1}\\
    \Phi_2 &= \mathrm{Tr}\{\hat{\mu}_- \mathcal{G}(t_3)[\hat{\mu}_+ \mathcal{G}(t_2)[\mathcal{G}(t_1) [\hat{\rho}_0 \hat{\mu}_-]\hat{\mu}_+]]\}, \\
    \Phi_3 &= \mathrm{Tr}\{\hat{\mu}_- \mathcal{G}(t_3) [\hat{\mu}_+ \mathcal{G}(t_2) [\hat{\mu}_+\mathcal{G}(t_1)[\hat{\rho}_0 \hat{\mu}_-]]]\}, \\
    \Phi_4 &= \mathrm{Tr}\{\hat{\mu}_- \mathcal{G}(t_3) [\mathcal{G}(t_2)[\mathcal{G}(t_1)[\hat{\mu}_+\hat{\rho}_0]\hat{\mu}_-]\hat{\mu}_+]\}, 
    \label{eq:def_phi4}\\
    \Phi_5 &= \mathrm{Tr}\{\hat{\mu}_- \mathcal{G}(t_3) [\hat{\mu}_+\mathcal{G}(t_2)[\hat{\mu}_-\mathcal{G}(t_1)[\hat{\mu}_+\hat{\rho}_0]]]\}, \\
     \Phi_6 &= \mathrm{Tr}\{\hat{\mu}_- \mathcal{G}(t_3) [\hat{\mu}_+\mathcal{G}(t_2)[\mathcal{G}(t_1)[\hat{\mu}_+\hat{\rho}_0]\hat{\mu}_-]]\},
     \label{eq:def_phi6}
\end{align}
\end{subequations}  
all of which are functions of the time intervals between the light-matter interactions, $t_3$, $t_2$, and $t_1$. Fig.~\ref{fig:feynman-diagrams} shows the Feynman diagrams for each of the six pathways in Eq.~\eqref{eq:def-phis}.\cite{Schl2011} One can interpret the response functions in terms of three physical processes:
stimulated emission (SE, $\Phi_1$ and $\Phi_4$), ground-state bleaching (GSB, $\Phi_2$ and $\Phi_5$), and excited-state absorption (ESA, $\Phi_3$ and $\Phi_6$).
These three processes cannot be distinguished experimentally -- they simply denote possible pathways through the excitation manifolds.
$\Phi_1$, $\Phi_2$, and $\Phi_3$ give the rephasing response function
\begin{equation}
    \label{eq:response_rp}
    R_\mathrm{rp} = \Phi_1 + \Phi_2 - \Phi_3,
\end{equation}
and $\Phi_4$, $\Phi_5$, and $\Phi_6$ the non-rephasing response function
\begin{equation}
    \label{eq:response_nr}
    R_\mathrm{nr} = \Phi_4 + \Phi_5 - \Phi_6.
\end{equation}
It is possible to measure $R_\mathrm{rp}$ and $R_\mathrm{nr}$ independently by placing the detector in different directions.\cite{Schl2011}
The rephasing response corresponds to emission in the direction $-\mathbf{k}_1 + \mathbf{k}_2 + \mathbf{k}_3$, and the non-rephasing response to emission in the direction $\mathbf{k}_1-\mathbf{k}_2+\mathbf{k}_3$ where $\mathbf{k}_1$, $\mathbf{k}_2$, and $\mathbf{k}_3$ are the wave vectors of the laser pulses at times  $0$, $t_1$, and $t_1+t_2$ respectively. 

The 2DES spectrum can be obtained from the rephasing and non-rephasing response functions as
\begin{align}
    \label{eq:ft_2des}
    S(\omega_3, t_2, \omega_1) &= \mathrm{Re}\int_0^\infty \dd t_1 \int_0^\infty \dd t_3\, \times \nonumber \\
    &\quad \left[ e^{i(\omega_3 t_3+\omega_1t_1)} R_\mathrm{nr}(t_3, t_2, t_1)\right.\nonumber \\
    &\quad \left. +\, e^{i(\omega_3 t_3-\omega_1t_1)} R_\mathrm{rp}(t_3, t_2, t_1) \right],
\end{align}
where $\omega_1$ and $\omega_3$ correspond to pump and probe frequencies, respectively, which are the absorption and emission axes of the spectrum. $t_2$ is the time delay between the laser pulses of a pump-probe experiment; one can obtain the resulting transient absorption (pump-probe) spectrum by integrating $S(\omega_3, t_2, \omega_1)$ over $\omega_1$. 

\section{Pure-state Ehrenfest Methods}
\label{sec:pure-state-methods}
Ehrenfest dynamics was originally introduced in the wavefunction picture\cite{McLa1964, Tull1998} but it can also be obtained in a density matrix formulation.\cite{Grun2009} The action of light-matter interaction operators on the density matrix leads to the construction and propagation of coherences. There are multiple ways to represent these coherences in terms of pure states, which lead to different accuracies and efficiencies within Ehrenfest theory\cite{Mont2016,Atsa2023}. In this section, we provide the background and notation needed to understand this issue, and introduce the pure-state decompositions that will be compared in Sec.~\ref{sec:model-systems}. 

In the Ehrenfest approach\cite{McLa1964, Tull1998} one separates the total Hamiltonian into a chosen partitioning of system and bath Hamiltonians, $\hat{H}_\mathrm{S}$ and  $\hat{H}_\mathrm{B}$, and the coupling between them, $\hat{H}_\mathrm{SB}$. Upon invoking a mean field interaction between the system and bath, and taking the classical limit of the bath degrees of freedom, one obtains 
\begin{equation}
    \hat{H}({\bs p},{\bs q}) = \hat{H}_\mathrm{S} + \hat{H}_\mathrm{SB}(\bs q)+H_{\rm B}(\bs p,\bs q)\hat{I},
\end{equation}
where $\hat{I}$ is the identity operator on the system Hilbert space and $\bs q$ and $\bs p$ are the mass-scaled positions and momenta of the bath degrees of freedom. The Ehrenfest equations of motion are
\begin{subequations}\label{eq:EOM}
    \begin{align}
        \dot{\ket{\Psi}} &= -i(\hat{{H}}_\mathrm{S} + \hat{{H}}_\mathrm{SB})\ket{\Psi}, \label{eq:TDSE}\\
        \dot{\bs p} &= - \bra{\Psi} \nabla_{\bs q} \hat{H} \ket{\Psi}, \\
        \dot{\bs q} &= \bs p,
    \end{align}
\end{subequations} 
where $\ket{\Psi}$ is the normalized wavefunction of the system which evolves under the time-dependent Schrödinger equation. The bath degrees of freedom ($\bs p,~\bs q$) evolve under classical equations of motion experiencing the average force from the quantum state $\ket{\Psi}$. To obtain the response functions in Eq.~\eqref{eq:def-phis}, the quantum trace is approximated by the mixed quantum-classical trace
\begin{equation}
\label{eq:qctrace}
    \mathrm{Tr}\{\cdot\} \approx \frac{1}{(2\pi)^{\mathcal{K}}} \int \dd \bs p \int \dd \bs q\; \mathrm{Tr}_\mathrm{S}\{\cdot\},
\end{equation}
where $\mathrm{Tr}_\mathrm{S}$ is a trace over the system degrees of freedom and $\mathcal{K}$ is the total number of bath degrees of freedom. We assume that the initial density operator can be factorized as
\begin{equation}
    \label{eq:initial_density}
    \hat{\rho}_0(\bs p, \bs q) =  \rho_\mathrm{B}(\bs p, \bs q)\hat{\rho}_{0,\rm S},
\end{equation}
where $\hat{\rho}_{0,\rm S}=\ket{0}\bra{0}$ is the system reduced density matrix and $\rho_{\rm B}({\bs p},{\bs q})$ is the bath density of the classical degrees of freedom. The latter can either be sampled from the classical Boltzmann distribution or from the Wigner distribution as described in Sec.~\ref{sec:frenkel_hamiltonian}.

\begin{figure}[b]
    \centering
    \begin{subfigure}[c]{.4\linewidth}
    \centering
        \includegraphics[width=\textwidth]{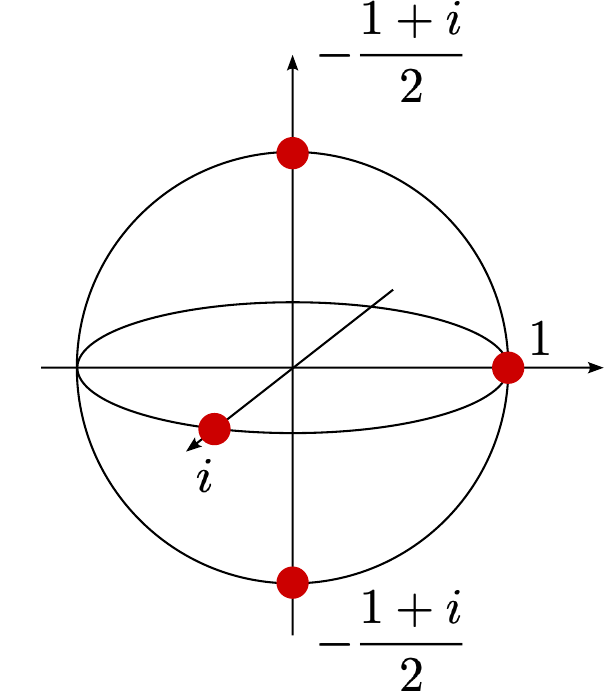}
        \subcaption{Polar}
        \label{subfig:sphere-polar}
    \end{subfigure}
    \begin{subfigure}[c]{.4\linewidth}
    \centering
        \includegraphics[width=\textwidth]{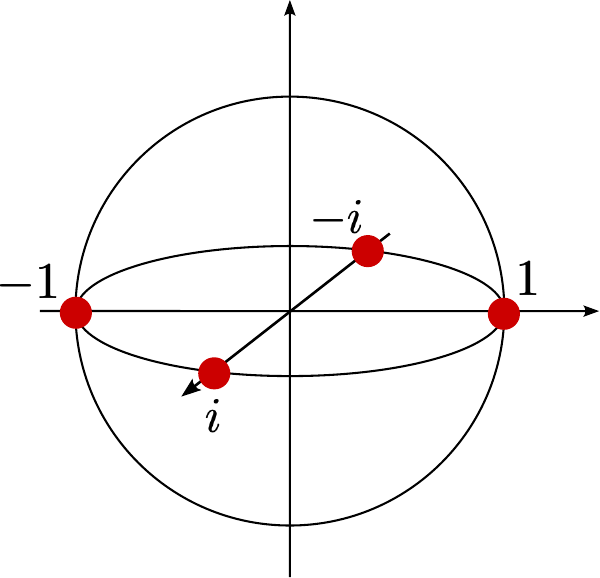}
        \subcaption{Equatorial}
        \label{subfig:sphere-equatorial}
    \end{subfigure}
    \caption{Bloch sphere visualization of (a) polar and (b) equatorial pure state decompositions. The states $\ket{a}$ and $\ket{b}$ are at the north and south poles in both cases. The 4 pure states in each decomposition, along with $\ket{a}$ and $\ket{b}$, are drawn as independently normalized. Each pure state is labeled by its weight in Eq.~\eqref{eq:def-pure-state-decomp-long} or \eqref{eq:def-pure-state-decomp-eq-weights}, with the factors of 1/2 in Eq.~\eqref{eq:def-pure-state-decomp-eq-weights} omitted for clarity.}
    \label{fig:sphere}
\end{figure}

Starting from the ground-state density in Eq.~\eqref{eq:initial_density}, the initial dipole interactions $\hat{\rho}_0\hat{\mu}_-$ and $\hat{\mu}_+ \hat{\rho}_0$ in Eq.~\eqref{eq:def-phis} yield the coherences
\begin{subequations}
\label{eq:ab0}
    \begin{align}
        \hat{\mu}_+ \hat{\rho}_0(\bs p, \bs q) = \ket{\mu} \rho_\mathrm{B}(\bs p, \bs q) \bra{0}, \\
        \hat{\rho}_0(\bs p, \bs q) \hat{\mu}_-  = \ket{0} \rho_\mathrm{B}(\bs p, \bs q) \bra{\mu},
    \end{align}
\end{subequations}
where $\ket{\mu} = \hat{\mu}_+ \ket{0}$ and $\bra{\mu} = \bra{0}\hat{\mu}_-$. These coherences present a challenge for Ehrenfest dynamics since they have zero trace and therefore cannot be propagated directly.\cite{Mont2016, Atsa2023} To overcome this issue, one can decompose each of them into a sum of pure states,
\begin{equation}
    \ket{a}\bra{b} = \sum_{j=0}^3 w_j \ket{j}\bra{j}, \label{eq:def-pure-state-decomp-general}
\end{equation}
where $\ket{a}\bra{b}$ is the coherence, $\ket{j}\bra{j}$ is a pure state, and $w_j$ is a scalar weight. There are many such pure-state decompositions. Here, we shall consider the particular decompositions depicted in Figs.~\ref{subfig:sphere-polar} and~\ref{subfig:sphere-equatorial}, which we shall refer to as the `polar' and `equatorial' decompositions respectively. 
 
\subsection{Polar Decomposition}
In the polar decomposition (Fig.~\ref{subfig:sphere-polar}), the pure states $|j\rangle=|\psi_j\rangle$ and their weights $w_j^{(\rm p)}$ are
\begin{subequations}
\begin{align}
    \ket{\psi_0} &= \ket{a}, & w_0^{(\rm p)} &= - \frac{1+i}{2}, \\
    \ket{\psi_1} &= \ket{b}, & w_1^{(\rm p)} &= -\frac{1+i}{2}, \\
    \ket{\psi_2} &= \frac{1}{\sqrt{2}} (\ket{a} + \ket{b}), & w_2^{(\rm p)} &= 1, \\
    \ket{\psi_3} &= \frac{1}{\sqrt{2}}( \ket{a} + i\ket{b}), & w_3^{(\rm p)} &= i.
\end{align}
\label{eq:def-pure-state-decomp-long}
\end{subequations}

This is the decomposition suggested by Atsango {\em et al.},\cite{Atsa2023} who calculated the response functions in Eq.~\eqref{eq:def-phis} as follows. The initial coherence is decomposed into 4 pure states, each of which is propagated with ${\mathcal G}(t_1)$. The second action of a dipole operator ($\hat{\mu}_+$ or $\hat{\mu}_-$) then generates another coherence, which is decomposed into a further 4 pure states, etc. The effort grows exponentially with the number of dipole interactions until a total of $4^3=64$ pure states are propagated with $G(t_3)$, acted on with the final dipole moment operator, and used to assemble the response function.

 The polar pure-state decomposition has been found to work well for a biexciton model with a low characteristic bath frequency $\omega_c$.\cite{Atsa2023} However, it degrades in accuracy when $\omega_c$ becomes comparable to the electronic excitation frequency, and it requires 64 branches of each trajectory to compute each response function. As we show in the supplementary information (SI, Sec.~S1), some of these branches are inconsistent with the Feynman diagrams in Fig.~\ref{fig:feynman-diagrams}, and others contribute zero to the final response function. These issues do not stem from the pure-state decomposition itself, but rather from the particular choice of the pure states in Eq.~\eqref{eq:def-pure-state-decomp-long}.
 
 Another problem with the polar pure states is that they contain unequal contributions from the two states in the coherence. For example, $\ket{\psi_0}$ contains purely state $\ket{a}$ with no contribution from $\ket{b}$. In the calculation of the rephasing response functions, we have $\ket{a} = \ket{0}$ and $\ket{b} = \ket{\mu}$ at the beginning of the $t_1$ evolution. If the elements of the dipole operator are very large, the pure states $\ket{\psi_2}$ and $\ket{\psi_3}$ will be dominated by the contribution from $\ket{\mu}$. When there is just one electronic state in the singly-excited manifold, the bath will therefore evolve on its potential energy surface. However, it is well known that evolution of the bath degrees of freedom on the average of the ground and excited potential energy surfaces would give a more accurate result. This average potential arises in the Wigner-averaged classical limit, which gives the exact linear (absorption and fluorescence) spectra when the ground and excited states have displaced harmonic potentials with the same frequency.\cite{Egor1999,Shi2005} It would thus be advantageous to develop a scheme in which each pure state contains equal contributions of the two states $\ket{a}$ and $\ket{b}$, giving a bath dynamics that more closely resembles that of the mean classical path approximation.\cite{Vegt2013} 

\subsection{Equatorial Decomposition}
\label{subsec:theory-eq-decomposition}

We shall show in this section that the proliferation of polar pure state branching trajectories can be avoided by defining the equatorial pure states 
\begin{align}
    \ket{\phi_j} &= \frac{1}{\sqrt{2}}(\ket{a} + e^{ij\frac{\pi}{2}} \ket{b}), 
    \label{eq:def-pure-state-decomp-eq-states}
\end{align}
with the weights
\begin{align}
    w_j^{(\rm e)} &= \frac{1}{2}e^{ij\frac{\pi}{2}},
    \label{eq:def-pure-state-decomp-eq-weights}
\end{align}
where the index $j$ again runs from 0 to 3 (see Fig.~\ref{subfig:sphere-equatorial}). To avoid the unbalanced contributions of $\ket{a}$ and $\ket{b}$ to the Ehrenfest force when one of them has a larger norm than the other, we also define the \emph{balanced} states
\def\phibar{\overline{\phi}}
\def\phibbar{\overline{\phibar}}
\begin{equation}
\label{eq:def-pure-state-decomp-eq-states-normalized}
    \ket{\phibar_j} = \frac{1}{\sqrt{2}} \left( \frac{\ket a}{\sqrt{\langle a | a \rangle}} + e^{ij\pihalf} \frac{\ket b}{\sqrt{\langle b | b \rangle}} \right),
\end{equation}
and their normalized versions
\begin{equation}
    \ket{\phibbar_j} = \langle\phibar_j | \phibar_j \rangle^{\;-1/2} \;\ket{\phibar_j},
\end{equation}
and use these to evaluate the force
\begin{equation}
\label{eq:eq-pure-state-force}
    F_j({\bs q}) = -\bra{\phibbar_j}\nabla_{\bs q}\hat{H}\ket{\phibbar_j}.
\end{equation}
However, we still need to use the states $\ket{\phi_j}$ to evaluate the response functions because the norms of their $\ket{a}$ and $\ket{b}$ components keep track of the strengths of the dipole interactions in Eq.~\eqref{eq:def-phis}.

To see why the equatorial pure states reduce the proliferation of trajectory branches, we first note that the weights $w_j^{\rm (e)}$ satisfy 
\begin{equation}
    \sum_{j=0}^3 w_j^{(\rm e)} = \sum_{j=0}^3 w_j^{(\rm e)} e^{ij\frac{\pi}{2}} = 0,
    \label{eq:eq-sumtozero}
\end{equation}
and
\begin{equation}
    \sum_{j=0}^3 w_j^{(\rm e)} e^{-ij\frac{\pi}{2}} = 2.
    \label{eq:eq-sumtotwo}
\end{equation}
If the Hamiltonian is block-diagonal with $\ket{a}$ and $\ket{b}$ in different manifolds such that $\bra{a} \hat{{H}} \ket{b} = 0$, as is the case during the $t_1$ and $t_3$ evolutions, the propagation of the equatorial pure states is particularly simple. In this case, the Ehrenfest force on each equatorial pure state is
\begin{align}
F_j(\bs q) &= - \bra{\phibbar_j} \nabla_{\bs q} \hat{{H}} \ket{\phibbar_j} \nonumber\\
&= - \frac{1}{2} \left[ \frac{\bra{a} \nabla_{\bs q} \hat{{H}} \ket{a}}{\langle a | a \rangle} + \frac{\bra{b} \nabla_{\bs q} \hat{{H}} \ket{b}}{\langle b | b \rangle}\right], 
\label{eq:force-averaged-ab}
\end{align}
which is independent of $j$. The bath modes therefore follow the same trajectories for all four states, which implies that we can write their time evolution as
\begin{equation}
    \ket{\phi_j(t)} = \frac{1}{\sqrt{2}} (\ket{a(t)} + e^{ij\pihalf} \ket{b(t)}),
\end{equation}
where $\ket{\psi(t)} = {\mathcal T}e^{-i\int_0^t \hat{{H}}(\bs q_\tau)\dd \tau} \ket{\psi}$, ${\mathcal T}$ is the time-ordering operator, and $\bs{q}_{\tau}$ evolves under the force in Eq.~\eqref{eq:force-averaged-ab}. Combining this with Eqs.~\eqref{eq:eq-sumtozero} and \eqref{eq:eq-sumtotwo} allows us to write the pure-state sum after the propagation as
\begin{align}
\label{eq:eq-resum-trick}
\mathcal{G}(t) \ket{a}\bra{b} &= \sum_{j=0}^3 w_j^{(\rm e)} \ket{\phi_j(t)} \bra{\phi_j(t)}\nonumber\\ &= \ket{a(t)} \bra{b(t)}.
\end{align}
This can be evaluated by propagating just {\em one} of the four pure states and extracting $\ket{a(t)}\bra{b(t)}$ from the result as 
\begin{equation}
\ket{a(t)}\bra{b(t)}=2e^{ij{\pi\over 2}}\hat{P}_a\ket{\phi_j(t)}\bra{\phi_j(t)}\hat{P}_b,
\label{eq:atbt}
\end{equation}
where $\hat{P}_a$ and $\hat{P}_b$ are the projection operators onto the separate manifolds of states containing $\ket{a}$ and $\ket{b}$.

Eq.~\eqref{eq:eq-resum-trick} is a simple but intuitive result. It is equivalent to propagating the ket and bra states separately along a trajectory that follows the force of the average potential, as is done in the mean classical path approximation.~\cite{Vegt2013}
In the polar pure-state decomposition, Eq.~\eqref{eq:eq-resum-trick} does not hold, because the polar pure states evolve along different nuclear trajectories. 
Neither does it hold for the $t_2$ evolution, even with equatorial pure states, because during $t_2$ the states $\ket{a(t)}$ and $\ket{b(t)}$ are in the same excitation manifold and $\bra{a} \hat{H} \ket{b}\not=0$ (see Fig.~\ref{fig:feynman-diagrams}). 

The equatorial decomposition therefore requires us to propagate 4 separate states $\ket{\phi_j(t)}$ during the $t_2$ evolution, each of which follows a different bath trajectory. This is the only remaining difference between the equatorial decomposition and the mean classical path approximation,\cite{Vegt2013} in which a single coherence is propagated on the average potential energy surface of its bra and ket states during all three time intervals. Each of the 4 equatorial pure states that is propagated through $t_2$ is converted into a coherence by the next dipole interaction, but since this can be dealt with using the resummation trick in Eq.~\eqref{eq:eq-resum-trick} it only produces one new state to be propagated through $t_3$. The net result is thus that, while the polar decomposition requires the propagation of 64 pure states through $t_3$, the equatorial decomposition only requires the propagation of 4. The $t_3$ evolution dominates the calculation because it has to be done separately for each value of $t_1$ and $t_2$ that are used to calculate the 2DES spectrum, so the resummations yield a reduction in the overall computational effort by a factor of 16.

We can also find a further factor of 2 by exploiting the symmetries of the rephasing (Eq.~\eqref{eq:response_rp}) and non-rephasing (Eq.~\eqref{eq:response_nr}) response functions. For each of the three physical processes captured by 2DES, SE ($\Phi_1$, $\Phi_4$), GSB ($\Phi_2$, $\Phi_5$), and ESA ($\Phi_3$, $\Phi_6$), one of the contributing terms is a rephasing pathway while the other is a non-rephasing pathway. The density matrix propagated through $t_2$ in each of these pairs is the adjoint of the other, e.g. $\mu_+ \ket{0(t_1)}\bra{\mu(t_1)}$ and $\ket{\mu(t_1)} \bra{0(t_1)} \mu_-$ in $\Phi_1$ and $\Phi_4$, respectively.
In the equatorial decomposition, one can exploit this symmetry simply by complex-conjugating the weights of the pure states,
\begin{align}
    \ket{b} \bra{a} &= \sum_{j=0}^3 w_j^{(\rm e) \ast} \ket{\phi_j} \bra{\phi_j},
\end{align}
where $\ket{\phi_j}$ are the pure states obtained from decomposing $\ket{a} \bra{b}$.
The non-rephasing response functions can therefore be obtained for free during the calculation of the rephasing functions, saving a further factor 2 in computational effort. With this additional saving, the equatorial pure state method only requires two times more force evaluations per initial trajectory than the mean classical path approximation of van der Vegte {\em et al.},\cite{Vegt2013} whereas the polar pure-state decomposition requires 64 times more.\cite{Atsa2023} \rev{One might imagine that it would also be possible to use the symmetry between the rephasing and non-rephasing response functions to reduce the number of trajectories by a factor of 2 in the mean classical path approximation, but we explain in the supplementary information why this is not the case (SI, Sec.~S2).}

\subsection{Use of Spin-Mapping to Improve Population Dynamics}
\label{subsec:spin-mapping-theory}

Ehrenfest dynamics has significant detailed-balance issues which lead to incorrectly predicted quantum state populations. We shall therefore investigate whether the population dynamics that occurs during the $t_2$ evolution can be improved by replacing Ehrenfest propagation with a more accurate method based on generalized spin mapping.\cite{Rune2020} 
We choose spin mapping\footnote{\rev{Note that we are referring here to fully linearized spin mapping, not the partially linearized spin-PLDM approach described in Ref.~29}} because it has been shown to give better population dynamics than Ehrenfest for the FMO model,~\cite{Rune2020} because it is straightforward to implement, and because it can use time steps of a similar size to Ehrenfest (as opposed to MASH, for example, which requires smaller times steps\cite{Rune2023}). 

Spin mapping differs from Ehrenfest dynamics in three crucial respects: the initialization of the system variables, the calculation of the reduced density matrix, and the equations of motion for the bath variables. We shall now explain how we implement these changes in the present context.

For each of the four pure states $\ket{{\phi}_j}\bra{{\phi}_j}$ created at the beginning of the $t_2$ evolution [Eq.~\eqref{eq:def-pure-state-decomp-eq-states}], we first normalize $\ket{{\phi}_j}$ to obtain $\ket{\tilde{\phi}_j}=\ket{\phi_j}\langle \phi_j|\phi_j\rangle^{-1/2}$, and keep $\langle\phi_j|\phi_j\rangle$ for later use. We then sample $M$ normalized system wavefunctions $\ket{c_m}$ from the density $\ket{\tilde{\phi}_j}\bra{\tilde{\phi}_j}$, where $M$ is the number of states in the relevant excitation manifold ($M=1$ for the GSB response functions and $M=N$ for SE and ESA, where $N$ is the number of singly excited states). The sampling is done in such a way that the initial spin-mapping estimator of the density operator is exactly equal to $\ket{\tilde{\phi}_j}\bra{\tilde{\phi}_j}$ when averaged over $m$ from 1 to $M$. The `doubly focused' sampling algorithm we use to do this is described in detail in the supplementary information (SI, Sec.~S3). \rev{(This algorithm can be viewed as an $M$-state generalization of the 2-state focused sampling algorithm described in the SI of Ref.~\cite{Mann2024}, in which two samples of the spin variables are taken at opposite points ($S_x,S_y$) and ($-S_x,-S_y$) on the polar circle at a fixed value of $S_z$. In the $M$-state case  samples are taken at $M$ equally-spaced angles on the appropriate generalization of this polar circle.)}

To calculate the spin-mapping density operator of each sample $\ket{c_m(t)}$ during the $t_2$ interval, we use the estimator\cite{Rune2020}
\begin{equation}
\label{eq:sm-density-matrix}
    \hat \rho_{{\rm S},m}(t) = \sqrt{M+1} \ket{c_m(t)} \bra{c_m(t)} - \frac{\sqrt{M+1}-1}{M} \hat I_M,
\end{equation}
where $\hat I_M$ is the identity operator within the manifold. In the ground state manifold where there is just one non-degenerate state $\ket{0}$, it is easy to verify that Eq.~\eqref{eq:sm-density-matrix} simplifies to the Ehrenfest reduced density matrix $\ket{0}\bra{0}$. Therefore, introducing spin mapping does not change the calculation of the GSB response functions. However, it will modify the reduced density matrix for the SE and ESA pathways when there is more than one coupled excited state. 

During the $t_2$ evolution, each of the $M$ system wavefunctions $\ket{c_m(t)}$ satisfies its own time-dependent Schr\"odinger equation [Eq.~\eqref{eq:TDSE}] and generates its own force
\begin{equation}
    F_m({\bs q}) = - {\rm Tr}\left\{\hat{\rho}_{{\rm S},m}\nabla_{\bs q}\hat{H}\right\}
\end{equation}
on the bath variables. The bath trajectories are therefore different for each of the $M$ wavefunctions. At the end of the $t_2$ evolution, we diagonalize each $\hat{\rho}_{\mathrm{S},m}(t_2)$ to obtain a weighted sum of $M$ pure states, 
\begin{equation}
\hat{\rho}_{{\rm S},m}(t_2) = \sum_{k=1}^M \rho_{k,m}(t_2)\ket{k,m(t_2)}\bra{k,m(t_2)},
\end{equation}
each of which can then be treated in the same way as $\ket{\phi_j(t_2)}\bra{\phi_j(t_2)}$ is treated in the equatorial pure state Ehrenfest calculation. The only differences are that, after the application of the next dipole operator, we now have $M^2$ pure states to propagate through $t_3$ instead of just one, and that the contribution of each of these to the target response function must be multiplied by an appropriate weight, $w_{j,k,m}=\langle \phi_j|\phi_j\rangle\rho_{k,m}(t_2)$.

In summary, using this implementation of spin mapping for the $t_2$ evolution yields an approach that leaves the GSB pathways unchanged but is expected to improve the population dynamics of the SE and ESA pathways, at the cost of having to run $N^2$ more trajectory branches for each of them.

\section{Application to Frenkel Exciton Hamiltonians}

We will now define the Frenkel exciton Hamiltonian we have used in our benchmark calculations and explain how the methods we have described above can be used to compute its linear and nonlinear response functions.

\subsection{Hamiltonian}
\label{sec:frenkel_hamiltonian}
The Frenkel exciton Hamiltonian consists of a set of $N$ sites, each with a two-level electronic system coupled to a bath of harmonic oscillators, 
\begin{equation}
    \hat{{H}} = \left( \begin{array}{ccc}
        \hat{H}^g & 0 & 0 \\
        0 & \hat{H}^s & 0 \\
         0 & 0 & \hat{H}^d 
    \end{array} \right) + \hat H_\mathrm{B}(\hat{\bs p}, \hat{\bs q}).\label{eq:fullH}
\end{equation}
The bath Hamiltonian $\hat H_\mathrm{B}(\hat{\bs p}, \hat{\bs q})$ consists of $K$ harmonic oscillators on each of the $N$ electronic sites, giving ${\mathcal K}=KN$ in Eq.~\eqref{eq:qctrace}:
\begin{equation}
    \hat H_\mathrm{B}(\hat{\bs p}, \hat{\bs q}) = \sum_{k=1}^K \sum_{n=1}^N \left( \frac{\hat p_{kn}^2}{2} + \frac{1}{2} \omega_k^2 \hat q_{kn}^2 \right).
    \label{eq:def-frenkel-bath-hamiltonian}
\end{equation}
Here $\omega_k$ are the bath mode frequencies and $\hat q_{kn}$ and $\hat p_{kn}$ are their mass-scaled position and momentum operators. 

The Hamiltonian $\hat{{H}}$ is block-diagonal in the ground, singly excited and doubly excited manifolds, with respective Hamiltonians $\hat{H}^g$, $\hat{H}^s$ and $\hat{H}^d$.
The ground state energy $\varepsilon_0$ is set to zero, giving $\hat{H}^g = \varepsilon_0 \ket{0}\bra{0} = 0$.
The Hamiltonian of the singly excited manifold is
\begin{equation}
    \hat{H}^s = \hat H^s_{\mathrm{S}} + \hat H^s_{\mathrm{SB}}(\hat{\bs q}), \label{eq:def-frenkel-singly-excited-hamiltonian}
\end{equation}
where $\hat H^s_{\mathrm{S}}$ is the Hamiltonian for the excitonic system and $\hat H^s_{\mathrm{SB}}$ is the system-bath coupling.
The system Hamiltonian is
\def\hc{\rm h.c.}
\begin{equation}
\label{eq:def-frenkel-system}
\hat{H}^s_{\mathrm{S}} = \sum_{n=1}^N \varepsilon_n |n\rangle\langle n|+
\sum_{m>n}^N J_{nm}\left(|n\rangle\langle m|+|m\rangle\langle n|\right),
\end{equation}
where $\varepsilon_n$ is the site excitation energy, $J_{nm}$ is the coupling between sites $n$ and $m$, and $N$ is the number of sites.
The system-bath coupling Hamiltonian is
\begin{equation}
    \hat{H}^s_{\mathrm{SB}}(\hat{\bs q}) = \sum_{k=1}^K \sum_{n=1}^N c_k \hat q_{kn} \ket{n}\bra{n}, \label{eq:def-frenkel-sysbath-hamiltonian}
\end{equation}
with coupling strengths $c_k$.

The Hamiltonian in the doubly excited manifold consists of states that are simultaneous excitations of two different sites. It does not matter which order they are excited in. We use the states $\ket{nm}$ with $m > n$ as a basis for $\hat H^d$, which can then be written as
\begin{equation}
    \hat H^d = \hat H^d_{\mathrm{S}} + \hat H^d_{\mathrm{SB}}(\hat{\bs q})
\end{equation}
with~\cite{Hein2012}
\begin{align}
\hat{H}^d_{\rm S} &= \sum_{m>n}^N (\varepsilon_n+\varepsilon_m)|nm\rangle\langle nm| \nonumber\\
&+ \sum_{m>n>l}^N J_{nm}\left(|ln\rangle\langle lm|+|lm\rangle\langle ln|\right)\nonumber\\ 
&+ \sum_{m>l>n}^N J_{nm}\left(|nl\rangle\langle lm|+|lm \rangle\langle nl|\right)\nonumber\\
&+ \sum_{l>m>n}^N J_{nm}\left(|nl\rangle\langle ml|+|ml\rangle\langle nl|\right)
\label{eq:def-frenkel-hamiltonian-Hs-doubly}
\end{align}
and
\begin{equation}
    \hat H^d_{\mathrm{SB}}(\hat{\bs q}) = \sum_{k=1}^K \sum_{m>n}^N  c_k (\hat q_{kn} + \hat q_{km}) \ket{nm}\bra{nm}.
\label{eq:def-frenkel-hamiltonian-Hsb-doubly}
\end{equation}

The dipole moment operator couples the ground state to the singly excited manifold, and the singly excited manifold to the doubly excited manifold. The excitation component of the dipole coupling between the ground and singly excited states is
\begin{equation}
    \hat{\mu}^{sg}_{+} = \sum_{n=1}^N \mu_n \ket{n}\bra{0},
    \label{eq:musg}
\end{equation}
and that between the single and double excitation manifolds is~\cite{Hein2012}
\begin{equation}
    \hat{\mu}^{ds}_{+} = \sum_{m>n}^N \left( \mu_m \ket{nm}\bra{n} + \mu_n \ket{nm}\bra{m} \right).
    \label{eq:muds}
\end{equation}
The full dipole excitation operator is thus
\begin{equation}
\label{eq:mu_up_tot}
    \hat{\mu}_+ = \left( \begin{array}{ccc}
         0 & 0 & 0 \\
        \hat{\mu}^{sg}_{+} & 0 & 0 \\
         0 & \hat{\mu}^{ds}_{+} & 0
    \end{array} \right),
\end{equation}
and the full dipole moment operator is $\hat{\mu} = \hat{\mu}_+ + \hat{\mu}_-$ with $\hat{\mu}_- = (\hat{\mu}_+)^\dagger$.

For simplicity, we use the same parameters $\{c_k, \omega_k\}$ for the bath associated with each of the $N$ electronic sites (Eqs.~\eqref{eq:def-frenkel-bath-hamiltonian} and \eqref{eq:def-frenkel-sysbath-hamiltonian}). Each bath has a Debye spectral density, 
\begin{equation}
    J(\omega) = \frac{2\lambda \omega_c \omega}{\omega^2 + \omega_c^2} = \frac{\pi}{2} \sum_{k=1}^K \frac{c_k^2}{\omega_k} \delta(\omega - \omega_k),
\end{equation}
where $\lambda$ is the reorganization energy and $\omega_c$ is the characteristic bath frequency. The bath relaxation time is $\tau_c = {\omega_c}^{-1}$.
We discretize the spectral density using a method\cite{Wang2001, Crai2007} that gives the correct reorganization energy, $\lambda = \frac{1}{\pi} \int_0^\infty \frac{J(\omega)}{\omega}\dd \omega$, for any choice of $K$.

The initial bath degrees of freedom can either be sampled from a classical Boltzmann distribution, or from a Wigner distribution to allow for thermal quantum effects. Both can be written in the form\cite{Wang1998}
\begin{equation}
    \rho_\mathrm{B}(\bs p, \bs q) = \prod_{k,n} \frac{\sigma_k}{\pi} \exp\left[ -\frac{2\sigma_k}{\omega_k} \left( \frac{p_{kn}^2}{2} + \frac{\omega_{k}^2 q_{kn}^2}{2} \right)\right],
    \label{eq:Wigner}
\end{equation}
where $\sigma_k = \tanh \left({\beta \omega_k}/{2}\right)$ in the Wigner case and $\beta\omega_k/2$ in the classical limit, with $\beta={1}/{k_{\rm B}T}$. The system density operator is initialized in the ground state, 
\begin{equation}
    \hat{\rho}_{0,\rm S} = \ket{0}\bra{0}.
\end{equation}

\subsection{Nonlinear Response Functions in the Equatorial Decomposition}
\label{subsec:pure-state-response-functions}

To show how the equatorial decomposition is used in more detail, we will now derive an explicit expression for the $\Phi_3$ response function, and then summarize how one obtains the analogous expressions for the remaining five response functions.

For $\Phi_3$, we begin by propagating the equatorial pure state $\ket{\phi} = \frac{1}{\sqrt{2}}(\ket{0} + \ket{\mu})$ through $t_1$, where we no longer include the pure state index, $j$, because only one state remains after summing over the pure states for the $t_1$ propagation. Because $\ket{0}$ and $\ket{\mu}$ belong to different manifolds, we can invoke Eq.~\eqref{eq:eq-resum-trick} in the form
\begin{equation}
    \mathcal{G}(t_1)(\hat{\rho}_0 \hat{\mu}_-) = \ket{0(t_1)} \bra{\mu(t_1)},
    \label{eq:results-t1}
\end{equation}
and
\begin{subequations}
\label{eq:phi3-derivation-projection-after-t1}
\begin{equation}
    \ket{0(t_1)} = \sqrt{2} \hat{P}_g \ket{\phi(t_1)},
\end{equation}
\begin{equation}
    \ket{\mu(t_1)} = \sqrt{2} \hat{P}_s \ket{\phi(t_1)},
\end{equation}
\end{subequations}
where $\hat{P}_g = \ket 0 \bra 0$ is the projection on the ground state and $\hat{P}_s = \sum_{n=1}^N \ket n \bra n$ is the projection on the singly excited manifold.
The nuclei experience the average force
\begin{equation}
    F(\bs q) = -\frac{1}{2} \left[ \bra{0} \nabla_{\bs q} \hat{{H}} \ket{0} + \frac{\bra{\mu} \nabla_{\bs q} \hat{{H}} \ket{\mu}}{\langle \mu | \mu \rangle} \right]
    \label{eq:force-t1}
\end{equation}
during the $t_1$ evolution.

The interaction with the second dipole moment operator acts on the ket $\ket{0(t_1)}$ to  give $\hat{\mu}_+\ket{0(t_1)}$. After this operation, the resulting states both lie in the singly excited manifold, $\hat{\mu}_+\ket{0(t_1)}\bra{\mu(t_1)}$. We then use Eq.~\eqref{eq:def-pure-state-decomp-general} to expand this in terms of four new equatorial pure states
\begin{equation}
    \ket{\phi_{3,j}^{(t_1)}} = \frac{1}{\sqrt{2}} \left[ \hat{\mu}_+\ket{0(t_1)} + e^{ij\pihalf} \ket{\mu(t_1)}\right], \label{eq:phijt1}
\end{equation}
where the subscript `3' indicates the target response function, $\Phi_3$.
The states $\ket{\phi_{3,j}^{(t_1)}}$ evolve along different nuclear trajectories during $t_2$ in accordance with Eq.~\eqref{eq:EOM} and under the force defined in Eq.~\eqref{eq:eq-pure-state-force}. The third dipole moment operator then acts on each pure state separately, but for each of them this operation creates only one new state to be propagated through $t_3$,
\begin{equation}
    \ket{\phi_{3,j}^{(t_2,t_1)}} = \frac{1}{\sqrt{2}} \left[\; \hat{\mu}_+ \ket{\phi_{3,j}^{(t_1)}(t_2)} + \ket{\phi_{3,j}^{(t_1)}(t_2)} \;\right].
\end{equation}

All that remains after the evolution through $t_3$ is to project out the necessary components of the time-evolved state $\ket{\phi_{3,j}^{(t_2, t_1)}(t_3)}$, apply the final dipole moment operator, and evaluate the trace in Eq.~\eqref{eq:def-phis}. The final result for $\Phi_3$ is thus
\begin{align}
\Phi_3(t_3, t_2, t_1) &= \int \dd \bs p \int \dd \bs q\, \rho_\mathrm{B}(\bs p, \bs q)\, \sum_{j=0}^3 2w_{j}^{(\rm e)} \times \nonumber\\
&\bra{\phi_{3,j}^{(t_2,t_1)}(t_3)} \hat{P}_s \hat{\mu}_- \hat{P}_d\ket{\phi_{3,j}^{(t_2,t_1)}(t_3)},    
\end{align}
where $\hat{P}_s$ and $\hat{P}_d$ are projection operators onto the singly and doubly excited manifolds, respectively, and the factor of 2 comes from Eq.~\eqref{eq:atbt}.
$\Phi_6$ can be obtained from the same expression simply by replacing the weights $w_{j}^{(\rm e)}$ with their complex conjugates.

$\Phi_1$ has the same Feynman diagram as $\Phi_3$ up to the third dipole interaction. We can therefore reuse $\ket{\phi_{3,j}^{(t_1)}(t_2)}$ to define the states for the $t_3$ propagation,
\begin{equation}
    \ket{\phi_{1,j}^{(t_2,t_1)}} = \frac{1}{\sqrt{2}} \left[ \ket{\phi_{3,j}^{(t_1)}(t_2)} + \hat{\mu}_-\ket{\phi_{3,j}^{(t_1)}(t_2)} \right].
\end{equation}
The final result is
\begin{align}
    \Phi_1(t_3, t_2, t_1) =& \int \dd \bs  p \int \dd \bs q \,\rho_\mathrm{B}(\bs p, \bs q) \sum_{j=0}^3 2w_{j}^{(\rm e)} \times \nonumber \\
    &\bra{\phi_{1,j}^{(t_2, t_1)}(t_3)} \hat{P}_g \hat{\mu}_- \hat{P}_s \ket{\phi_{1,j}^{(t_2, t_1)}(t_3)}.
\label{eq:phi-1-final-summed}
\end{align}
By replacing the weights in Eq.~\ref{eq:phi-1-final-summed} with their complex conjugates, we obtain $\Phi_4$.

For $\Phi_2$, the analogue of Eq.~\eqref{eq:phijt1} is
\begin{equation}
    \ket{\phi_{2,j}^{(t_1)}} = \frac{1}{\sqrt{2}} \left[ \ket{0(t_1)} + e^{ij\pihalf} \hat \mu_- \ket{\mu(t_1)} \right]
\end{equation}
Since this is proportional to the ground state $\ket{0}$, it can be propagated through $t_2$ (to within an irrelevant overall phase factor) simply by evolving the nuclei on the ground state potential (under the influence of $H_{\rm B}(\bs p,\bs q)$).
We can then propagate the states
\begin{equation}
    \label{eq:phi2-derivation-t3-states}
    \ket{\phi_{2,j}^{(t_2, t_1)}} = \frac{1}{\sqrt{2}} \left[ \hat \mu_+ \ket{\phi_{2,j}^{(t_1)}(t_2)} + \ket{\phi_{2,j}^{(t_1)}(t_2)} \right]
\end{equation}
through $t_3$ and obtain $\Phi_2$ as
\begin{align}
\label{eq:phi2-final}
\Phi_2(t_3, t_2, t_1) =& \int \dd \bs  p \int \dd \bs q\, \rho_\mathrm{B}(\bs p, \bs q) \sum_{j=0}^3 2w_{j}^{(\rm e)} \times \nonumber \\
&\bra{\phi_{2,j}^{(t_2, t_1)}(t_3)} \hat{P}_g \hat{\mu}_- \hat{P}_s \ket{\phi_{2,j}^{(t_2,t_1)}(t_3)}.
\end{align}
This time we get $\Phi_5$ by complex conjugating the weights.

The key simplification in this algorithm is the summation trick in Eq.~\eqref{eq:eq-resum-trick}, which is only possible because the pure states used in the $t_1$ and $t_3$ evolutions lie on the equator and have identical nuclear motion paths.
More generally, pure states at any given latitude on the Bloch sphere have the same nuclear trajectories. However, equal contributions to the forces on the nuclei from the states in the $\ket{a}$ and $\ket{b}$ manifolds only occur on the equator.
With anything other than equatorial pure states, we would populate the wrong excitation manifolds and produce dynamics that are inconsistent with the Feynman diagrams.
In the supplementary information, we compare the polar and equatorial wavefunction dynamics in detail (SI, Sec.~S2). 

\subsection{Linear Absorption Spectroscopy}
\label{subsec:linear-spectra}

Here we compare how the polar and equatorial Ehrenfest methods, our spin-mapping adaptation, and the mean classical path approximation treat the linear absorption spectrum, which is proportional to the Fourier transform 
\begin{equation}
    I(\omega) = \mathrm{Re}\int_{0}^\infty e^{-i\omega t} R^{(1)}(t)\, \dd t
\end{equation}
of the first-order response function
\begin{equation}
    R^{(1)}(t) = \mathrm{Tr}\{\hat{\mu}_- \mathcal{G}(t) [\hat{\mu}_+ \hat{\rho}_0]\}.
    \label{eq:def-response-linear}
\end{equation}
This response function only depends on one time variable: it is equivalent to calculating the third-order response of a non-rephasing pathway through $t_1$ only (e.g., Eq.~\eqref{eq:def_phi4}). The equatorial Ehrenfest, mean classical path, and spin mapping methods all treat the $t_1$ evolution in the same way, and therefore yield the same result when applied to the linear spectrum. However, the polar Ehrenfest method yields a slightly different result because of a subtle difference in the way it treats the evolution of the bath degrees of freedom.  

In the polar Ehrenfest method, four pure states are propagated up to time $t$. However, the polar states $\ket{\psi_0} = \ket{\mu}$ and $\ket{\psi_1} = \ket{0}$ do not contribute to $R^{(1)}(t)$, since they vanish in the final trace with $\hat{\mu}_-$.
Only the coherences $\ket{\psi_2}=\frac{1}{\sqrt{2}}(\ket{\mu}+\ket{0})$ and $\ket{\psi_3}=\frac{1}{\sqrt{2}}(\ket{\mu}+i\ket{0})$ contribute, giving
\begin{align}
R^{(1)}(t) &= \mathrm{Tr}\left\{\hat{\mu}_- \left(\ket{\psi_2(t)}\bra{\psi_2(t)}+i\ket{\psi_3(t)}\bra{\psi_3(t)}\right)\right\}\nonumber\\
&= \mathrm{Tr}\left\{\hat{\mu}_- \ket{\mu(t)} \bra{0(t)}\right\}.
\end{align}
This is almost the same answer as one obtains in the equatorial decomposition [see Eq.~\eqref{eq:results-t1}], except that the nuclear degrees of freedom in the polar Ehrenfest method evolve under the force
\begin{equation}
\label{eq:polar-force-t1}
	F_\mathrm{pol}(\bs q) = -\frac{1}{1 + c_\mu} (\bra \mu \nabla_{\bs q} \hat{{H}} \ket{\mu} + \bra 0 \nabla_{\bs q} \hat{ \mathcal H} \ket 0),
\end{equation}
where $c_\mu = \langle\mu|\mu\rangle = \sum_n \mu_n^2$.
This is subtly different from the equatorial force in Eq.~\eqref{eq:force-t1} because, motivated by the success of the Wigner-averaged classical limit for linear spectroscopy,\cite{Egor1999,Shi2005} we have chosen to use balanced forces in the equatorial Ehrenfest method as described in Sec.~III.B. This results in a small difference between the predictions of the polar and equatorial methods for the linear spectrum of the biexciton model, as we will see in Fig.~5 below.

\section{Results and Discussion}
\label{sec:model-systems}

In this section, we compare the results obtained for the 2DES, pump-probe, and linear absorption spectra of a 2-site biexciton model and a 7-site FMO Frenkel exciton model using the polar and equatorial Ehrenfest methods, as well as the spin mapping variation. Throughout, these methods will be labeled as `polar', `equatorial', and `spin mapping', respectively.

\subsection{Biexciton Model}
\label{subsec:biexciton-results}

\begin{figure}
    \centering
    \includegraphics[width=0.48\textwidth]{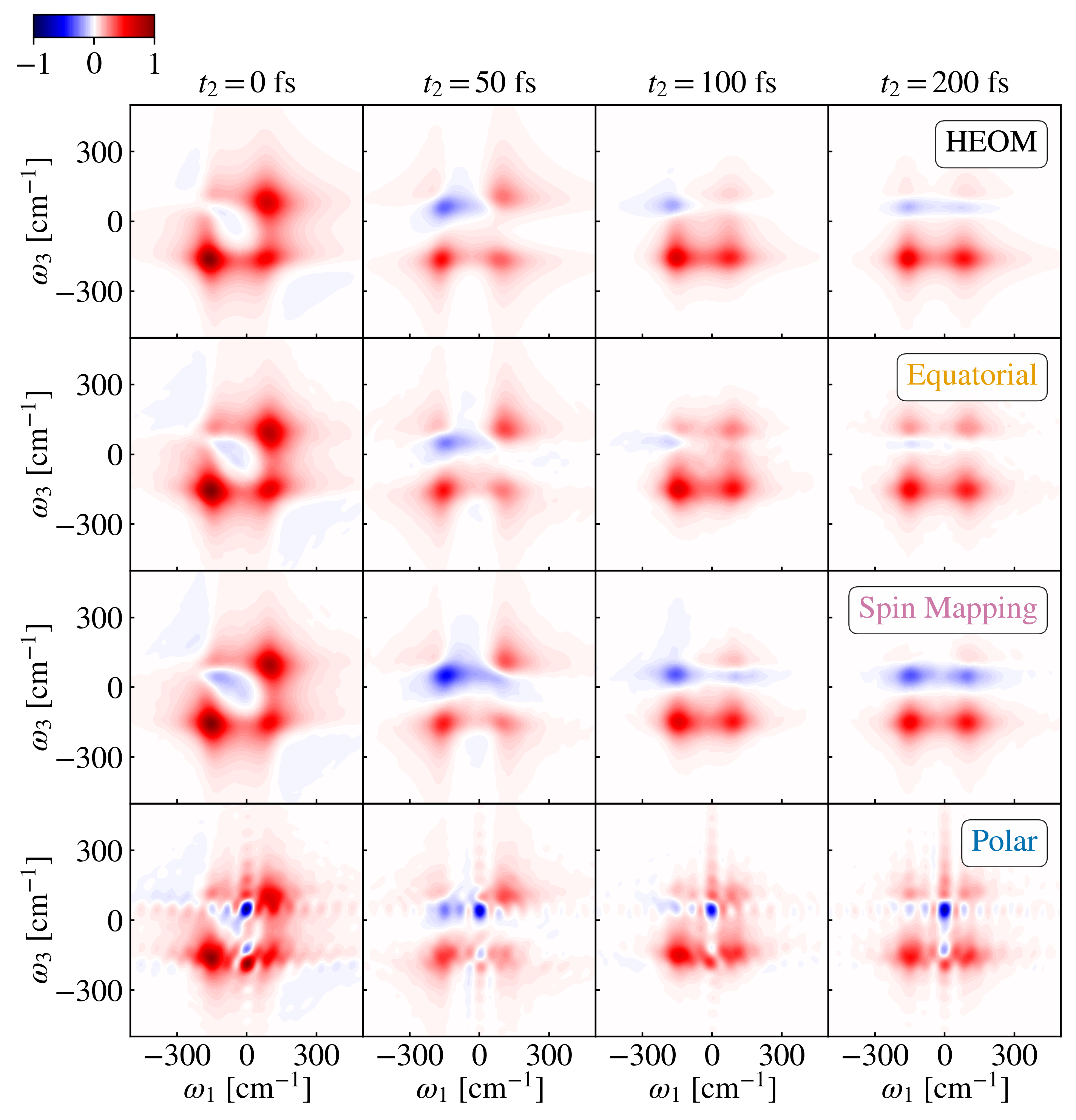}
    \caption{Comparison of polar, equatorial, and spin mapping 2DES spectra for the biexciton model with numerically exact HEOM results, as a function of the delay time $t_2$. The three quantum-classical spectra were computed with 96,000 trajectories. All spectra are normalized to a maximum value of 1 at $t_2=0$.}
    \label{fig:biexc-spectra}
\end{figure}

Figure~\ref{fig:biexc-spectra} compares the 2DES spectra of the biexciton model obtained from the polar, equatorial, and spin mapping approaches with the exact results from HEOM. Each 2DES spectrum is normalized by dividing by its maximum absolute value at $t_2=0$ fs to provide a comparison of the relative peak intensities and their time evolution. The biexciton parameters (SI, Sec.~S4) were selected to be in a regime where the polar approach is known to be inaccurate\cite{Atsa2023}. This is evident in Fig.~\ref{fig:biexc-spectra}, which shows that the equatorial spectra are significantly closer to the exact results. Spin mapping also yields good agreement with the HEOM results, except for overestimating the negative (blue) feature at $\omega_3 \approx 0~\cm$, particularly at longer $t_2$ delay times.  This feature is underestimated by the equatorial approach.

\begin{figure*}[t]
    \centering
    \includegraphics[width=0.6\textwidth]{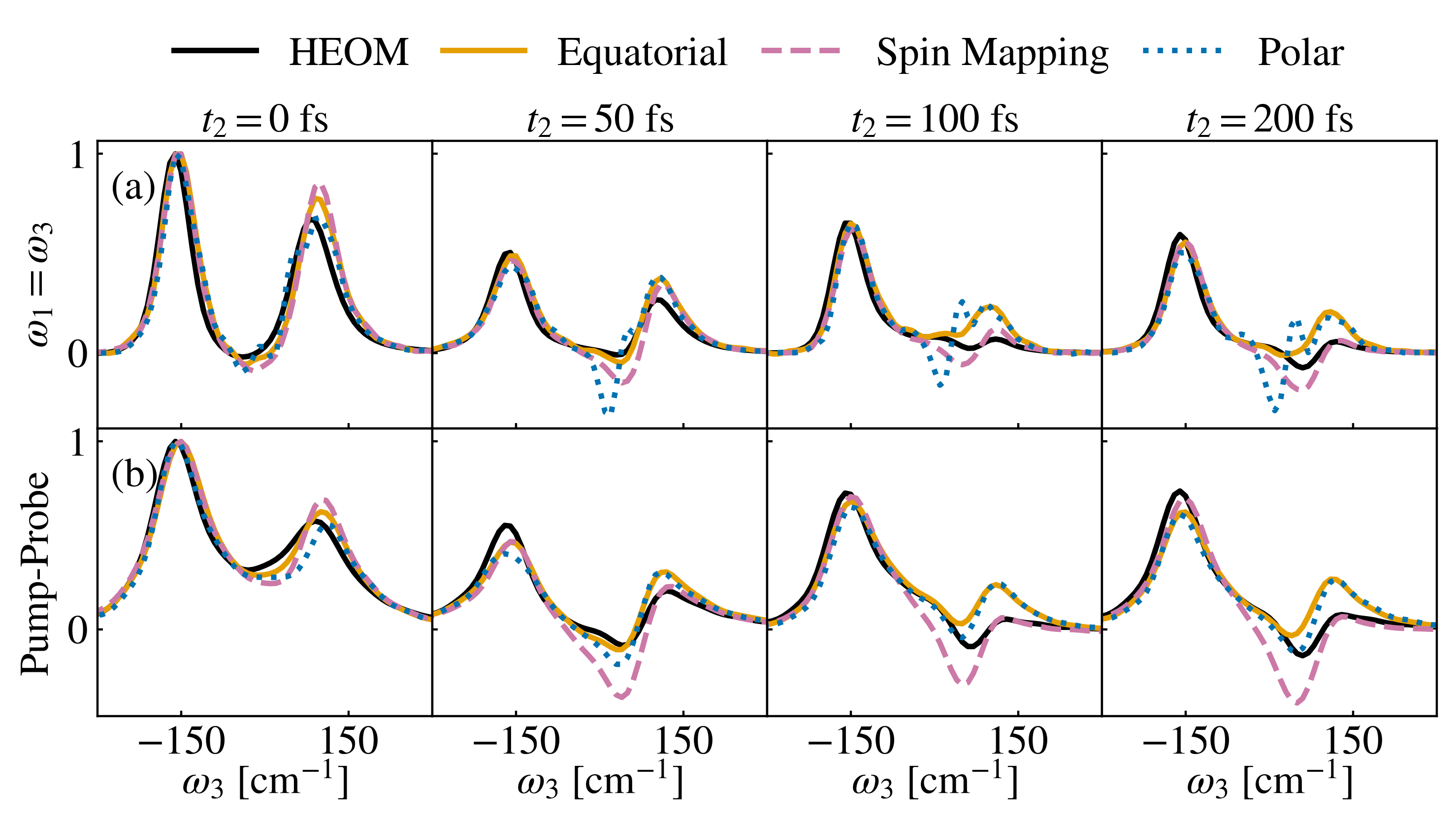}
    \caption{Comparison of polar, equatorial, and spin mapping diagonal cuts (top) and pump-probe spectra (bottom) for the biexciton model with numerically exact HEOM results, as a function of the delay time $t_{2}$. All spectra are normalized to a maximum value of 1 at $t_2=0$.}
    \label{fig:biexc-diag-pp}
\end{figure*}

The top panels of Fig.~\ref{fig:biexc-diag-pp} show diagonal ($\omega_1=\omega_3$) slices through the 2DES spectra, and the bottom panels show the pump-probe spectra, as a function of the delay time $t_2$. The diagonal slices are often used to assess how excitation at an initial frequency ($\omega_1$) correlates with emission at the same frequency ($\omega_3=\omega_1$) following the $t_2$ delay time.
The pump-probe spectrum is related to the integral of the 2DES signal $S(\omega_1, t_2, \omega_3)$ [Eq.~\eqref{eq:ft_2des}] by 
\begin{equation}
    \label{eq:def-pump-probe}
    \sigma_{\rm PP}(t_2, \omega_3) = \int_0^\infty \dd \omega_1 S(\omega_1, t_2, \omega_3).
\end{equation}

For both the diagonal slices and the pump-probe spectra, all the approaches agree well at $t_2=0$~fs, but there are differences in their peak intensities at later times. In particular, the diagonal slice of the polar approach reveals an overestimated negative feature at $\omega_1=0~\cm$ that appears by $t_2=50$~ fs and persists through $t_2=200$~fs. The equatorial and spin-mapping approaches capture this region more accurately; however, the spin-mapping approach overestimates a negative feature around the same frequency in the pump-probe spectrum. It should be noted that this negative region in the pump-probe spectrum is also observed in the HEOM calculation. It arises from the ESA pathways, which transfer energy from a singly excited state to a higher-lying doubly excited state, in contrast to the SE and GSB pathways, which transfer energy from a singly excited state back to the ground state. In other words, the negative intensity in the pump-probe spectrum arises from a physically relevant process that spin mapping simply overemphasizes. However, spin mapping performs significantly better in capturing the intensity of the peak near 150 $\cm$ at longer $t_2$, which both the equatorial and polar approaches overestimate. This deviation from the HEOM results likely arises from the inability of the Ehrenfest methods to correctly capture long-time state populations. This deficiency of Ehrenfest dynamics is well established, as is the fact that it can be exacerbated by zero-point energy leakage due to the Wigner sampling of the bath degrees of freedom, which is appropriate here since $\omega_c=300~\cm > k_{\rm B}T=208~\cm$. We have checked that using classical Boltzmann sampling does not change the results in this case (SI, Sec.~S5), which allows us to identify the incorrect long-time Ehrenfest populations as the source of the error.  

\begin{figure}[b]
    \centering
    \includegraphics[width=0.35\textwidth]{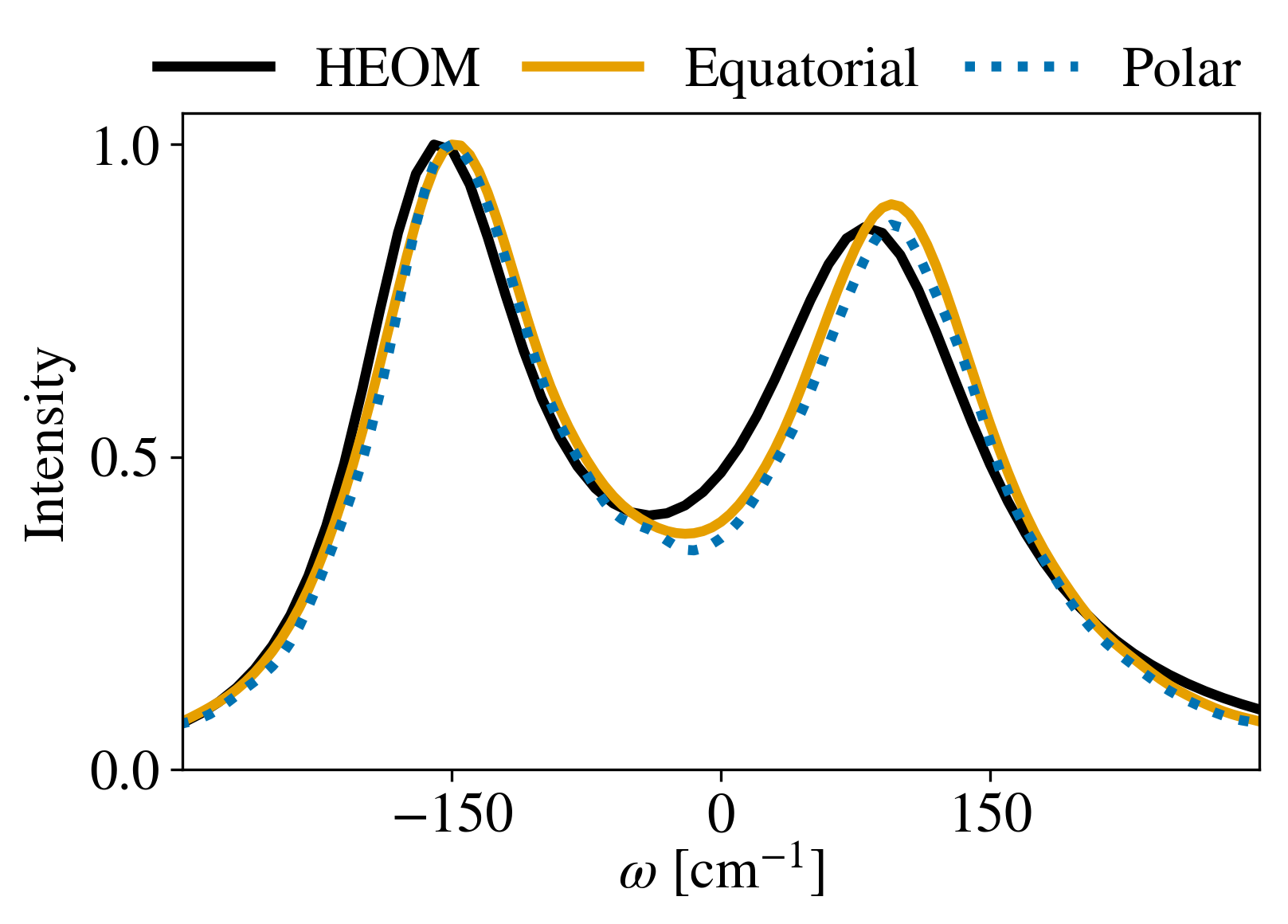}
    \caption{Linear absorption spectra of the biexciton model obtained using the polar and equatorial approaches, compared with the numerically exact HEOM benchmark. The spin-mapping approach is identical to the equatorial approach for the linear spectrum and therefore not shown as a separate curve.}
    \label{fig:biexc-linear}
\end{figure}

Figure~\ref{fig:biexc-linear} shows the linear absorption spectra calculated using the equatorial and polar approaches. Since the linear absorption spectrum does not involve a $t_2$ evolution, the equatorial and spin-mapping approaches give identical results. This spectroscopic observable is known to be easier to simulate correctly using approximate methods\cite{Feth2017, Atsa2023} and, indeed, both the equatorial and polar approaches agree well with the HEOM benchmark and are almost indistinguishable from one another.

\begin{figure}
    \centering
    \includegraphics[width=0.45\textwidth]{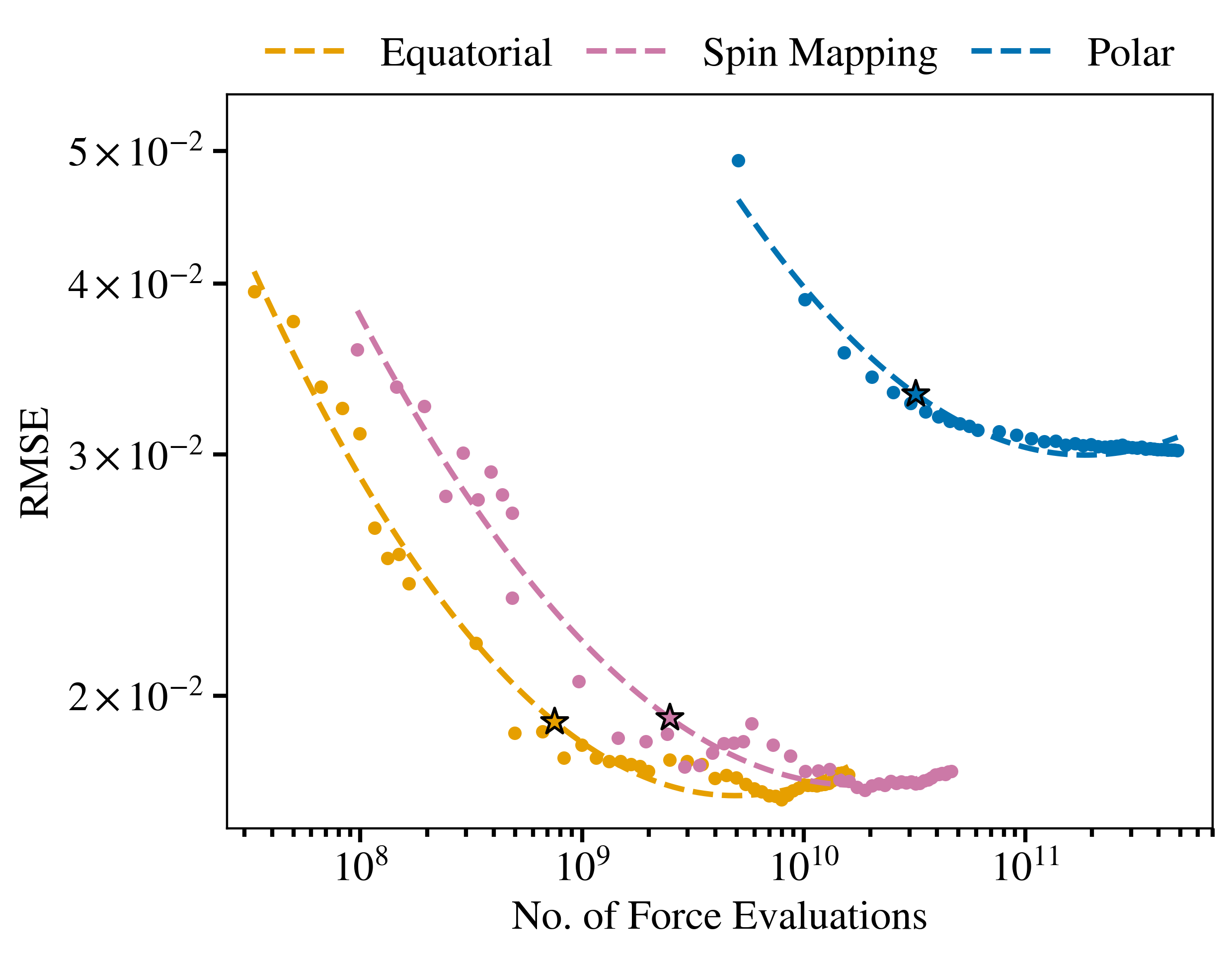}
    \caption{Root mean square errors in the 2DES spectra of the biexciton model relative to the exact HEOM reference, as a function of the number of force evaluations used in each  method. The dashed curves are a guide to the eye and the stars mark the points at which each RMSE is within 10\% of its limiting value.}
    \label{fig:biexc-rmse}
\end{figure}

To provide a quantitative assessment of the accuracy of the methods and how quickly they converge, Fig.~\ref{fig:biexc-rmse} shows the root mean square error (RMSE) in the 2DES spectra compared to the exact HEOM reference of the each of the present approaches as function of the number of force evaluations. The RMSE errors were calculated by averaging over $\Delta S(\omega_1, t_2, \omega_3)^2$ for $\omega_1, ~ \omega_3 \in [-800, 800~\cm]$ and $t_2 \in \{0, 50, 100, 150, 200~\mathrm{fs}\}$ before taking the square root. We use the number of force evaluations as a metric of the overall computational cost required to converge the 2DES spectra because the more commonly used number of trajectories hides the fact that each method splits its trajectories into a different number of pure states after each dipole interaction (see Sec.~\ref{sec:pure-state-methods} and SI, Sec.~S6). In addition, if these methods were to be combined with more advanced (beyond harmonic) descriptions of the electronic surfaces, such as empirical force fields, machine learned potentials, or {\it ab initio} forces, the force evaluations would dominate computational cost. 

A dashed line of best fit is shown for each method in Fig.~\ref{fig:biexc-rmse} as a guide to the eye. Note that, because the reference is the exact HEOM result, the RMSE tends to a non-zero limit which quantifies the accuracy of each method. For the biexciton model, the limiting RMSEs are 0.0175, 0.0176, and 0.0302 for the equatorial, spin-mapping, and polar approaches, respectively. This quantitatively reflects the lower quality of the polar approach and shows that equatorial and spin mapping approaches achieve very similar overall accuracy for this problem. Spin mapping captures the long-time population dynamics more accurately while the equatorial method does a better job of capturing the negative ESA features in Fig.~\ref{fig:biexc-diag-pp}. The star on each fit line in Fig.~\ref{fig:biexc-rmse} marks the point at which the RMSE is within 10\% of its limiting value. These stars correspond to $7.5\times10^8$, $2.5\times10^9$, and $3.2\times10^{10}$ force evaluations, or 4,500, 5,100, and 6,300 trajectories, for the equatorial, spin-mapping, and polar approaches, respectively. While not fully converged, a 10\% error results in spectra that retain all the key features of the fully converged results (SI, Sec.~S7).   

The results in Fig.~\ref{fig:biexc-rmse} confirm that the present (equatorial) version of Ehrenfest dynamics is both more accurate and more efficient than the original (polar) version.\cite{Atsa2023} For each trajectory, we see a factor of 30.6 reduction in the number of force evaluations for the equatorial approach compared to the polar approach. This is slightly below the factor of 32 we anticipated in Sec.~\ref{subsec:theory-eq-decomposition} because the $t_3$ evolution was skipped for all but the few $t_2$ values we plotted and analyzed, which reduced the total number of force evaluations differently for each method (SI, Sec.~S6). Likewise, each spin mapping trajectory was found to require a factor of 3.88 more force evaluations for the SE and ESA pathways than the equatorial approach, just below the estimated factor of $N^2=4$ expected for the biexciton model.

\subsection{Fenna-Matthews-Olson Complex}
\label{subsec:fmo-results}

The parameters of the FMO Hamiltonian and dipole moment operator are given in SI, Sec.~S4. 
Since we have already established that the polar approach is less accurate and significantly more computationally expensive than the equatorial approach, we will only show equatorial and spin mapping results for this model.

The dipole operator depends on the orientation of the FMO complex relative to an incident laser pulse, so to accurately replicate the signal of many proteins dissolved in solution, one must average over a uniform distribution FMO orientations. Here, we assume that all three light pulses are colinear.\cite{Tian2003,Bisw2022} Averaging over orientations then becomes equivalent to averaging over light polarizations as described in the supplementary information (SI, Sec.~S4). The HEOM results shown here were generated by Jonathan Mannouch for $t_2=0$, 200, and 600 fs, and were presented in Ref.~\onlinecite{Rune2022}, which also contains the HEOM simulation details.

\begin{figure}
    \centering\includegraphics[width=0.45\textwidth]{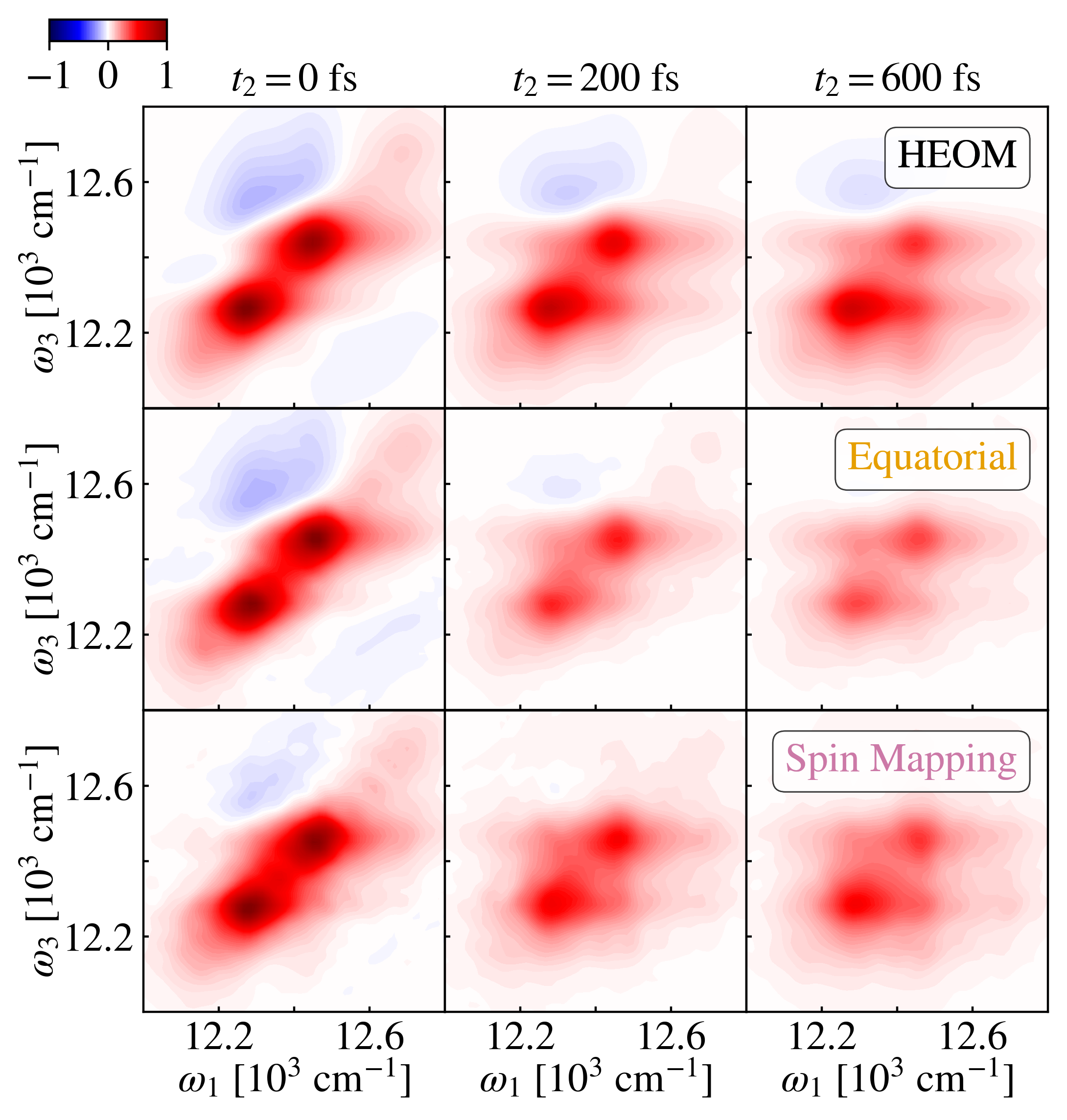}
    \caption{Comparison of equatorial and spin mapping 2DES spectra for the FMO complex with numerically exact HEOM results, as a function of the delay time $t_2$. Both the equatorial and spin-mapping results were calculated using 20,800 trajectories.}
    \label{fig:fmo-spectra}
\end{figure}

Figure~\ref{fig:fmo-spectra} shows the 2DES spectra of the FMO model at 300\,K obtained using the equatorial and spin-mapping approaches. The energies of the 7 states in the FMO model are grouped such that only two dominant peaks are resolvable in the linear and 2DES spectra at this temperature. Both the equatorial method and spin mapping are seen to give reasonable overall agreement with the HEOM spectrum. However, the survival of the spectral intensity is clearly underestimated by both methods at long delay times, as are the (negative) intensities of the weak off-diagonal HEOM minima. The first of these problems is more noticeable in the equatorial Ehrenfest results and the second is more apparent in spin mapping.

\begin{figure}[b]
    \centering
    \includegraphics[width=0.45\textwidth]{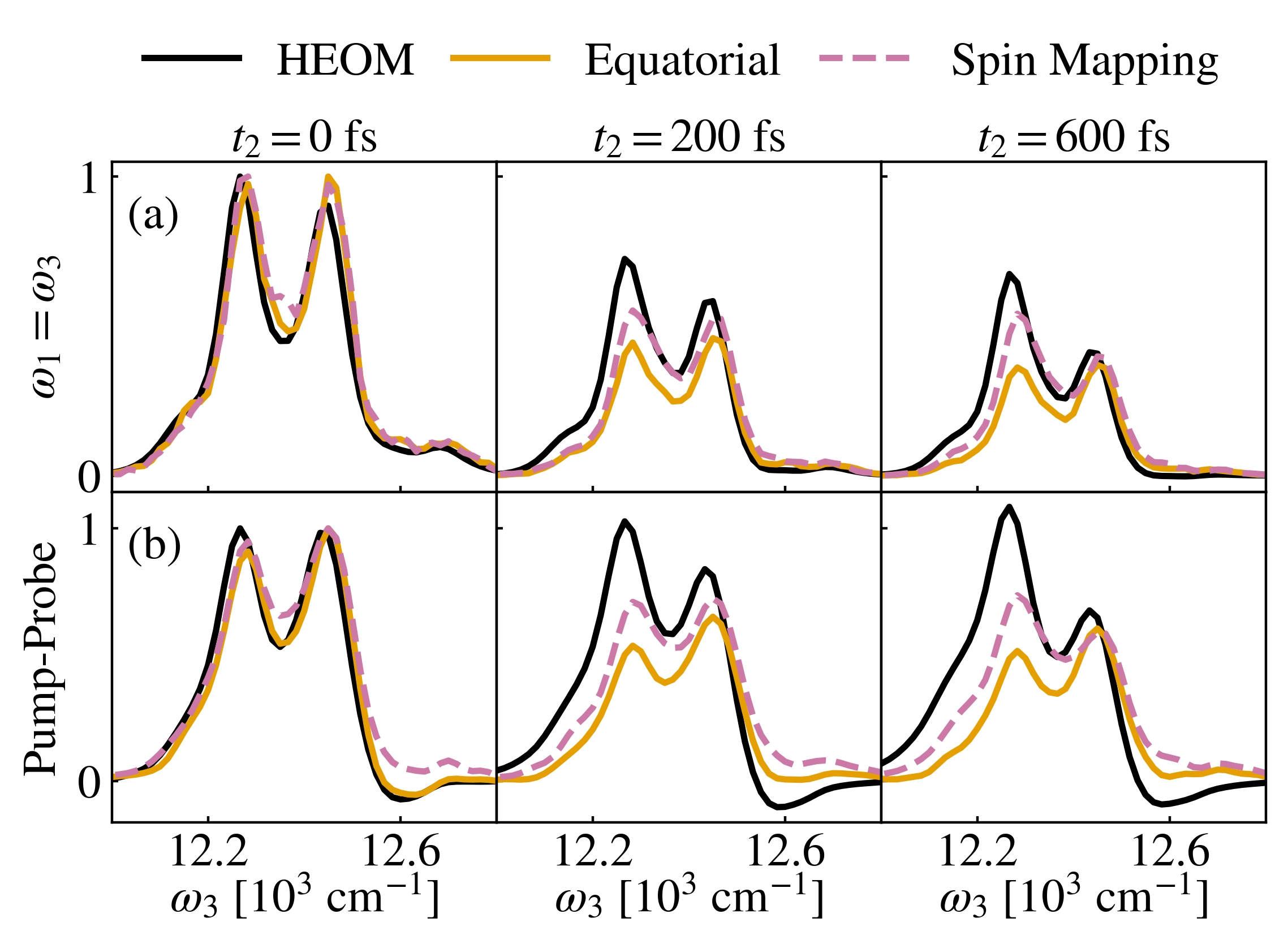}
    \caption{Comparison of equatorial and spin mapping diagonal cuts (top) and pump-probe spectra (bottom) for the FMO model with numerically exact HEOM results, as a function of the delay time $t_{2}$.}
    \label{fig:fmo-diag-pp}
\end{figure}

Figure~\ref{fig:fmo-diag-pp} shows the diagonal ($\omega_1=\omega_3$) slices and pump-probe spectra obtained from the FMO 2DES. The diagonal slices from both approaches are in good agreement with the HEOM benchmark at $t_2 = 0$, but at longer delay times, both the equatorial and spin-mapping approaches underestimate the signal intensity, especially in the peak at $\sim$12,200~$\cm$. In both the diagonal slices and the pump-probe spectra, this disagreement with HEOM is less pronounced in spin-mapping than in the equatorial approach, consistent with results for the biexciton model in Fig.~\ref{fig:biexc-diag-pp}. The problematic populations obtained from the equatorial approach are again due to the failure of Ehrenfest dynamics to satisfy detailed balance rather than zero point energy leakage: replacing the Wigner bath with a classical Boltzmann distribution gives graphically indistinguishable results (SI, Sec.~S5). 

\rev{We should point out in passing that the spin-mapping results do not agree precisely with the equatorial Ehrenfest results in Fig.~4 at $t_2=0$, which is before the spin mapping variables have even begun to evolve. This is not a convergence issue because both results are well converged: it is an undesirable feature of the present spin-mapping implementation. The doubly focused sampling algorithm described in Sec.~III.C and the subsequent regeneration of pure states from Eq.~(27) convert the original Ehrenfest state $|\phi_j\rangle\langle\phi_j|$ into a sum of $M^2$ new pure states. These then evolve differently from the Ehrenfest state during $t_3$, even when no evolution has occurred during $t_2$. The difference this makes to the resulting spectra is rather minor. It is clearest in the $t_2=0$ pump-probe results for the FMO model in Fig.~8, but barely noticeable in either the two $t_2=0$ panels for the biexciton model in Fig.~4. The fact that there is a difference at all is interesting nevertheless, as it suggests that it may be possible to develop a better version of spin mapping than the one we have implemented here.}

\begin{figure}
    \centering
    \includegraphics[width=0.45\textwidth]{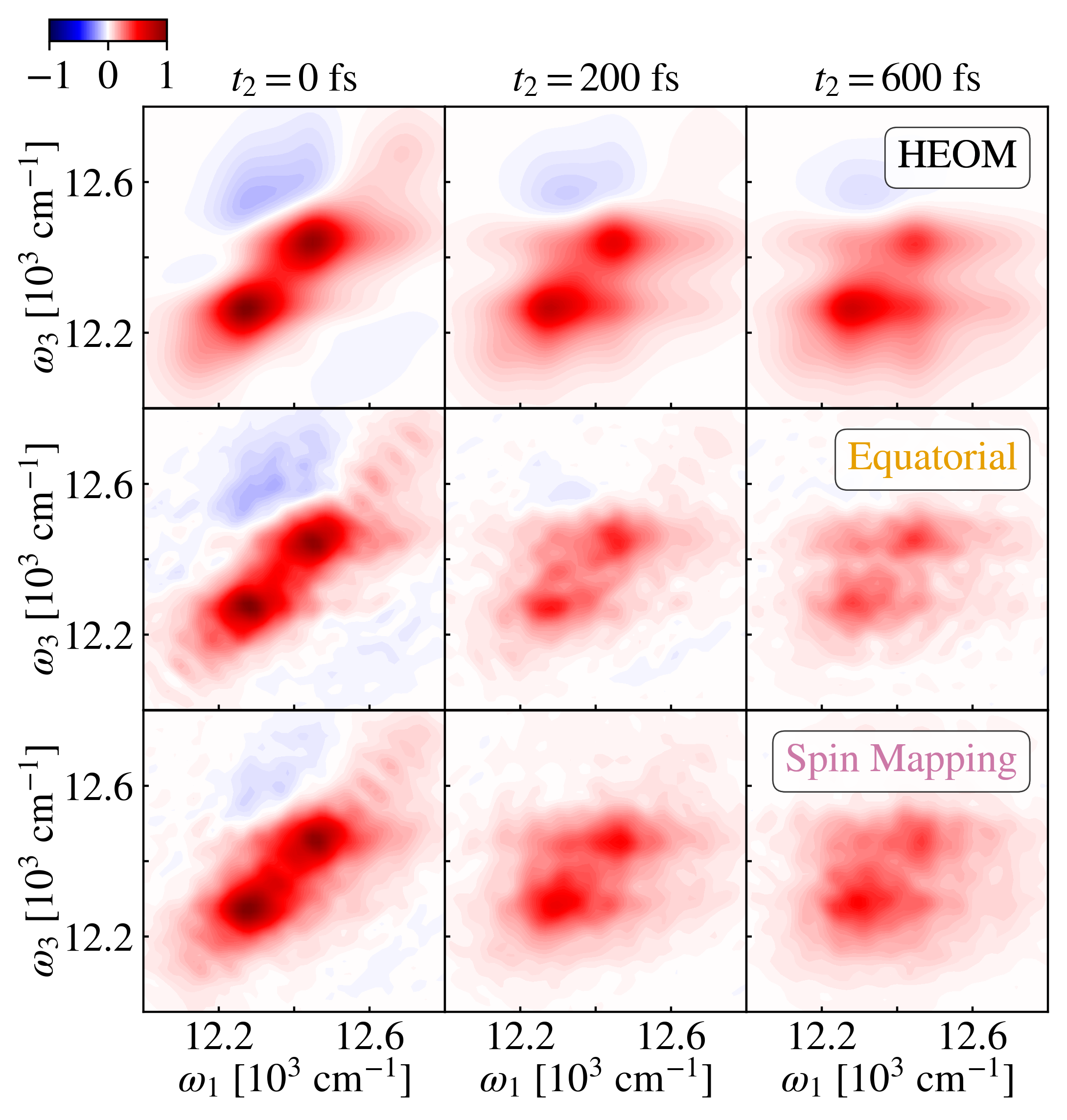}
    \caption{As in Fig.~\ref{fig:fmo-spectra}, but with just 2,000 trajectories in the equatorial and spin mapping calculations.}
    \label{fig:fmo-2d-2000traj}
\end{figure}

To demonstrate the efficiency of the equatorial and spin-mapping approaches, Fig.~\ref{fig:fmo-2d-2000traj} shows the 2DES spectra for the FMO model calculated using only 2,000 trajectories. While significantly noisier than the fully converged results in Fig.~\ref{fig:fmo-spectra}, which utilized 20,800 trajectories, these spectra still retain the key features. SI, Sec.~S7 shows that the $\omega_1=\omega_3$ slice and pump-probe spectra also retain the same structure and patterns with this smaller number of trajectories, which are  thus able to provide almost the same physical insights as would be obtained from a fully converged calculation. 

\begin{figure}
    \centering
    \includegraphics[width=0.45\textwidth]{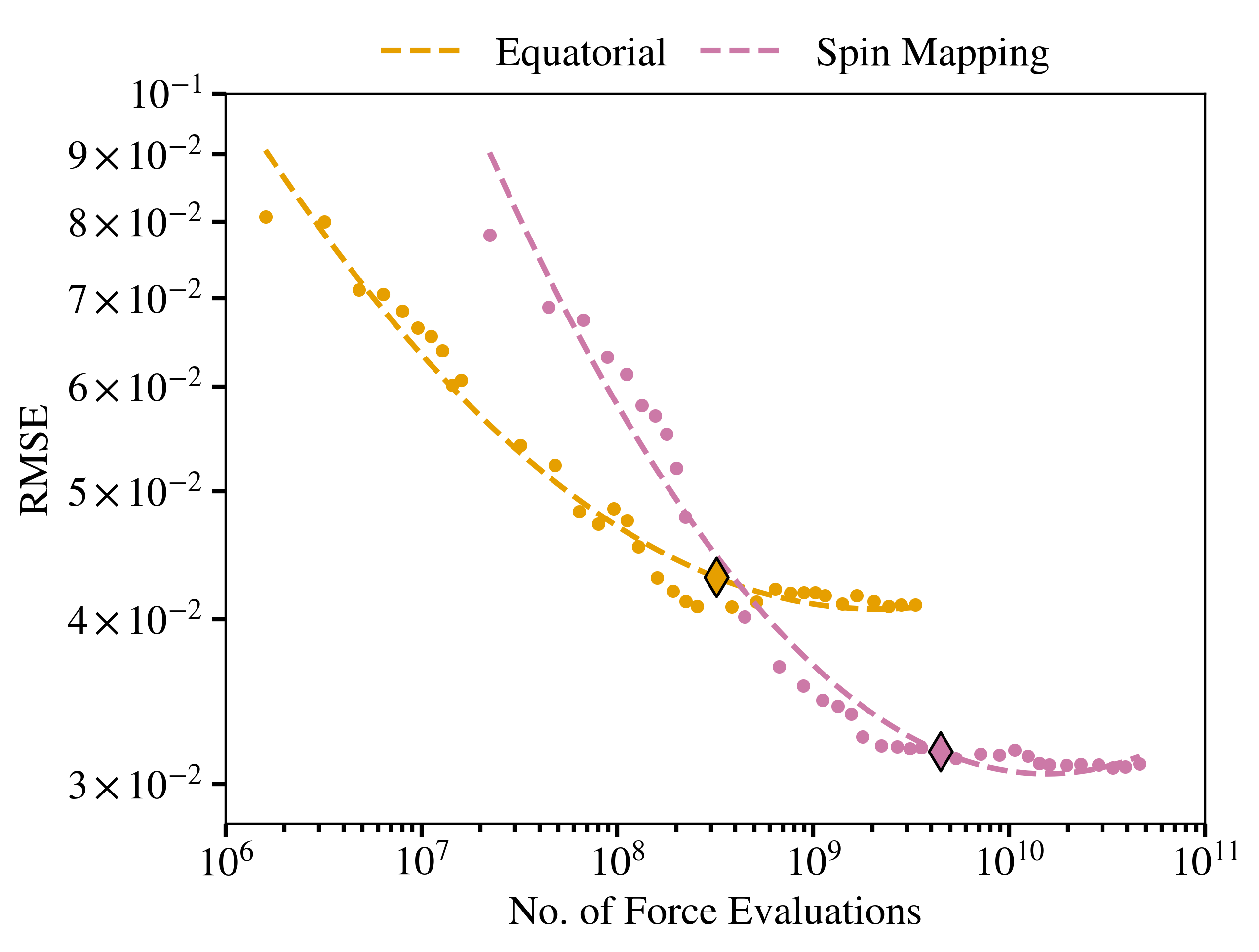}
    \caption{Root mean square errors in the 2DES spectra of the FMO model relative to the exact HEOM reference, as a function of the number of force evaluations used in the equatorial and spin mapping methods. The dashed curves are a guide to the eye and the diamonds mark the number of force evaluations needed to run 2,000  trajectories.}
    \label{fig:fmo-2d-convergence}
\end{figure}

Figure~\ref{fig:fmo-2d-convergence} shows the RMSE convergence of the polar and spin mapping approaches for FMO as the number of force evaluations is increased. As discussed for the biexciton model, force evaluation is the most expensive component of simulations performed with atomistic potentials, so the $x$ axis of Fig.~\ref{fig:fmo-2d-convergence} acts as a proxy for the total cost of such simulations. Because of the differences in trajectory decomposition in the equatorial and spin-mapping approaches, a different number of force evaluations is required for each. For FMO, our spin mapping calculations required 26.9 times force evaluations than our equatorial calculations, which was below the expected $N^2=49$-fold increase for the reasons described in SI, Sec.~S6. The diamond symbols in Fig.~\ref{fig:fmo-2d-convergence} correspond to using 2,000 trajectories for the equatorial and spin-mapping approaches, which then involve $\sim3.2\times10^8$ and $\sim8.6\times10^9$ force evaluations, respectively. This many force evaluations would be prohibitively expensive using most \textit{ab initio} electronic structure methods, but is potentially within in the realm of feasibility for machine learned potentials and empirical force fields.

\rev{
\subsection{Comparison with the Mean Classical Path Approximation}}

\rev{We have already argued that the mean classical path and equatorial Ehrenfest approaches are closely related. For the linear absorption spectrum, they are equivalent. However, this is no longer true for 2DES spectra. Since the mean classical path approach does not invoke a pure state decomposition of the coherence at $t_2=0$, both the $t_2$ evolution and the subsequent $t_3$ evolution differ in the two methods. We shall therefore now compare the mean classical path and equatorial Ehrenfest spectra for the FMO model and present a similar comparison for the biexciton model in the supplementary material (SI, Sec.~S8).}

\begin{figure}[t]
	\centering
	\includegraphics[width=.45\textwidth]{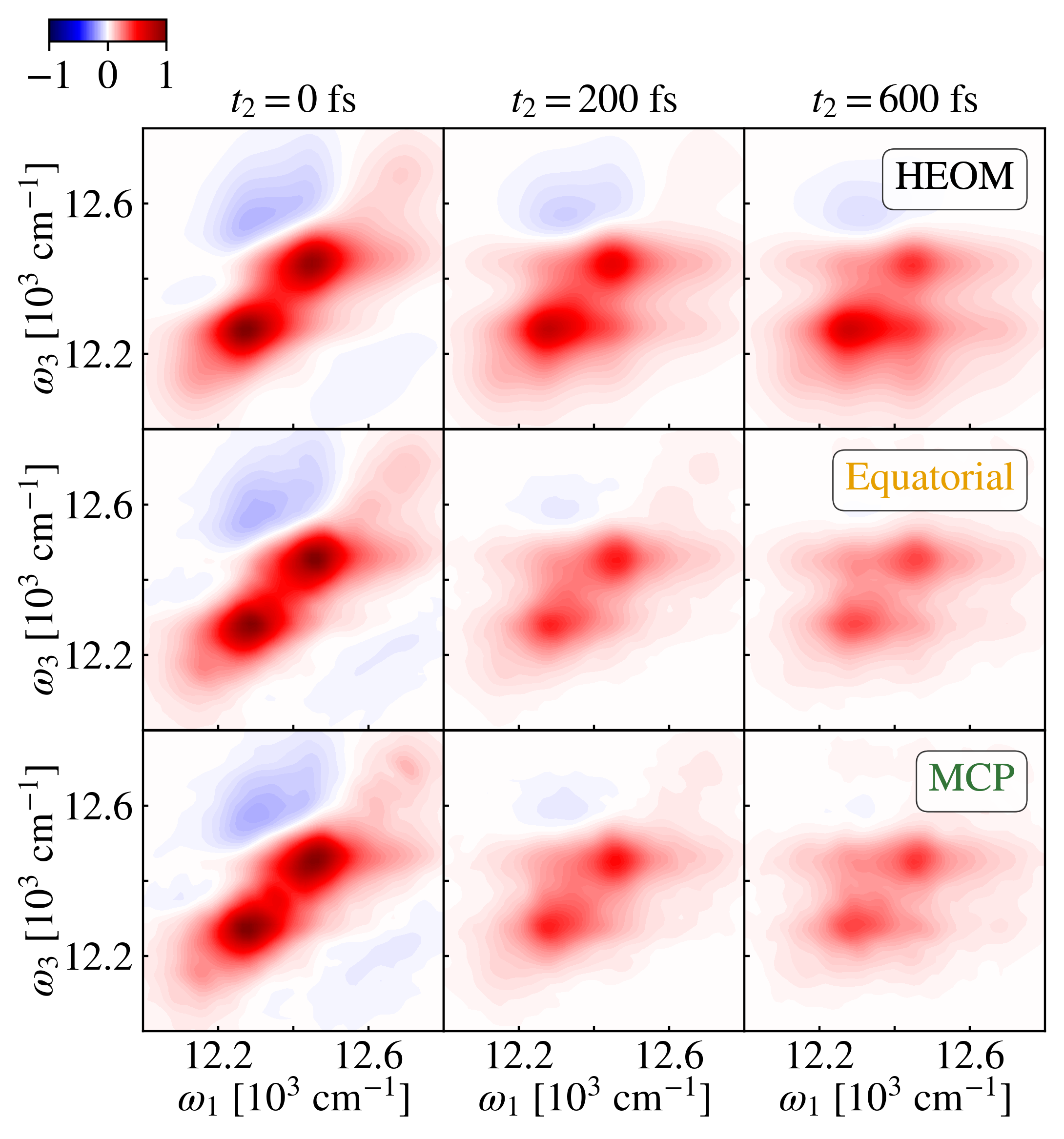}
	\caption{\rev{Comparison of equatorial Ehrenfest and mean classical path 2DES spectra of the FMO complex with numerically exact HEOM results, as a function of the delay time $t_2$.
   Both sets of quantum-classical results were obtained with 21,000 trajectories.}}
    \label{fig:mcp_eq_2d_fmo}
\end{figure}

\begin{figure}[t]
	\centering
	\includegraphics[width=.45\textwidth]{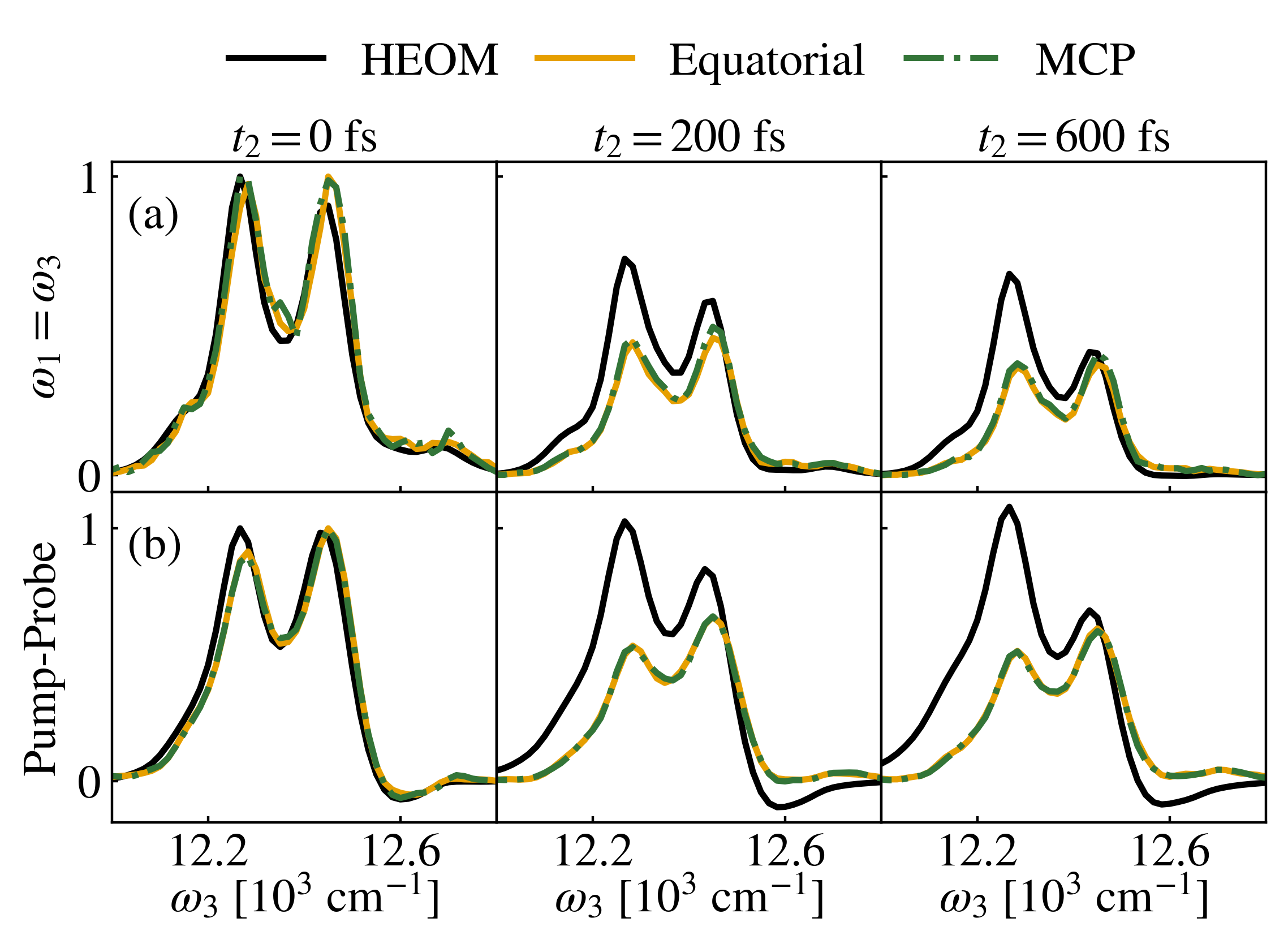}
	\caption{\rev{Comparison of equatorial Ehrenfest and mean classical path diagonal cuts (top) and pump-probe spectra (bottom) for FMO, at different delay times $t_2$.}}
    \label{fig:mcp_eq_diag_pp_fmo}
\end{figure}

\rev{Fig.~\ref{fig:mcp_eq_2d_fmo} compares the 2DES spectra of these methods with the exact HEOM benchmark results for the FMO complex. One sees that the two variants of Ehrenfest dynamics produce spectra that are almost indistinguishable. The diagonal slices and the pump-probe spectra in Fig.~\ref{fig:mcp_eq_diag_pp_fmo} paint the same picture, as do the results for the biexciton model in SI, Sec.~S8. This is interesting because we had no reason to believe it would be the case before doing these calculations. We have shown analytically in Sec.~III.B that the equatorial Ehrenfest and mean classical path approximations are equivalent when the states $|a\rangle$ and $|b\rangle$ in the initial coherence are orthogonal -- as is the case during the $t_1$ evolution -- and now we have found numerically that they give almost identical results even when the states $|a\rangle$ and $|b\rangle$ are non-orthogonal -- as is generally the case during the $t_2$ evolution. The two methods do treat the evolution differently in this case -- via the standard Ehrenfest evolution of four pure states or the somewhat less well justified (but nevertheless balanced and symmetric) Ehrenfest evolution of a single coherence. However, this difference does not appear to matter in practice for either of the standard model problems we have considered in this paper.} 

\rev{Since the mean classical path approximation is just as accurate as equatorial Ehrenfest for these model problems, since it is the cheapest of the methods we have considered, and since it is also the simplest to implement, we expect it will become the most popular method going forward. Especially when the forces on the nuclei are obtained from (machine-learned) {\em ab initio} potentials, which will make the calculations more expensive than those we have performed here. Given the increased cost of each force evaluation, the relatively modest improvement in accuracy afforded by spin mapping is unlikely to be worth the extra effort it involves (see Fig.~10).} 

\section{Conclusion}

In this paper, we have shown how one can develop a more accurate and computationally efficient pure-state Ehrenfest method\cite{Atsa2023} by decomposing coherences into equatorial pure states. Unlike the original polar pure states,\cite{Atsa2023} these equatorial pure states are compatible with a resummation trick that significantly streamlines the calculation. We have benchmarked the accuracy of this approach by using it to simulate the 2DES and pump probe spectra of two Frenkel exciton models, a 2-site biexciton model and a 7-site FMO complex. For the biexciton model, we have shown that the number of force evaluations is reduced by a factor approaching 32 compared with the previous polar Ehrenfest approach,\cite{Atsa2023} while at the same time giving more accurate results. 

All methods based on Ehrenfest dynamics, including the present pure state methods and the mean classical path approximation, have well known issues associated with their failure to satisfy detailed balance. To address this problem, we have used the equatorial pure state framework to introduce spin mapping variables for the delay time evolution, and shown that these variables do indeed somewhat improve the accuracy of the computed spectra, although at an additional computational cost.

\rev{More importantly, we have shown that the improved (equatorial) pure state Ehrenfest method is exactly equivalent to making the mean classical path approximation to the initial evolution of the coherences in the nonlinear response functions, and that it gives almost identical results to this approximation when used throughout. This provides some fresh justification for the mean classical path approximation:\cite{Vegt2013} since it agrees so well with the standard Ehrenfest evolution of pure states, the way this approximation evolves coherences cannot be introducing any larger error than that in Ehrenfest dynamics itself (for the model problems we have considered here). This is important because the mean classical path approach is both the simplest and the cheapest of the methods we have considered in this paper.}

We have shown that spectra good enough for analysis purposes can be obtained from just 2,000 equatorial Ehrenfest or mean classical path trajectories (see Fig.~9 and SI, Sec.~S8). This is true for both of the models we have considered, which require around $3\times10^8$ pure state force evaluations to reach this level of accuracy (SI, Sec.~S7). Recently, a machine learned potential trained on {\em ab initio} data using an equivariant transformer\cite{Vasw2017, Thol2021} architecture was found to generate energies and forces at a rate of 25 configurations per second, using the TorchMD-Net package\cite{Pela2024} on a single GPU, for a Nile blue chromophore solvated by 175 ethanol molecules (1619 atoms).\cite{Kell2025} Calculating $3\times 10^8$ forces from this machine learned potential would take over a year on a single GPU, but with 24 GPUs the calculation could be completed within 3 weeks. Thus, either the equatorial Ehrenfest method, or the simpler and cheaper mean classical path approximation, in conjunction with ongoing advancements in machine learned potentials, could offer a path to fully atomistic simulations of condensed phase 2DES spectra based on {\em ab initio} data. 

We have, of course, not been able to eliminate the failures of Ehrenfest dynamics itself. Our introduction of spin mapping variables does not fully satisfy detailed balance and therefore still gives inaccurate peak intensities in the 2DES spectrum at long time delays. In some systems, this problem could be exacerbated by zero-point energy leakage from high-frequency bath modes when using an initial Wigner distribution for the vibrational bath. The failure of spin mapping to exactly satisfy detailed balance could be fixed by replacing the delay time evolution with a method that does, such as MASH.\cite{Rune2023} The zero-point energy leakage from Wigner initial conditions could be addressed by using a variational polaron transformation\cite{Yark1976,Wang2020} to eliminate the high-frequency bath modes from the calculation, as has recently been done in a study of exciton energy transfer in biological light-harvesting complexes.~\cite{Rune2025} We may explore these possibilities in future work.

\section*{Supplementary Material}

The supplementary material contains the following sections. S1: An analysis of the wave function evolution pathways in the polar and equatorial Ehrenfest methods. \rev{S2: An analysis of how the symmetry of the response functions affects the mean classical path calculation.} S3: A derivation of the `doubly focused' sampling of spin-mapping variables. S4: A summary of the simulation details for the biexciton and FMO models. S5: A comparison of Wigner and classical initial conditions for each model. S6: A discussion of the numerical effort required by the pure state Ehrenfest and spin-mapping methods. S7: Various convergence tests. \rev{S8: More comparisons of equatorial Ehrenfest and mean classical path results.}

\section*{Acknowledgments}
We thank Jonathan Mannouch for providing the HEOM spectra for the FMO model, for his help in understanding the rotational averages described in SI, Sec.~S4, and for many other helpful discussions. This work was funded by the National Science Foundation Grant No. CHE-2154291 to T.E.M. Annina Z. Lieberherr was supported by a Berrow Foundation Lord Florey Scholarship. Joseph Kelly was supported by a John Stauffer Memorial Award. Johan Runeson was supported by a Mobility Fellowship from the Swiss National Science Foundation (grant no. P500PN-206641/1).

\section*{Author declarations}

\subsection*{Conflict of interest}

The authors have no conflicts to declare.

\section*{Data availability}

The data that support the findings of this study are available within the article and its supplementary material.

\bibliography{nonlinear}

\begin{thebibliography}{7}%
\makeatletter
\providecommand \@ifxundefined [1]{%
 \@ifx{#1\undefined}
}%
\providecommand \@ifnum [1]{%
 \ifnum #1\expandafter \@firstoftwo
 \else \expandafter \@secondoftwo
 \fi
}%
\providecommand \@ifx [1]{%
 \ifx #1\expandafter \@firstoftwo
 \else \expandafter \@secondoftwo
 \fi
}%
\providecommand \natexlab [1]{#1}%
\providecommand \enquote  [1]{``#1''}%
\providecommand \bibnamefont  [1]{#1}%
\providecommand \bibfnamefont [1]{#1}%
\providecommand \citenamefont [1]{#1}%
\providecommand \href@noop [0]{\@secondoftwo}%
\providecommand \href [0]{\begingroup \@sanitize@url \@href}%
\providecommand \@href[1]{\@@startlink{#1}\@@href}%
\providecommand \@@href[1]{\endgroup#1\@@endlink}%
\providecommand \@sanitize@url [0]{\catcode `\\12\catcode `\$12\catcode `\&12\catcode `\#12\catcode `\^12\catcode `\_12\catcode `\%12\relax}%
\providecommand \@@startlink[1]{}%
\providecommand \@@endlink[0]{}%
\providecommand \url  [0]{\begingroup\@sanitize@url \@url }%
\providecommand \@url [1]{\endgroup\@href {#1}{\urlprefix }}%
\providecommand \urlprefix  [0]{URL }%
\providecommand \Eprint [0]{\href }%
\providecommand \doibase [0]{http://dx.doi.org/}%
\providecommand \selectlanguage [0]{\@gobble}%
\providecommand \bibinfo  [0]{\@secondoftwo}%
\providecommand \bibfield  [0]{\@secondoftwo}%
\providecommand \translation [1]{[#1]}%
\providecommand \BibitemOpen [0]{}%
\providecommand \bibitemStop [0]{}%
\providecommand \bibitemNoStop [0]{.\EOS\space}%
\providecommand \EOS [0]{\spacefactor3000\relax}%
\providecommand \BibitemShut  [1]{\csname bibitem#1\endcsname}%
\let\auto@bib@innerbib\@empty
\bibitem [{\citenamefont {Runeson}\ and\ \citenamefont {Richardson}(2020)}]{Rune2020}%
  \BibitemOpen
  \bibfield  {author} {\bibinfo {author} {\bibfnamefont {J.~E.}\ \bibnamefont {Runeson}}\ and\ \bibinfo {author} {\bibfnamefont {J.~O.}\ \bibnamefont {Richardson}},\ }\bibfield  {title} {\enquote {\bibinfo {title} {Generalized spin mapping for quantum-classical dynamics},}\ }\href {\doibase 10.1063/1.5143412} {\bibfield  {journal} {\bibinfo  {journal} {J. Chem. Phys.}\ }\textbf {\bibinfo {volume} {152}},\ \bibinfo {pages} {084110} (\bibinfo {year} {2020})}\BibitemShut {NoStop}%
\bibitem [{\citenamefont {Wang}, \citenamefont {Thoss},\ and\ \citenamefont {Miller}(2001)}]{Wang2001}%
  \BibitemOpen
  \bibfield  {author} {\bibinfo {author} {\bibfnamefont {H.}~\bibnamefont {Wang}}, \bibinfo {author} {\bibfnamefont {M.}~\bibnamefont {Thoss}}, \ and\ \bibinfo {author} {\bibfnamefont {W.~H.}\ \bibnamefont {Miller}},\ }\bibfield  {title} {\enquote {\bibinfo {title} {Systematic convergence in the dynamical hybrid approach for complex systems: A numerically exact methodology},}\ }\href {\doibase 10.1063/1.1385561} {\bibfield  {journal} {\bibinfo  {journal} {J. Chem. Phys.}\ }\textbf {\bibinfo {volume} {115}},\ \bibinfo {pages} {2979–2990} (\bibinfo {year} {2001})}\BibitemShut {NoStop}%
\bibitem [{\citenamefont {Craig}, \citenamefont {Thoss},\ and\ \citenamefont {Wang}(2007)}]{Crai2007}%
  \BibitemOpen
  \bibfield  {author} {\bibinfo {author} {\bibfnamefont {I.~R.}\ \bibnamefont {Craig}}, \bibinfo {author} {\bibfnamefont {M.}~\bibnamefont {Thoss}}, \ and\ \bibinfo {author} {\bibfnamefont {H.}~\bibnamefont {Wang}},\ }\bibfield  {title} {\enquote {\bibinfo {title} {Proton transfer reactions in model condensed-phase environments: Accurate quantum dynamics using the multilayer multiconfiguration time-dependent {H}artree approach},}\ }\href {\doibase 10.1063/1.2772265} {\bibfield  {journal} {\bibinfo  {journal} {J. Chem. Phys.}\ }\textbf {\bibinfo {volume} {127}},\ \bibinfo {pages} {144503} (\bibinfo {year} {2007})}\BibitemShut {NoStop}%
\bibitem [{\citenamefont {Atsango}, \citenamefont {Montoya-Castillo},\ and\ \citenamefont {Markland}(2023)}]{Atsa2023}%
  \BibitemOpen
  \bibfield  {author} {\bibinfo {author} {\bibfnamefont {A.~O.}\ \bibnamefont {Atsango}}, \bibinfo {author} {\bibfnamefont {A.}~\bibnamefont {Montoya-Castillo}}, \ and\ \bibinfo {author} {\bibfnamefont {T.~E.}\ \bibnamefont {Markland}},\ }\bibfield  {title} {\enquote {\bibinfo {title} {An accurate and efficient {E}hrenfest dynamics approach for calculating linear and nonlinear electronic spectra},}\ }\href {\doibase 10.1063/5.0138671} {\bibfield  {journal} {\bibinfo  {journal} {J. Chem. Phys.}\ }\textbf {\bibinfo {volume} {158}},\ \bibinfo {pages} {074107} (\bibinfo {year} {2023})}\BibitemShut {NoStop}%
\bibitem [{\citenamefont {Berkelbach}\ \emph {et~al.}(2020)\citenamefont {Berkelbach}, \citenamefont {Fetherolf}, \citenamefont {Shih},\ and\ \citenamefont {{Iansdunn}}}]{Berk2020}%
  \BibitemOpen
  \bibfield  {author} {\bibinfo {author} {\bibfnamefont {T.}~\bibnamefont {Berkelbach}}, \bibinfo {author} {\bibfnamefont {J.}~\bibnamefont {Fetherolf}}, \bibinfo {author} {\bibfnamefont {P.}~\bibnamefont {Shih}}, \ and\ \bibinfo {author} {\bibnamefont {{Iansdunn}}},\ }\href {\doibase 10.5281/ZENODO.4015527} {\enquote {\bibinfo {title} {berkelbach-group/pyrho v1.0},}\ } (\bibinfo {year} {2020})\BibitemShut {NoStop}%
\bibitem [{\citenamefont {Kramer}\ \emph {et~al.}(2018)\citenamefont {Kramer}, \citenamefont {Noack}, \citenamefont {Reinefeld}, \citenamefont {Rodríguez},\ and\ \citenamefont {Zelinskyy}}]{Kram2018}%
  \BibitemOpen
  \bibfield  {author} {\bibinfo {author} {\bibfnamefont {T.}~\bibnamefont {Kramer}}, \bibinfo {author} {\bibfnamefont {M.}~\bibnamefont {Noack}}, \bibinfo {author} {\bibfnamefont {A.}~\bibnamefont {Reinefeld}}, \bibinfo {author} {\bibfnamefont {M.}~\bibnamefont {Rodríguez}}, \ and\ \bibinfo {author} {\bibfnamefont {Y.}~\bibnamefont {Zelinskyy}},\ }\bibfield  {title} {\enquote {\bibinfo {title} {Efficient calculation of open quantum system dynamics and time‐resolved spectroscopy with distributed memory {HEOM} ({DM‐HEOM})},}\ }\href {\doibase 10.1002/jcc.25354} {\bibfield  {journal} {\bibinfo  {journal} {J. Comput. Chem.}\ }\textbf {\bibinfo {volume} {39}},\ \bibinfo {pages} {1779--1794} (\bibinfo {year} {2018})}\BibitemShut {NoStop}%
\bibitem [{\citenamefont {Hein}\ \emph {et~al.}(2012)\citenamefont {Hein}, \citenamefont {Kreisbeck}, \citenamefont {Kramer},\ and\ \citenamefont {Rodríguez}}]{Hein2012}%
  \BibitemOpen
  \bibfield  {author} {\bibinfo {author} {\bibfnamefont {B.}~\bibnamefont {Hein}}, \bibinfo {author} {\bibfnamefont {C.}~\bibnamefont {Kreisbeck}}, \bibinfo {author} {\bibfnamefont {T.}~\bibnamefont {Kramer}}, \ and\ \bibinfo {author} {\bibfnamefont {M.}~\bibnamefont {Rodríguez}},\ }\bibfield  {title} {\enquote {\bibinfo {title} {Modelling of oscillations in two-dimensional echo-spectra of the {F}enna–{M}atthews–{O}lson complex},}\ }\href {\doibase 10.1088/1367-2630/14/2/023018} {\bibfield  {journal} {\bibinfo  {journal} {New J. Phys.}\ }\textbf {\bibinfo {volume} {14}},\ \bibinfo {pages} {023018} (\bibinfo {year} {2012})}\BibitemShut {NoStop}%
\end{thebibliography}%


\begin{thebibliography}{59}%
\makeatletter
\providecommand \@ifxundefined [1]{%
 \@ifx{#1\undefined}
}%
\providecommand \@ifnum [1]{%
 \ifnum #1\expandafter \@firstoftwo
 \else \expandafter \@secondoftwo
 \fi
}%
\providecommand \@ifx [1]{%
 \ifx #1\expandafter \@firstoftwo
 \else \expandafter \@secondoftwo
 \fi
}%
\providecommand \natexlab [1]{#1}%
\providecommand \enquote  [1]{``#1''}%
\providecommand \bibnamefont  [1]{#1}%
\providecommand \bibfnamefont [1]{#1}%
\providecommand \citenamefont [1]{#1}%
\providecommand \href@noop [0]{\@secondoftwo}%
\providecommand \href [0]{\begingroup \@sanitize@url \@href}%
\providecommand \@href[1]{\@@startlink{#1}\@@href}%
\providecommand \@@href[1]{\endgroup#1\@@endlink}%
\providecommand \@sanitize@url [0]{\catcode `\\12\catcode `\$12\catcode `\&12\catcode `\#12\catcode `\^12\catcode `\_12\catcode `\%12\relax}%
\providecommand \@@startlink[1]{}%
\providecommand \@@endlink[0]{}%
\providecommand \url  [0]{\begingroup\@sanitize@url \@url }%
\providecommand \@url [1]{\endgroup\@href {#1}{\urlprefix }}%
\providecommand \urlprefix  [0]{URL }%
\providecommand \Eprint [0]{\href }%
\providecommand \doibase [0]{http://dx.doi.org/}%
\providecommand \selectlanguage [0]{\@gobble}%
\providecommand \bibinfo  [0]{\@secondoftwo}%
\providecommand \bibfield  [0]{\@secondoftwo}%
\providecommand \translation [1]{[#1]}%
\providecommand \BibitemOpen [0]{}%
\providecommand \bibitemStop [0]{}%
\providecommand \bibitemNoStop [0]{.\EOS\space}%
\providecommand \EOS [0]{\spacefactor3000\relax}%
\providecommand \BibitemShut  [1]{\csname bibitem#1\endcsname}%
\let\auto@bib@innerbib\@empty
\bibitem [{\citenamefont {Hybl}\ \emph {et~al.}(1998)\citenamefont {Hybl}, \citenamefont {Albrecht}, \citenamefont {Faeder},\ and\ \citenamefont {Jonas}}]{Hybl1998}%
  \BibitemOpen
  \bibfield  {author} {\bibinfo {author} {\bibfnamefont {J.~D.}\ \bibnamefont {Hybl}}, \bibinfo {author} {\bibfnamefont {A.~W.}\ \bibnamefont {Albrecht}}, \bibinfo {author} {\bibfnamefont {S.~M.~G.}\ \bibnamefont {Faeder}}, \ and\ \bibinfo {author} {\bibfnamefont {D.~M.}\ \bibnamefont {Jonas}},\ }\bibfield  {title} {\enquote {\bibinfo {title} {Two-dimensional electronic spectroscopy},}\ }\href {\doibase 10.1016/S0009-2614(98)01140-3} {\bibfield  {journal} {\bibinfo  {journal} {Chem. Phys. Lett.}\ }\textbf {\bibinfo {volume} {297}},\ \bibinfo {pages} {307--313} (\bibinfo {year} {1998})}\BibitemShut {NoStop}%
\bibitem [{\citenamefont {Mukamel}(2000)}]{Muka2000}%
  \BibitemOpen
  \bibfield  {author} {\bibinfo {author} {\bibfnamefont {S.}~\bibnamefont {Mukamel}},\ }\bibfield  {title} {\enquote {\bibinfo {title} {Multidimensional femtosecond correlation spectroscopies of electronic and vibrational excitations},}\ }\href {\doibase 10.1146/ANNUREV.PHYSCHEM.51.1.691/CITE/REFWORKS} {\bibfield  {journal} {\bibinfo  {journal} {Annu. Rev. Phys. Chem.}\ }\textbf {\bibinfo {volume} {51}},\ \bibinfo {pages} {691--729} (\bibinfo {year} {2000})}\BibitemShut {NoStop}%
\bibitem [{\citenamefont {Hybl}, \citenamefont {Ferro},\ and\ \citenamefont {Jonas}(2001)}]{Hybl2001}%
  \BibitemOpen
  \bibfield  {author} {\bibinfo {author} {\bibfnamefont {J.~D.}\ \bibnamefont {Hybl}}, \bibinfo {author} {\bibfnamefont {A.~A.}\ \bibnamefont {Ferro}}, \ and\ \bibinfo {author} {\bibfnamefont {D.~M.}\ \bibnamefont {Jonas}},\ }\bibfield  {title} {\enquote {\bibinfo {title} {Two-dimensional {F}ourier transform electronic spectroscopy},}\ }\href {\doibase 10.1063/1.1398579} {\bibfield  {journal} {\bibinfo  {journal} {J. Chem. Phys.}\ }\textbf {\bibinfo {volume} {115}},\ \bibinfo {pages} {6606--6622} (\bibinfo {year} {2001})}\BibitemShut {NoStop}%
\bibitem [{\citenamefont {Brixner}\ \emph {et~al.}(2005)\citenamefont {Brixner}, \citenamefont {Stenger}, \citenamefont {Vaswani}, \citenamefont {Cho}, \citenamefont {Blankenship},\ and\ \citenamefont {Fleming}}]{Brix2005}%
  \BibitemOpen
  \bibfield  {author} {\bibinfo {author} {\bibfnamefont {T.}~\bibnamefont {Brixner}}, \bibinfo {author} {\bibfnamefont {J.}~\bibnamefont {Stenger}}, \bibinfo {author} {\bibfnamefont {H.~M.}\ \bibnamefont {Vaswani}}, \bibinfo {author} {\bibfnamefont {M.}~\bibnamefont {Cho}}, \bibinfo {author} {\bibfnamefont {R.~E.}\ \bibnamefont {Blankenship}}, \ and\ \bibinfo {author} {\bibfnamefont {G.~R.}\ \bibnamefont {Fleming}},\ }\bibfield  {title} {\enquote {\bibinfo {title} {Two-dimensional spectroscopy of electronic couplings in photosynthesis},}\ }\href {\doibase 10.1038/nature03429} {\bibfield  {journal} {\bibinfo  {journal} {Nature}\ }\textbf {\bibinfo {volume} {434}},\ \bibinfo {pages} {625–628} (\bibinfo {year} {2005})}\BibitemShut {NoStop}%
\bibitem [{\citenamefont {Engel}\ \emph {et~al.}(2007)\citenamefont {Engel}, \citenamefont {Calhoun}, \citenamefont {Read}, \citenamefont {Ahn}, \citenamefont {Mančal}, \citenamefont {Cheng}, \citenamefont {Blankenship},\ and\ \citenamefont {Fleming}}]{Enge2007}%
  \BibitemOpen
  \bibfield  {author} {\bibinfo {author} {\bibfnamefont {G.~S.}\ \bibnamefont {Engel}}, \bibinfo {author} {\bibfnamefont {T.~R.}\ \bibnamefont {Calhoun}}, \bibinfo {author} {\bibfnamefont {E.~L.}\ \bibnamefont {Read}}, \bibinfo {author} {\bibfnamefont {T.-K.}\ \bibnamefont {Ahn}}, \bibinfo {author} {\bibfnamefont {T.}~\bibnamefont {Mančal}}, \bibinfo {author} {\bibfnamefont {Y.-C.}\ \bibnamefont {Cheng}}, \bibinfo {author} {\bibfnamefont {R.~E.}\ \bibnamefont {Blankenship}}, \ and\ \bibinfo {author} {\bibfnamefont {G.~R.}\ \bibnamefont {Fleming}},\ }\bibfield  {title} {\enquote {\bibinfo {title} {Evidence for wavelike energy transfer through quantum coherence in photosynthetic systems},}\ }\href {\doibase 10.1038/nature05678} {\bibfield  {journal} {\bibinfo  {journal} {Nature}\ }\textbf {\bibinfo {volume} {446}},\ \bibinfo {pages} {782--786} (\bibinfo {year} {2007})}\BibitemShut {NoStop}%
\bibitem [{\citenamefont {Smith}\ and\ \citenamefont {Jonas}(2011)}]{Smit2011}%
  \BibitemOpen
  \bibfield  {author} {\bibinfo {author} {\bibfnamefont {E.~R.}\ \bibnamefont {Smith}}\ and\ \bibinfo {author} {\bibfnamefont {D.~M.}\ \bibnamefont {Jonas}},\ }\bibfield  {title} {\enquote {\bibinfo {title} {Alignment, vibronic level splitting, and coherent coupling effects on the pump-probe polarization anisotropy},}\ }\href {\doibase 10.1021/jp201928s} {\bibfield  {journal} {\bibinfo  {journal} {The Journal of Physical Chemistry A}\ }\textbf {\bibinfo {volume} {115}},\ \bibinfo {pages} {4101–4113} (\bibinfo {year} {2011})}\BibitemShut {NoStop}%
\bibitem [{\citenamefont {Cao}\ \emph {et~al.}(2020)\citenamefont {Cao}, \citenamefont {Cogdell}, \citenamefont {Coker}, \citenamefont {Duan}, \citenamefont {Hauer}, \citenamefont {Kleinekathöfer}, \citenamefont {Jansen}, \citenamefont {Mančal}, \citenamefont {Miller}, \citenamefont {Ogilvie}, \citenamefont {Prokhorenko}, \citenamefont {Renger}, \citenamefont {Tan}, \citenamefont {Tempelaar}, \citenamefont {Thorwart}, \citenamefont {Thyrhaug}, \citenamefont {Westenhoff},\ and\ \citenamefont {Zigmantas}}]{Cao2020}%
  \BibitemOpen
  \bibfield  {author} {\bibinfo {author} {\bibfnamefont {J.}~\bibnamefont {Cao}}, \bibinfo {author} {\bibfnamefont {R.~J.}\ \bibnamefont {Cogdell}}, \bibinfo {author} {\bibfnamefont {D.~F.}\ \bibnamefont {Coker}}, \bibinfo {author} {\bibfnamefont {H.-G.}\ \bibnamefont {Duan}}, \bibinfo {author} {\bibfnamefont {J.}~\bibnamefont {Hauer}}, \bibinfo {author} {\bibfnamefont {U.}~\bibnamefont {Kleinekathöfer}}, \bibinfo {author} {\bibfnamefont {T.~L.~C.}\ \bibnamefont {Jansen}}, \bibinfo {author} {\bibfnamefont {T.}~\bibnamefont {Mančal}}, \bibinfo {author} {\bibfnamefont {R.~J.~D.}\ \bibnamefont {Miller}}, \bibinfo {author} {\bibfnamefont {J.~P.}\ \bibnamefont {Ogilvie}}, \bibinfo {author} {\bibfnamefont {V.~I.}\ \bibnamefont {Prokhorenko}}, \bibinfo {author} {\bibfnamefont {T.}~\bibnamefont {Renger}}, \bibinfo {author} {\bibfnamefont {H.-S.}\ \bibnamefont {Tan}}, \bibinfo {author} {\bibfnamefont {R.}~\bibnamefont {Tempelaar}}, \bibinfo {author} {\bibfnamefont {M.}~\bibnamefont {Thorwart}}, \bibinfo {author}
  {\bibfnamefont {E.}~\bibnamefont {Thyrhaug}}, \bibinfo {author} {\bibfnamefont {S.}~\bibnamefont {Westenhoff}}, \ and\ \bibinfo {author} {\bibfnamefont {D.}~\bibnamefont {Zigmantas}},\ }\bibfield  {title} {\enquote {\bibinfo {title} {Quantum biology revisited},}\ }\href {\doibase 10.1126/sciadv.aaz4888} {\bibfield  {journal} {\bibinfo  {journal} {Sci. Adv.}\ }\textbf {\bibinfo {volume} {6}},\ \bibinfo {pages} {eaaz4888} (\bibinfo {year} {2020})}\BibitemShut {NoStop}%
\bibitem [{\citenamefont {Schultz}\ \emph {et~al.}(2024)\citenamefont {Schultz}, \citenamefont {Yuly}, \citenamefont {Arsenault}, \citenamefont {Parker}, \citenamefont {Chowdhury}, \citenamefont {Dani}, \citenamefont {Kundu}, \citenamefont {Nuomin}, \citenamefont {Zhang}, \citenamefont {Valdiviezo}, \citenamefont {Zhang}, \citenamefont {Orcutt}, \citenamefont {Jang}, \citenamefont {Fleming}, \citenamefont {Makri}, \citenamefont {Ogilvie}, \citenamefont {Therien}, \citenamefont {Wasielewski},\ and\ \citenamefont {Beratan}}]{Schu2024}%
  \BibitemOpen
  \bibfield  {author} {\bibinfo {author} {\bibfnamefont {J.~D.}\ \bibnamefont {Schultz}}, \bibinfo {author} {\bibfnamefont {J.~L.}\ \bibnamefont {Yuly}}, \bibinfo {author} {\bibfnamefont {E.~A.}\ \bibnamefont {Arsenault}}, \bibinfo {author} {\bibfnamefont {K.}~\bibnamefont {Parker}}, \bibinfo {author} {\bibfnamefont {S.~N.}\ \bibnamefont {Chowdhury}}, \bibinfo {author} {\bibfnamefont {R.}~\bibnamefont {Dani}}, \bibinfo {author} {\bibfnamefont {S.}~\bibnamefont {Kundu}}, \bibinfo {author} {\bibfnamefont {H.}~\bibnamefont {Nuomin}}, \bibinfo {author} {\bibfnamefont {Z.}~\bibnamefont {Zhang}}, \bibinfo {author} {\bibfnamefont {J.}~\bibnamefont {Valdiviezo}}, \bibinfo {author} {\bibfnamefont {P.}~\bibnamefont {Zhang}}, \bibinfo {author} {\bibfnamefont {K.}~\bibnamefont {Orcutt}}, \bibinfo {author} {\bibfnamefont {S.~J.}\ \bibnamefont {Jang}}, \bibinfo {author} {\bibfnamefont {G.~R.}\ \bibnamefont {Fleming}}, \bibinfo {author} {\bibfnamefont {N.}~\bibnamefont {Makri}}, \bibinfo {author} {\bibfnamefont {J.~P.}\
  \bibnamefont {Ogilvie}}, \bibinfo {author} {\bibfnamefont {M.~J.}\ \bibnamefont {Therien}}, \bibinfo {author} {\bibfnamefont {M.~R.}\ \bibnamefont {Wasielewski}}, \ and\ \bibinfo {author} {\bibfnamefont {D.~N.}\ \bibnamefont {Beratan}},\ }\bibfield  {title} {\enquote {\bibinfo {title} {Coherence in chemistry: Foundations and frontiers},}\ }\href {\doibase 10.1021/acs.chemrev.3c00643} {\bibfield  {journal} {\bibinfo  {journal} {Chem. Rev.}\ }\textbf {\bibinfo {volume} {124}},\ \bibinfo {pages} {11641--11766} (\bibinfo {year} {2024})}\BibitemShut {NoStop}%
\bibitem [{\citenamefont {McLachlan}(1964)}]{McLa1964}%
  \BibitemOpen
  \bibfield  {author} {\bibinfo {author} {\bibfnamefont {A.}~\bibnamefont {McLachlan}},\ }\bibfield  {title} {\enquote {\bibinfo {title} {A variational solution of the time-dependent {S}chrodinger equation},}\ }\href {\doibase 10.1080/00268976400100041} {\bibfield  {journal} {\bibinfo  {journal} {Mol. Phys.}\ }\textbf {\bibinfo {volume} {8}},\ \bibinfo {pages} {39--44} (\bibinfo {year} {1964})}\BibitemShut {NoStop}%
\bibitem [{\citenamefont {Tully}(1998)}]{Tull1998}%
  \BibitemOpen
  \bibfield  {author} {\bibinfo {author} {\bibfnamefont {J.~C.}\ \bibnamefont {Tully}},\ }\bibfield  {title} {\enquote {\bibinfo {title} {Mixed quantum–classical dynamics},}\ }\href {\doibase 10.1039/A801824C} {\bibfield  {journal} {\bibinfo  {journal} {Faraday Discuss.}\ }\textbf {\bibinfo {volume} {110}},\ \bibinfo {pages} {407--419} (\bibinfo {year} {1998})}\BibitemShut {NoStop}%
\bibitem [{\citenamefont {Grunwald}, \citenamefont {Kelly},\ and\ \citenamefont {Kapral}(2009)}]{Grun2009}%
  \BibitemOpen
  \bibfield  {author} {\bibinfo {author} {\bibfnamefont {R.}~\bibnamefont {Grunwald}}, \bibinfo {author} {\bibfnamefont {A.}~\bibnamefont {Kelly}}, \ and\ \bibinfo {author} {\bibfnamefont {R.}~\bibnamefont {Kapral}},\ }\bibfield  {title} {\enquote {\bibinfo {title} {Quantum dynamics in almost classical environments},}\ }in\ \href@noop {} {\emph {\bibinfo {booktitle} {Energy Transfer Dynamics in Biomaterial Systems}}}\ (\bibinfo  {publisher} {Springer},\ \bibinfo {year} {2009})\BibitemShut {NoStop}%
\bibitem [{\citenamefont {Tully}(1990)}]{Tull1990}%
  \BibitemOpen
  \bibfield  {author} {\bibinfo {author} {\bibfnamefont {J.~C.}\ \bibnamefont {Tully}},\ }\bibfield  {title} {\enquote {\bibinfo {title} {Molecular dynamics with electronic transitions},}\ }\href {\doibase 10.1063/1.459170} {\bibfield  {journal} {\bibinfo  {journal} {J. Chem. Phys.}\ }\textbf {\bibinfo {volume} {93}},\ \bibinfo {pages} {1061--1071} (\bibinfo {year} {1990})}\BibitemShut {NoStop}%
\bibitem [{\citenamefont {Meyer}\ and\ \citenamefont {Miller}(1979)}]{Meye1979}%
  \BibitemOpen
  \bibfield  {author} {\bibinfo {author} {\bibfnamefont {H.-D.}\ \bibnamefont {Meyer}}\ and\ \bibinfo {author} {\bibfnamefont {W.~H.}\ \bibnamefont {Miller}},\ }\bibfield  {title} {\enquote {\bibinfo {title} {A classical analog for electronic degrees of freedom in nonadiabatic collision processes},}\ }\href {\doibase 10.1063/1.437910} {\bibfield  {journal} {\bibinfo  {journal} {J. Chem. Phys.}\ }\textbf {\bibinfo {volume} {70}},\ \bibinfo {pages} {3214--3223} (\bibinfo {year} {1979})}\BibitemShut {NoStop}%
\bibitem [{\citenamefont {Stock}\ and\ \citenamefont {Thoss}(1997)}]{Stoc1997}%
  \BibitemOpen
  \bibfield  {author} {\bibinfo {author} {\bibfnamefont {G.}~\bibnamefont {Stock}}\ and\ \bibinfo {author} {\bibfnamefont {M.}~\bibnamefont {Thoss}},\ }\bibfield  {title} {\enquote {\bibinfo {title} {Semiclassical description of nonadiabatic quantum dynamics},}\ }\href {\doibase 10.1103/physrevlett.78.578} {\bibfield  {journal} {\bibinfo  {journal} {Phys. Rev. Lett.}\ }\textbf {\bibinfo {volume} {78}},\ \bibinfo {pages} {578--581} (\bibinfo {year} {1997})}\BibitemShut {NoStop}%
\bibitem [{\citenamefont {Runeson}\ and\ \citenamefont {Richardson}(2020)}]{Rune2020}%
  \BibitemOpen
  \bibfield  {author} {\bibinfo {author} {\bibfnamefont {J.~E.}\ \bibnamefont {Runeson}}\ and\ \bibinfo {author} {\bibfnamefont {J.~O.}\ \bibnamefont {Richardson}},\ }\bibfield  {title} {\enquote {\bibinfo {title} {Generalized spin mapping for quantum-classical dynamics},}\ }\href {\doibase 10.1063/1.5143412} {\bibfield  {journal} {\bibinfo  {journal} {J. Chem. Phys.}\ }\textbf {\bibinfo {volume} {152}},\ \bibinfo {pages} {084110} (\bibinfo {year} {2020})}\BibitemShut {NoStop}%
\bibitem [{\citenamefont {Miller}\ and\ \citenamefont {Cotton}(2016)}]{Mill2016}%
  \BibitemOpen
  \bibfield  {author} {\bibinfo {author} {\bibfnamefont {W.~H.}\ \bibnamefont {Miller}}\ and\ \bibinfo {author} {\bibfnamefont {S.~J.}\ \bibnamefont {Cotton}},\ }\bibfield  {title} {\enquote {\bibinfo {title} {Classical molecular dynamics simulation of electronically non-adiabatic processes},}\ }\href {\doibase 10.1039/c6fd00181e} {\bibfield  {journal} {\bibinfo  {journal} {Faraday Discuss.}\ }\textbf {\bibinfo {volume} {195}},\ \bibinfo {pages} {9--30} (\bibinfo {year} {2016})}\BibitemShut {NoStop}%
\bibitem [{\citenamefont {Mannouch}\ and\ \citenamefont {Richardson}(2023)}]{Mann2023}%
  \BibitemOpen
  \bibfield  {author} {\bibinfo {author} {\bibfnamefont {J.~R.}\ \bibnamefont {Mannouch}}\ and\ \bibinfo {author} {\bibfnamefont {J.~O.}\ \bibnamefont {Richardson}},\ }\bibfield  {title} {\enquote {\bibinfo {title} {A mapping approach to surface hopping},}\ }\href {\doibase 10.1063/5.0139734} {\bibfield  {journal} {\bibinfo  {journal} {J. Chem. Phys.}\ }\textbf {\bibinfo {volume} {158}},\ \bibinfo {pages} {104111} (\bibinfo {year} {2023})}\BibitemShut {NoStop}%
\bibitem [{\citenamefont {Runeson}\ and\ \citenamefont {Manolopoulos}(2023)}]{Rune2023}%
  \BibitemOpen
  \bibfield  {author} {\bibinfo {author} {\bibfnamefont {J.~E.}\ \bibnamefont {Runeson}}\ and\ \bibinfo {author} {\bibfnamefont {D.~E.}\ \bibnamefont {Manolopoulos}},\ }\bibfield  {title} {\enquote {\bibinfo {title} {A multi-state mapping approach to surface hopping},}\ }\href {\doibase 10.1063/5.0158147} {\bibfield  {journal} {\bibinfo  {journal} {J. Chem. Phys.}\ }\textbf {\bibinfo {volume} {159}},\ \bibinfo {pages} {094115} (\bibinfo {year} {2023})}\BibitemShut {NoStop}%
\bibitem [{\citenamefont {Mannouch}\ and\ \citenamefont {Kelly}(2024)}]{Mann2024}%
  \BibitemOpen
  \bibfield  {author} {\bibinfo {author} {\bibfnamefont {J.~R.}\ \bibnamefont {Mannouch}}\ and\ \bibinfo {author} {\bibfnamefont {A.}~\bibnamefont {Kelly}},\ }\bibfield  {title} {\enquote {\bibinfo {title} {Towards a correct description of initial electronic coherence in nonadiabatic dynamics simulations},}\ }\href {\doibase 10.1021/ACS.JPCLETT.4C02418/SUPPL_FILE/JZ4C02418_SI_001.PDF} {\bibfield  {journal} {\bibinfo  {journal} {J. Phys. Chem. Lett}\ }\textbf {\bibinfo {volume} {23}},\ \bibinfo {pages} {11687--11695} (\bibinfo {year} {2024})}\BibitemShut {NoStop}%
\bibitem [{\citenamefont {Crespo-Otero}\ and\ \citenamefont {Barbatti}(2018)}]{Cres2018}%
  \BibitemOpen
  \bibfield  {author} {\bibinfo {author} {\bibfnamefont {R.}~\bibnamefont {Crespo-Otero}}\ and\ \bibinfo {author} {\bibfnamefont {M.}~\bibnamefont {Barbatti}},\ }\bibfield  {title} {\enquote {\bibinfo {title} {Recent advances and perspectives on nonadiabatic mixed quantum–classical dynamics},}\ }\href {\doibase 10.1021/acs.chemrev.7b00577} {\bibfield  {journal} {\bibinfo  {journal} {Chemical Reviews}\ }\textbf {\bibinfo {volume} {118}},\ \bibinfo {pages} {7026–7068} (\bibinfo {year} {2018})}\BibitemShut {NoStop}%
\bibitem [{\citenamefont {Ceotto}\ \emph {et~al.}(2009)\citenamefont {Ceotto}, \citenamefont {Atahan}, \citenamefont {Shim}, \citenamefont {Tantardini},\ and\ \citenamefont {Aspuru-Guzik}}]{Ceot2009}%
  \BibitemOpen
  \bibfield  {author} {\bibinfo {author} {\bibfnamefont {M.}~\bibnamefont {Ceotto}}, \bibinfo {author} {\bibfnamefont {S.}~\bibnamefont {Atahan}}, \bibinfo {author} {\bibfnamefont {S.}~\bibnamefont {Shim}}, \bibinfo {author} {\bibfnamefont {G.~F.}\ \bibnamefont {Tantardini}}, \ and\ \bibinfo {author} {\bibfnamefont {A.}~\bibnamefont {Aspuru-Guzik}},\ }\bibfield  {title} {\enquote {\bibinfo {title} {First-principles semiclassical initial value representation molecular dynamics},}\ }\href {\doibase 10.1039/B820785B} {\bibfield  {journal} {\bibinfo  {journal} {Phys. Chem. Chem. Phys.}\ }\textbf {\bibinfo {volume} {11}},\ \bibinfo {pages} {3861--3867} (\bibinfo {year} {2009})}\BibitemShut {NoStop}%
\bibitem [{\citenamefont {Aieta}\ \emph {et~al.}(2025)\citenamefont {Aieta}, \citenamefont {Cazzaniga}, \citenamefont {Moscato}, \citenamefont {Lanzi}, \citenamefont {Bocchi}, \citenamefont {Costanza}, \citenamefont {Ceotto},\ and\ \citenamefont {Conte}}]{Aiet2025}%
  \BibitemOpen
  \bibfield  {author} {\bibinfo {author} {\bibfnamefont {C.}~\bibnamefont {Aieta}}, \bibinfo {author} {\bibfnamefont {M.}~\bibnamefont {Cazzaniga}}, \bibinfo {author} {\bibfnamefont {D.}~\bibnamefont {Moscato}}, \bibinfo {author} {\bibfnamefont {C.}~\bibnamefont {Lanzi}}, \bibinfo {author} {\bibfnamefont {L.}~\bibnamefont {Bocchi}}, \bibinfo {author} {\bibfnamefont {M.~M.}\ \bibnamefont {Costanza}}, \bibinfo {author} {\bibfnamefont {M.}~\bibnamefont {Ceotto}}, \ and\ \bibinfo {author} {\bibfnamefont {R.}~\bibnamefont {Conte}},\ }\bibfield  {title} {\enquote {\bibinfo {title} {Quantum dynamics through a handful of semiclassical trajectories},}\ }\href {\doibase 10.1007/S12210-025-01326-4/FIGURES/4} {\bibfield  {journal} {\bibinfo  {journal} {Rend. Fis. Acc. Lincei}\ }\textbf {\bibinfo {volume} {36}},\ \bibinfo {pages} {445–455} (\bibinfo {year} {2025})}\BibitemShut {NoStop}%
\bibitem [{\citenamefont {Scheidegger}\ and\ \citenamefont {Vaníček}(2025)}]{Sche2025}%
  \BibitemOpen
  \bibfield  {author} {\bibinfo {author} {\bibfnamefont {A.}~\bibnamefont {Scheidegger}}\ and\ \bibinfo {author} {\bibfnamefont {J.~J.~L.}\ \bibnamefont {Vaníček}},\ }\bibfield  {title} {\enquote {\bibinfo {title} {Ehrenfest dynamics accelerated with speed},}\ }\href {\doibase 10.1063/5.0276025} {\bibfield  {journal} {\bibinfo  {journal} {The Journal of Chemical Physics}\ }\textbf {\bibinfo {volume} {163}} (\bibinfo {year} {2025}),\ 10.1063/5.0276025}\BibitemShut {NoStop}%
\bibitem [{\citenamefont {Lawrence}\ \emph {et~al.}(2024)\citenamefont {Lawrence}, \citenamefont {Ansari}, \citenamefont {Mannouch}, \citenamefont {Manae}, \citenamefont {Asnaashari}, \citenamefont {Kelly},\ and\ \citenamefont {Richardson}}]{Lawr2024b}%
  \BibitemOpen
  \bibfield  {author} {\bibinfo {author} {\bibfnamefont {J.~E.}\ \bibnamefont {Lawrence}}, \bibinfo {author} {\bibfnamefont {I.~M.}\ \bibnamefont {Ansari}}, \bibinfo {author} {\bibfnamefont {J.~R.}\ \bibnamefont {Mannouch}}, \bibinfo {author} {\bibfnamefont {M.~A.}\ \bibnamefont {Manae}}, \bibinfo {author} {\bibfnamefont {K.}~\bibnamefont {Asnaashari}}, \bibinfo {author} {\bibfnamefont {A.}~\bibnamefont {Kelly}}, \ and\ \bibinfo {author} {\bibfnamefont {J.~O.}\ \bibnamefont {Richardson}},\ }\bibfield  {title} {\enquote {\bibinfo {title} {A {MASH} simulation of the photoexcited dynamics of cyclobutanone},}\ }\href {\doibase 10.1063/5.0203695} {\bibfield  {journal} {\bibinfo  {journal} {J. Chem. Phys.}\ }\textbf {\bibinfo {volume} {160}},\ \bibinfo {pages} {174306} (\bibinfo {year} {2024})}\BibitemShut {NoStop}%
\bibitem [{\citenamefont {Mukamel}(1999)}]{Muka1999}%
  \BibitemOpen
  \bibfield  {author} {\bibinfo {author} {\bibfnamefont {S.}~\bibnamefont {Mukamel}},\ }\href@noop {} {\emph {\bibinfo {title} {Principles of Nonlinear Optical Spectroscopy}}}\ (\bibinfo  {publisher} {Oxford University Press},\ \bibinfo {year} {1999})\BibitemShut {NoStop}%
\bibitem [{\citenamefont {Hamm}(2005)}]{Hamm2005}%
  \BibitemOpen
  \bibfield  {author} {\bibinfo {author} {\bibfnamefont {P.}~\bibnamefont {Hamm}},\ }\href@noop {} {\emph {\bibinfo {title} {Principles of Nonlinear Optical Spectroscopy: A Practical Approach}}}\ (\bibinfo  {publisher} {Lecture Notes, University of Zurich},\ \bibinfo {year} {2005})\BibitemShut {NoStop}%
\bibitem [{\citenamefont {Provazza}\ and\ \citenamefont {Coker}(2018)}]{Prov2018}%
  \BibitemOpen
  \bibfield  {author} {\bibinfo {author} {\bibfnamefont {J.}~\bibnamefont {Provazza}}\ and\ \bibinfo {author} {\bibfnamefont {D.~F.}\ \bibnamefont {Coker}},\ }\bibfield  {title} {\enquote {\bibinfo {title} {Communication: Symmetrical quasi-classical analysis of linear optical spectroscopy},}\ }\href {\doibase 10.1063/1.5031788} {\bibfield  {journal} {\bibinfo  {journal} {J. Chem. Phys.}\ }\textbf {\bibinfo {volume} {148}},\ \bibinfo {pages} {181102} (\bibinfo {year} {2018})}\BibitemShut {NoStop}%
\bibitem [{\citenamefont {Provazza}\ \emph {et~al.}(2018)\citenamefont {Provazza}, \citenamefont {Segatta}, \citenamefont {Garavelli},\ and\ \citenamefont {Coker}}]{Prov2018a}%
  \BibitemOpen
  \bibfield  {author} {\bibinfo {author} {\bibfnamefont {J.}~\bibnamefont {Provazza}}, \bibinfo {author} {\bibfnamefont {F.}~\bibnamefont {Segatta}}, \bibinfo {author} {\bibfnamefont {M.}~\bibnamefont {Garavelli}}, \ and\ \bibinfo {author} {\bibfnamefont {D.~F.}\ \bibnamefont {Coker}},\ }\bibfield  {title} {\enquote {\bibinfo {title} {Semiclassical path integral calculation of nonlinear optical spectroscopy},}\ }\href {\doibase 10.1021/acs.jctc.7b01063} {\bibfield  {journal} {\bibinfo  {journal} {J. Chem. Theory Comput.}\ }\textbf {\bibinfo {volume} {14}},\ \bibinfo {pages} {856--866} (\bibinfo {year} {2018})}\BibitemShut {NoStop}%
\bibitem [{\citenamefont {Mannouch}\ and\ \citenamefont {Richardson}(2022)}]{Mann2022}%
  \BibitemOpen
  \bibfield  {author} {\bibinfo {author} {\bibfnamefont {J.~R.}\ \bibnamefont {Mannouch}}\ and\ \bibinfo {author} {\bibfnamefont {J.~O.}\ \bibnamefont {Richardson}},\ }\bibfield  {title} {\enquote {\bibinfo {title} {A partially linearized spin-mapping approach for simulating nonlinear optical spectra},}\ }\href {\doibase 10.1063/5.0077744} {\bibfield  {journal} {\bibinfo  {journal} {J. Chem. Phys.}\ }\textbf {\bibinfo {volume} {156}},\ \bibinfo {pages} {024108} (\bibinfo {year} {2022})}\BibitemShut {NoStop}%
\bibitem [{\citenamefont {Sun}, \citenamefont {Wang},\ and\ \citenamefont {Miller}(1998)}]{Sun1998}%
  \BibitemOpen
  \bibfield  {author} {\bibinfo {author} {\bibfnamefont {X.}~\bibnamefont {Sun}}, \bibinfo {author} {\bibfnamefont {H.}~\bibnamefont {Wang}}, \ and\ \bibinfo {author} {\bibfnamefont {W.~H.}\ \bibnamefont {Miller}},\ }\bibfield  {title} {\enquote {\bibinfo {title} {Semiclassical theory of electronically nonadiabatic dynamics: Results of a linearized approximation to the initial value representation},}\ }\href {\doibase 10.1063/1.477389} {\bibfield  {journal} {\bibinfo  {journal} {J. Chem. Phys.}\ }\textbf {\bibinfo {volume} {109}},\ \bibinfo {pages} {7064--7074} (\bibinfo {year} {1998})}\BibitemShut {NoStop}%
\bibitem [{\citenamefont {Gao}\ and\ \citenamefont {Geva}(2020)}]{Gao2020}%
  \BibitemOpen
  \bibfield  {author} {\bibinfo {author} {\bibfnamefont {X.}~\bibnamefont {Gao}}\ and\ \bibinfo {author} {\bibfnamefont {E.}~\bibnamefont {Geva}},\ }\bibfield  {title} {\enquote {\bibinfo {title} {A nonperturbative methodology for simulating multidimensional spectra of multiexcitonic molecular systems via quasiclassical mapping {H}amiltonian methods},}\ }\href {\doibase 10.1021/acs.jctc.0c00843} {\bibfield  {journal} {\bibinfo  {journal} {J. Chem. Theory Comput.}\ }\textbf {\bibinfo {volume} {16}},\ \bibinfo {pages} {6491--6502} (\bibinfo {year} {2020})}\BibitemShut {NoStop}%
\bibitem [{\citenamefont {Seidner}, \citenamefont {Stock},\ and\ \citenamefont {Domcke}(1995)}]{Seid1995}%
  \BibitemOpen
  \bibfield  {author} {\bibinfo {author} {\bibfnamefont {L.}~\bibnamefont {Seidner}}, \bibinfo {author} {\bibfnamefont {G.}~\bibnamefont {Stock}}, \ and\ \bibinfo {author} {\bibfnamefont {W.}~\bibnamefont {Domcke}},\ }\bibfield  {title} {\enquote {\bibinfo {title} {Nonperturbative approach to femtosecond spectroscopy: General theory and application to multidimensional nonadiabatic photoisomerization processes},}\ }\href {\doibase 10.1063/1.469586} {\bibfield  {journal} {\bibinfo  {journal} {J. Chem. Phys.}\ }\textbf {\bibinfo {volume} {103}},\ \bibinfo {pages} {3998--4011} (\bibinfo {year} {1995})}\BibitemShut {NoStop}%
\bibitem [{\citenamefont {Krumland}\ \emph {et~al.}(2024)\citenamefont {Krumland}, \citenamefont {Guerrini}, \citenamefont {De~Sio}, \citenamefont {Lienau},\ and\ \citenamefont {Cocchi}}]{Krum2024}%
  \BibitemOpen
  \bibfield  {author} {\bibinfo {author} {\bibfnamefont {J.}~\bibnamefont {Krumland}}, \bibinfo {author} {\bibfnamefont {M.}~\bibnamefont {Guerrini}}, \bibinfo {author} {\bibfnamefont {A.}~\bibnamefont {De~Sio}}, \bibinfo {author} {\bibfnamefont {C.}~\bibnamefont {Lienau}}, \ and\ \bibinfo {author} {\bibfnamefont {C.}~\bibnamefont {Cocchi}},\ }\bibfield  {title} {\enquote {\bibinfo {title} {Two-dimensional electronic spectroscopy from first principles},}\ }\href {\doibase 10.1063/5.0172621} {\bibfield  {journal} {\bibinfo  {journal} {Appl. Phys. Rev.}\ }\textbf {\bibinfo {volume} {11}},\ \bibinfo {pages} {011305} (\bibinfo {year} {2024})}\BibitemShut {NoStop}%
\bibitem [{\citenamefont {van~der Vegte}\ \emph {et~al.}(2013)\citenamefont {van~der Vegte}, \citenamefont {Dijkstra}, \citenamefont {Knoester},\ and\ \citenamefont {Jansen}}]{Vegt2013}%
  \BibitemOpen
  \bibfield  {author} {\bibinfo {author} {\bibfnamefont {C.~P.}\ \bibnamefont {van~der Vegte}}, \bibinfo {author} {\bibfnamefont {A.~G.}\ \bibnamefont {Dijkstra}}, \bibinfo {author} {\bibfnamefont {J.}~\bibnamefont {Knoester}}, \ and\ \bibinfo {author} {\bibfnamefont {T.~L.~C.}\ \bibnamefont {Jansen}},\ }\bibfield  {title} {\enquote {\bibinfo {title} {Calculating two-dimensional spectra with the mixed quantum-classical {E}hrenfest method},}\ }\href {\doibase 10.1021/jp311668r} {\bibfield  {journal} {\bibinfo  {journal} {J. Phys. Chem. A}\ }\textbf {\bibinfo {volume} {117}},\ \bibinfo {pages} {5970--5980} (\bibinfo {year} {2013})}\BibitemShut {NoStop}%
\bibitem [{Note1()}]{Note1}%
  \BibitemOpen
  \bibinfo {note} {\protect \color {black}Note that this is not the same as the other ``classical path approximation'' that is sometimes used in mixed quantum-classical simulations, which is rather more drastic. In that, the back-action of the electronic system on the nuclear motion is neglected entirely, whereas the mean classical path approximation used in Ref.~34 includes the average back action from the bra and ket states of an electronic coherence \protect \color {black}}\BibitemShut {NoStop}%
\bibitem [{\citenamefont {Atsango}, \citenamefont {Montoya-Castillo},\ and\ \citenamefont {Markland}(2023)}]{Atsa2023}%
  \BibitemOpen
  \bibfield  {author} {\bibinfo {author} {\bibfnamefont {A.~O.}\ \bibnamefont {Atsango}}, \bibinfo {author} {\bibfnamefont {A.}~\bibnamefont {Montoya-Castillo}}, \ and\ \bibinfo {author} {\bibfnamefont {T.~E.}\ \bibnamefont {Markland}},\ }\bibfield  {title} {\enquote {\bibinfo {title} {An accurate and efficient {E}hrenfest dynamics approach for calculating linear and nonlinear electronic spectra},}\ }\href {\doibase 10.1063/5.0138671} {\bibfield  {journal} {\bibinfo  {journal} {J. Chem. Phys.}\ }\textbf {\bibinfo {volume} {158}},\ \bibinfo {pages} {074107} (\bibinfo {year} {2023})}\BibitemShut {NoStop}%
\bibitem [{\citenamefont {Kramer}\ \emph {et~al.}(2018)\citenamefont {Kramer}, \citenamefont {Noack}, \citenamefont {Reinefeld}, \citenamefont {Rodríguez},\ and\ \citenamefont {Zelinskyy}}]{Kram2018}%
  \BibitemOpen
  \bibfield  {author} {\bibinfo {author} {\bibfnamefont {T.}~\bibnamefont {Kramer}}, \bibinfo {author} {\bibfnamefont {M.}~\bibnamefont {Noack}}, \bibinfo {author} {\bibfnamefont {A.}~\bibnamefont {Reinefeld}}, \bibinfo {author} {\bibfnamefont {M.}~\bibnamefont {Rodríguez}}, \ and\ \bibinfo {author} {\bibfnamefont {Y.}~\bibnamefont {Zelinskyy}},\ }\bibfield  {title} {\enquote {\bibinfo {title} {Efficient calculation of open quantum system dynamics and time‐resolved spectroscopy with distributed memory {HEOM} ({DM‐HEOM})},}\ }\href {\doibase 10.1002/jcc.25354} {\bibfield  {journal} {\bibinfo  {journal} {J. Comput. Chem.}\ }\textbf {\bibinfo {volume} {39}},\ \bibinfo {pages} {1779--1794} (\bibinfo {year} {2018})}\BibitemShut {NoStop}%
\bibitem [{\citenamefont {Tanimura}(2020)}]{Tani2020}%
  \BibitemOpen
  \bibfield  {author} {\bibinfo {author} {\bibfnamefont {Y.}~\bibnamefont {Tanimura}},\ }\bibfield  {title} {\enquote {\bibinfo {title} {Numerically “exact” approach to open quantum dynamics: The hierarchical equations of motion (heom)},}\ }\href {\doibase 10.1063/5.0011599} {\bibfield  {journal} {\bibinfo  {journal} {J. Chem. Phys.}\ }\textbf {\bibinfo {volume} {153}},\ \bibinfo {pages} {020901} (\bibinfo {year} {2020})}\BibitemShut {NoStop}%
\bibitem [{\citenamefont {Cho}(2008)}]{Cho2008}%
  \BibitemOpen
  \bibfield  {author} {\bibinfo {author} {\bibfnamefont {M.}~\bibnamefont {Cho}},\ }\bibfield  {title} {\enquote {\bibinfo {title} {Coherent two-dimensional optical spectroscopy},}\ }\href {\doibase 10.1021/cr078377b} {\bibfield  {journal} {\bibinfo  {journal} {Chemical Reviews}\ }\textbf {\bibinfo {volume} {108}},\ \bibinfo {pages} {1331–1418} (\bibinfo {year} {2008})}\BibitemShut {NoStop}%
\bibitem [{\citenamefont {Schlau-Cohen}, \citenamefont {Ishizaki},\ and\ \citenamefont {Fleming}(2011)}]{Schl2011}%
  \BibitemOpen
  \bibfield  {author} {\bibinfo {author} {\bibfnamefont {G.}~\bibnamefont {Schlau-Cohen}}, \bibinfo {author} {\bibfnamefont {A.}~\bibnamefont {Ishizaki}}, \ and\ \bibinfo {author} {\bibfnamefont {G.~R.}\ \bibnamefont {Fleming}},\ }\bibfield  {title} {\enquote {\bibinfo {title} {Two-dimensional electronic spectroscopy and photosynthesis: Fundamentals and applications to photosynthetic light-harvesting},}\ }\href@noop {} {\bibfield  {journal} {\bibinfo  {journal} {Chem. Phys.}\ }\textbf {\bibinfo {volume} {386}},\ \bibinfo {pages} {1--22} (\bibinfo {year} {2011})}\BibitemShut {NoStop}%
\bibitem [{\citenamefont {Montoya-Castillo}\ and\ \citenamefont {Reichman}(2016)}]{Mont2016}%
  \BibitemOpen
  \bibfield  {author} {\bibinfo {author} {\bibfnamefont {A.}~\bibnamefont {Montoya-Castillo}}\ and\ \bibinfo {author} {\bibfnamefont {D.~R.}\ \bibnamefont {Reichman}},\ }\bibfield  {title} {\enquote {\bibinfo {title} {Approximate but accurate quantum dynamics from the {M}ori formalism: {I}. {N}onequilibrium dynamics},}\ }\href {\doibase 10.1063/1.4948408} {\bibfield  {journal} {\bibinfo  {journal} {J. Chem. Phys.}\ }\textbf {\bibinfo {volume} {144}},\ \bibinfo {pages} {184104} (\bibinfo {year} {2016})}\BibitemShut {NoStop}%
\bibitem [{\citenamefont {Egorov}, \citenamefont {Rabani},\ and\ \citenamefont {Berne}(1999)}]{Egor1999}%
  \BibitemOpen
  \bibfield  {author} {\bibinfo {author} {\bibfnamefont {S.~A.}\ \bibnamefont {Egorov}}, \bibinfo {author} {\bibfnamefont {E.}~\bibnamefont {Rabani}}, \ and\ \bibinfo {author} {\bibfnamefont {B.~J.}\ \bibnamefont {Berne}},\ }\bibfield  {title} {\enquote {\bibinfo {title} {Nonradiative relaxation processes in condensed phases: Quantum versus classical baths},}\ }\href {\doibase 10.1063/1.478420} {\bibfield  {journal} {\bibinfo  {journal} {J. Chem. Phys.}\ }\textbf {\bibinfo {volume} {110}},\ \bibinfo {pages} {5238--5248} (\bibinfo {year} {1999})}\BibitemShut {NoStop}%
\bibitem [{\citenamefont {Shi}\ and\ \citenamefont {Geva}(2005)}]{Shi2005}%
  \BibitemOpen
  \bibfield  {author} {\bibinfo {author} {\bibfnamefont {Q.}~\bibnamefont {Shi}}\ and\ \bibinfo {author} {\bibfnamefont {E.}~\bibnamefont {Geva}},\ }\bibfield  {title} {\enquote {\bibinfo {title} {A comparison between different semiclassical approximations for optical response functions in nonpolar liquid solutions},}\ }\href {\doibase 10.1063/1.1843813} {\bibfield  {journal} {\bibinfo  {journal} {J. Chem. Phys.}\ }\textbf {\bibinfo {volume} {122}},\ \bibinfo {pages} {064506} (\bibinfo {year} {2005})}\BibitemShut {NoStop}%
\bibitem [{Note2()}]{Note2}%
  \BibitemOpen
  \bibinfo {note} {\protect \color {black}Note that we are referring here to fully linearized spin mapping, not the partially linearized spin-PLDM approach described in Ref.~29 \protect \color {black}}\BibitemShut {NoStop}%
\bibitem [{\citenamefont {Hein}\ \emph {et~al.}(2012)\citenamefont {Hein}, \citenamefont {Kreisbeck}, \citenamefont {Kramer},\ and\ \citenamefont {Rodríguez}}]{Hein2012}%
  \BibitemOpen
  \bibfield  {author} {\bibinfo {author} {\bibfnamefont {B.}~\bibnamefont {Hein}}, \bibinfo {author} {\bibfnamefont {C.}~\bibnamefont {Kreisbeck}}, \bibinfo {author} {\bibfnamefont {T.}~\bibnamefont {Kramer}}, \ and\ \bibinfo {author} {\bibfnamefont {M.}~\bibnamefont {Rodríguez}},\ }\bibfield  {title} {\enquote {\bibinfo {title} {Modelling of oscillations in two-dimensional echo-spectra of the {F}enna–{M}atthews–{O}lson complex},}\ }\href {\doibase 10.1088/1367-2630/14/2/023018} {\bibfield  {journal} {\bibinfo  {journal} {New J. Phys.}\ }\textbf {\bibinfo {volume} {14}},\ \bibinfo {pages} {023018} (\bibinfo {year} {2012})}\BibitemShut {NoStop}%
\bibitem [{\citenamefont {Wang}, \citenamefont {Thoss},\ and\ \citenamefont {Miller}(2001)}]{Wang2001}%
  \BibitemOpen
  \bibfield  {author} {\bibinfo {author} {\bibfnamefont {H.}~\bibnamefont {Wang}}, \bibinfo {author} {\bibfnamefont {M.}~\bibnamefont {Thoss}}, \ and\ \bibinfo {author} {\bibfnamefont {W.~H.}\ \bibnamefont {Miller}},\ }\bibfield  {title} {\enquote {\bibinfo {title} {Systematic convergence in the dynamical hybrid approach for complex systems: A numerically exact methodology},}\ }\href {\doibase 10.1063/1.1385561} {\bibfield  {journal} {\bibinfo  {journal} {J. Chem. Phys.}\ }\textbf {\bibinfo {volume} {115}},\ \bibinfo {pages} {2979–2990} (\bibinfo {year} {2001})}\BibitemShut {NoStop}%
\bibitem [{\citenamefont {Craig}, \citenamefont {Thoss},\ and\ \citenamefont {Wang}(2007)}]{Crai2007}%
  \BibitemOpen
  \bibfield  {author} {\bibinfo {author} {\bibfnamefont {I.~R.}\ \bibnamefont {Craig}}, \bibinfo {author} {\bibfnamefont {M.}~\bibnamefont {Thoss}}, \ and\ \bibinfo {author} {\bibfnamefont {H.}~\bibnamefont {Wang}},\ }\bibfield  {title} {\enquote {\bibinfo {title} {Proton transfer reactions in model condensed-phase environments: Accurate quantum dynamics using the multilayer multiconfiguration time-dependent {H}artree approach},}\ }\href {\doibase 10.1063/1.2772265} {\bibfield  {journal} {\bibinfo  {journal} {J. Chem. Phys.}\ }\textbf {\bibinfo {volume} {127}},\ \bibinfo {pages} {144503} (\bibinfo {year} {2007})}\BibitemShut {NoStop}%
\bibitem [{\citenamefont {Wang}, \citenamefont {Sun},\ and\ \citenamefont {Miller}(1998)}]{Wang1998}%
  \BibitemOpen
  \bibfield  {author} {\bibinfo {author} {\bibfnamefont {H.}~\bibnamefont {Wang}}, \bibinfo {author} {\bibfnamefont {X.}~\bibnamefont {Sun}}, \ and\ \bibinfo {author} {\bibfnamefont {W.~H.}\ \bibnamefont {Miller}},\ }\bibfield  {title} {\enquote {\bibinfo {title} {Semiclassical approximations for the calculation of thermal rate constants for chemical reactions in complex molecular systems},}\ }\href {\doibase 10.1063/1.476447} {\bibfield  {journal} {\bibinfo  {journal} {J. Chem. Phys.}\ }\textbf {\bibinfo {volume} {108}},\ \bibinfo {pages} {9726--9736} (\bibinfo {year} {1998})}\BibitemShut {NoStop}%
\bibitem [{\citenamefont {Fetherolf}\ and\ \citenamefont {Berkelbach}(2017)}]{Feth2017}%
  \BibitemOpen
  \bibfield  {author} {\bibinfo {author} {\bibfnamefont {J.~H.}\ \bibnamefont {Fetherolf}}\ and\ \bibinfo {author} {\bibfnamefont {T.~C.}\ \bibnamefont {Berkelbach}},\ }\bibfield  {title} {\enquote {\bibinfo {title} {Linear and nonlinear spectroscopy from quantum master equations},}\ }\href {\doibase 10.1063/1.5006824} {\bibfield  {journal} {\bibinfo  {journal} {J. Chem. Phys.}\ }\textbf {\bibinfo {volume} {147}},\ \bibinfo {pages} {244109} (\bibinfo {year} {2017})}\BibitemShut {NoStop}%
\bibitem [{\citenamefont {Tian}\ \emph {et~al.}(2003)\citenamefont {Tian}, \citenamefont {Keusters}, \citenamefont {Suzaki},\ and\ \citenamefont {Warren}}]{Tian2003}%
  \BibitemOpen
  \bibfield  {author} {\bibinfo {author} {\bibfnamefont {P.}~\bibnamefont {Tian}}, \bibinfo {author} {\bibfnamefont {D.}~\bibnamefont {Keusters}}, \bibinfo {author} {\bibfnamefont {Y.}~\bibnamefont {Suzaki}}, \ and\ \bibinfo {author} {\bibfnamefont {W.~S.}\ \bibnamefont {Warren}},\ }\bibfield  {title} {\enquote {\bibinfo {title} {Femtosecond phase-coherent two-dimensional spectroscopy},}\ }\href {\doibase 10.1126/science.1083433} {\bibfield  {journal} {\bibinfo  {journal} {Science}\ }\textbf {\bibinfo {volume} {300}},\ \bibinfo {pages} {1553–1555} (\bibinfo {year} {2003})}\BibitemShut {NoStop}%
\bibitem [{\citenamefont {Biswas}\ \emph {et~al.}(2022)\citenamefont {Biswas}, \citenamefont {Kim}, \citenamefont {Zhang},\ and\ \citenamefont {Scholes}}]{Bisw2022}%
  \BibitemOpen
  \bibfield  {author} {\bibinfo {author} {\bibfnamefont {S.}~\bibnamefont {Biswas}}, \bibinfo {author} {\bibfnamefont {J.}~\bibnamefont {Kim}}, \bibinfo {author} {\bibfnamefont {X.}~\bibnamefont {Zhang}}, \ and\ \bibinfo {author} {\bibfnamefont {G.~D.}\ \bibnamefont {Scholes}},\ }\bibfield  {title} {\enquote {\bibinfo {title} {Coherent two-dimensional and broadband electronic spectroscopies},}\ }\href {\doibase 10.1021/acs.chemrev.1c00623} {\bibfield  {journal} {\bibinfo  {journal} {Chemical Reviews}\ }\textbf {\bibinfo {volume} {122}},\ \bibinfo {pages} {4257–4321} (\bibinfo {year} {2022})}\BibitemShut {NoStop}%
\bibitem [{\citenamefont {Runeson}\ \emph {et~al.}(2022)\citenamefont {Runeson}, \citenamefont {Mannouch}, \citenamefont {Amati}, \citenamefont {Fiechter},\ and\ \citenamefont {Richardson}}]{Rune2022}%
  \BibitemOpen
  \bibfield  {author} {\bibinfo {author} {\bibfnamefont {J.~E.}\ \bibnamefont {Runeson}}, \bibinfo {author} {\bibfnamefont {J.~R.}\ \bibnamefont {Mannouch}}, \bibinfo {author} {\bibfnamefont {G.}~\bibnamefont {Amati}}, \bibinfo {author} {\bibfnamefont {M.~R.}\ \bibnamefont {Fiechter}}, \ and\ \bibinfo {author} {\bibfnamefont {J.~O.}\ \bibnamefont {Richardson}},\ }\bibfield  {title} {\enquote {\bibinfo {title} {Spin-mapping methods for simulating ultrafast nonadiabatic dynamics},}\ }\href {\doibase 10.2533/chimia.2022.582} {\bibfield  {journal} {\bibinfo  {journal} {Chimia}\ }\textbf {\bibinfo {volume} {76}},\ \bibinfo {pages} {582} (\bibinfo {year} {2022})}\BibitemShut {NoStop}%
\bibitem [{\citenamefont {Vaswani}\ \emph {et~al.}(2017)\citenamefont {Vaswani}, \citenamefont {Shazeer}, \citenamefont {Parmar}, \citenamefont {Uszkoreit}, \citenamefont {Jones}, \citenamefont {Gomez}, \citenamefont {Kaiser},\ and\ \citenamefont {Polosukhin}}]{Vasw2017}%
  \BibitemOpen
  \bibfield  {author} {\bibinfo {author} {\bibfnamefont {A.}~\bibnamefont {Vaswani}}, \bibinfo {author} {\bibfnamefont {N.}~\bibnamefont {Shazeer}}, \bibinfo {author} {\bibfnamefont {N.}~\bibnamefont {Parmar}}, \bibinfo {author} {\bibfnamefont {J.}~\bibnamefont {Uszkoreit}}, \bibinfo {author} {\bibfnamefont {L.}~\bibnamefont {Jones}}, \bibinfo {author} {\bibfnamefont {A.}~\bibnamefont {Gomez}}, \bibinfo {author} {\bibfnamefont {L.}~\bibnamefont {Kaiser}}, \ and\ \bibinfo {author} {\bibfnamefont {I.}~\bibnamefont {Polosukhin}},\ }\bibfield  {title} {\enquote {\bibinfo {title} {Attention is all you need},}\ }\href@noop {} {\bibfield  {journal} {\bibinfo  {journal} {31st Conference on Neural Information Processing Systems}\ } (\bibinfo {year} {2017})}\BibitemShut {NoStop}%
\bibitem [{\citenamefont {Tholke}\ and\ \citenamefont {De~Fabritiis}(2021)}]{Thol2021}%
  \BibitemOpen
  \bibfield  {author} {\bibinfo {author} {\bibfnamefont {P.}~\bibnamefont {Tholke}}\ and\ \bibinfo {author} {\bibfnamefont {G.}~\bibnamefont {De~Fabritiis}},\ }\bibfield  {title} {\enquote {\bibinfo {title} {Equivariant transformers for neural network based molecular potentials},}\ }\href@noop {} {\bibfield  {journal} {\bibinfo  {journal} {Fourth Workshop on Machine Learning and the Physical Sciences}\ } (\bibinfo {year} {2021})}\BibitemShut {NoStop}%
\bibitem [{\citenamefont {Pelaez}\ \emph {et~al.}(2024)\citenamefont {Pelaez}, \citenamefont {Simeon}, \citenamefont {Galvelis}, \citenamefont {Mirarchi}, \citenamefont {Eastman}, \citenamefont {Doerr}, \citenamefont {Th\"{o}lke}, \citenamefont {Markland},\ and\ \citenamefont {De~Fabritiis}}]{Pela2024}%
  \BibitemOpen
  \bibfield  {author} {\bibinfo {author} {\bibfnamefont {R.~P.}\ \bibnamefont {Pelaez}}, \bibinfo {author} {\bibfnamefont {G.}~\bibnamefont {Simeon}}, \bibinfo {author} {\bibfnamefont {R.}~\bibnamefont {Galvelis}}, \bibinfo {author} {\bibfnamefont {A.}~\bibnamefont {Mirarchi}}, \bibinfo {author} {\bibfnamefont {P.}~\bibnamefont {Eastman}}, \bibinfo {author} {\bibfnamefont {S.}~\bibnamefont {Doerr}}, \bibinfo {author} {\bibfnamefont {P.}~\bibnamefont {Th\"{o}lke}}, \bibinfo {author} {\bibfnamefont {T.~E.}\ \bibnamefont {Markland}}, \ and\ \bibinfo {author} {\bibfnamefont {G.}~\bibnamefont {De~Fabritiis}},\ }\bibfield  {title} {\enquote {\bibinfo {title} {Torchmd-net 2.0: Fast neural network potentials for molecular simulations},}\ }\href {\doibase 10.1021/acs.jctc.4c00253} {\bibfield  {journal} {\bibinfo  {journal} {J. Chem. Theory Comput.}\ }\textbf {\bibinfo {volume} {20}},\ \bibinfo {pages} {4076–4087} (\bibinfo {year} {2024})}\BibitemShut {NoStop}%
\bibitem [{\citenamefont {Kelly}\ \emph {et~al.}(2025)\citenamefont {Kelly}, \citenamefont {Hu}, \citenamefont {Damiani}, \citenamefont {Chen}, \citenamefont {Snider}, \citenamefont {Son}, \citenamefont {Lee}, \citenamefont {Gupta}, \citenamefont {Montoya-Castillo}, \citenamefont {Zuehlsdorff}, \citenamefont {Schlau-Cohen}, \citenamefont {Isborn},\ and\ \citenamefont {Markland}}]{Kell2025}%
  \BibitemOpen
  \bibfield  {author} {\bibinfo {author} {\bibfnamefont {J.}~\bibnamefont {Kelly}}, \bibinfo {author} {\bibfnamefont {F.}~\bibnamefont {Hu}}, \bibinfo {author} {\bibfnamefont {A.}~\bibnamefont {Damiani}}, \bibinfo {author} {\bibfnamefont {M.~S.}\ \bibnamefont {Chen}}, \bibinfo {author} {\bibfnamefont {A.}~\bibnamefont {Snider}}, \bibinfo {author} {\bibfnamefont {M.}~\bibnamefont {Son}}, \bibinfo {author} {\bibfnamefont {A.}~\bibnamefont {Lee}}, \bibinfo {author} {\bibfnamefont {P.}~\bibnamefont {Gupta}}, \bibinfo {author} {\bibfnamefont {A.}~\bibnamefont {Montoya-Castillo}}, \bibinfo {author} {\bibfnamefont {T.~J.}\ \bibnamefont {Zuehlsdorff}}, \bibinfo {author} {\bibfnamefont {G.~S.}\ \bibnamefont {Schlau-Cohen}}, \bibinfo {author} {\bibfnamefont {C.~M.}\ \bibnamefont {Isborn}}, \ and\ \bibinfo {author} {\bibfnamefont {T.~E.}\ \bibnamefont {Markland}},\ }\bibfield  {title} {\enquote {\bibinfo {title} {Two-dimensional electronic spectroscopy in the condensed phase using equivariant transformer accelerated
  molecular dynamics simulations},}\ }\href {\doibase 10.1021/acs.jpclett.5c00911} {\bibfield  {journal} {\bibinfo  {journal} {J. Phys. Chem. Lett.}\ }\textbf {\bibinfo {volume} {16}},\ \bibinfo {pages} {5561–5569} (\bibinfo {year} {2025})}\BibitemShut {NoStop}%
\bibitem [{\citenamefont {Yarkony}\ and\ \citenamefont {Silbey}(1976)}]{Yark1976}%
  \BibitemOpen
  \bibfield  {author} {\bibinfo {author} {\bibfnamefont {D.}~\bibnamefont {Yarkony}}\ and\ \bibinfo {author} {\bibfnamefont {R.}~\bibnamefont {Silbey}},\ }\bibfield  {title} {\enquote {\bibinfo {title} {Comments on exciton phonon coupling: Temperature dependence},}\ }\href {\doibase 10.1063/1.433182} {\bibfield  {journal} {\bibinfo  {journal} {J. Chem. Phys.}\ }\textbf {\bibinfo {volume} {65}},\ \bibinfo {pages} {1042--1052} (\bibinfo {year} {1976})}\BibitemShut {NoStop}%
\bibitem [{\citenamefont {Wang}\ and\ \citenamefont {Zhao}(2020)}]{Wang2020}%
  \BibitemOpen
  \bibfield  {author} {\bibinfo {author} {\bibfnamefont {Y.-C.}\ \bibnamefont {Wang}}\ and\ \bibinfo {author} {\bibfnamefont {Y.}~\bibnamefont {Zhao}},\ }\bibfield  {title} {\enquote {\bibinfo {title} {Variational polaron transformation approach toward the calculation of thermopower in organic crystals},}\ }\href {\doibase 10.1103/physrevb.101.075205} {\bibfield  {journal} {\bibinfo  {journal} {Phys. Rev. B}\ }\textbf {\bibinfo {volume} {101}},\ \bibinfo {pages} {075205} (\bibinfo {year} {2020})}\BibitemShut {NoStop}%
\bibitem [{\citenamefont {Runeson}\ and\ \citenamefont {Manolopoulos}(2025)}]{Rune2025}%
  \BibitemOpen
  \bibfield  {author} {\bibinfo {author} {\bibfnamefont {J.~E.}\ \bibnamefont {Runeson}}\ and\ \bibinfo {author} {\bibfnamefont {D.~E.}\ \bibnamefont {Manolopoulos}},\ }\bibfield  {title} {\enquote {\bibinfo {title} {Nuclear quantum effects slow down the energy transfer in biological light-harvesting complexes},}\ }\href {\doibase 10.1126/sciadv.adw4798} {\bibfield  {journal} {\bibinfo  {journal} {Sci. Adv.}\ }\textbf {\bibinfo {volume} {11}},\ \bibinfo {pages} {eadw4798} (\bibinfo {year} {2025})}\BibitemShut {NoStop}%
\end{thebibliography}%

\end{document}


\title{Two-dimensional electronic spectra from trajectory-based dynamics: pure-state Ehrenfest, spin-mapping, and mean classical path approaches}

\author{Annina Z. Lieberherr}
\thanks{These authors contributed equally to this work.}
\affiliation{Department of Chemistry, University of Oxford, Physical and Theoretical Chemistry Laboratory, South Parks Road, Oxford OX1 3QZ, United Kingdom}

\author{Joseph Kelly}
\thanks{These authors contributed equally to this work.}
\affiliation{Department of Chemistry, Stanford University, Stanford, California, 94305, USA}

\author{Johan E. Runeson}
\affiliation{Department of Chemistry, University of Oxford, Physical and Theoretical Chemistry Laboratory, South Parks Road, Oxford OX1 3QZ, United Kingdom}

\author{Thomas E. Markland}
\email{tmarkland@stanford.edu}
\affiliation{Department of Chemistry, Stanford University, Stanford, California, 94305, USA}

\author{David E. Manolopoulos}
\email{david.manolopoulos@chem.ox.ac.uk}
\affiliation{Department of Chemistry, University of Oxford, Physical and Theoretical Chemistry Laboratory, South Parks Road, Oxford OX1 3QZ, United Kingdom}

\date{\today}

\pacs{}

\maketitle
\tableofcontents

\newpage
\section{Polar and equatorial wave function evolution}
\label{sisec:wave-function-evolution}

\begin{figure}[tbp]
    \centering
    \begin{subfigure}{\textwidth}
    \centering
    \begin{tikzpicture}
        \node at (0,0) {\includegraphics[width=\textwidth]{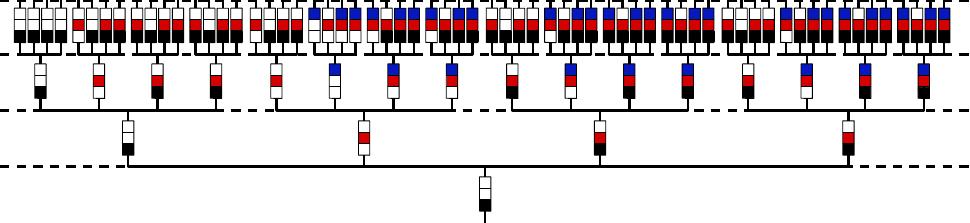}};
        \node at (-8.3,-.5) {$t_1$};
        \node at (-8.3,.5) {$t_2$};
        \node at (-8.3,1.4) {$t_3$};
    \end{tikzpicture}
    \subcaption{}
    \label{subfig:si-paths-polar}
    \end{subfigure}
    \begin{subfigure}{\textwidth}
    \centering
    \begin{tikzpicture}
        \node at (0,0) {\includegraphics[width=\textwidth]{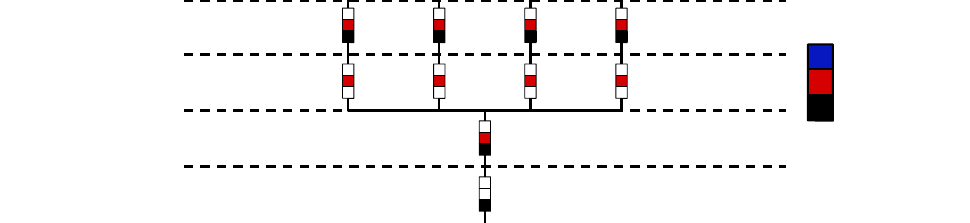}};
        \node at (-8.3,-.5) {$t_1$};
        \node at (-8.3,.5) {$t_2$};
        \node at (-8.3,1.4) {$t_3$};
        \node[align=left] at (6.3, 0.05) {$|g\rangle$};
        \node[align=left] at (6.3, 0.525) {$|s\rangle$};
        \node[align=left] at (6.3, 1.) {$|d\rangle$};
    \end{tikzpicture}
    \subcaption{}
    \end{subfigure}
    \caption{Wave function evolution for the response function $\Phi_1$ in (a) the polar decomposition and (b) the equatorial decomposition. The solid lines represent different pure states and the three boxes represent the electronic coefficients in the ground (black), singly excited (red) and doubly excited manifolds (blue).
    If a box is colored in, the electronic state has non-zero contributions from the corresponding manifold.}
    \label{fig:si-paths}
\end{figure}

Figure~\ref{fig:si-paths} shows the time evolution of the electronic wave function during the calculation of $\Phi_1$ with the polar and equatorial Ehrenfest methods.
The polar Ehrenfest graph is much wider and more complicated because it does not take advantage of the summation trick.
During the $t_2$ evolution, the Feynman diagram in Fig.~1 in the main text shows that the wave function is fully in the singly excited manifold. This is reflected in the equatorial Ehrenfest path.
In contrast, there are contributions from the ground and doubly excited manifolds in the polar Ehrenfest evolution.
Furthermore, even though we propagate 16 times as many states in the polar Ehrenfest method, many of them do not contribute to the final result.
This is because the response function involves a trace with the final dipole moment de-excitation operator. Since there are some polar pure states that only have contributions from a single manifold during $t_3$ (for example, the six states to the very left in the $t_3$ evolution in Fig.~\ref{subfig:si-paths-polar}), these will trace out to zero.

\newpage
\phantom{This is needed to make the next newpage work properly. }

\newpage
\rev{
\section{More on the symmetry of the response functions}}

\rev{The following argument explains why the symmetry of the rephasing and non-rephasing response functions cannot be exploited in the same way in a mean classical path calculation as it can in an equatorial Ehrenfest calculation.}

\rev{Consider the $\Phi_1$ and $\Phi_4$ response functions (for example). At the end of the $t_2$ evolution, the coherences that contribute to these response functions form a hermitian conjugate pair: ${\cal G}(t_2)[\hat{\mu}_+{\cal G}(t_1)[\hat{\rho}_0\hat{\mu}_-]]=|a\rangle\langle b|$ for $\Phi_1$ and ${\cal G}(t_2)[{\cal G}(t_1)[\hat{\mu}_+\hat{\rho}_0]\hat{\mu}_-]=|b\rangle\langle a|$ for $\Phi_4$. Both can be obtained by running a single mean classical path trajectory through $t_1$ and $t_2$, because the mean classical path force from a coherence and its hermitian conjugate are the same. So far, so good. However, a $\hat{\mu}_+$ operator now operates on both coherences from the right. Since $|a\rangle\langle b|\hat{\mu}_+$ is no longer the hermitian conjugate of $|b\rangle\langle a|\hat{\mu}_+$, the forces experienced by the mean classical path trajectories that contribute to $\Phi_1$ and $\Phi_4$ are different during the $t_3$ evolution, which is what dominates the cost of the calculation. (Note that the $t_3$ evolution must be done separately for each required $t_1$ and $t_2$, and for sufficiently many time steps to obtain the 2DES spectrum from a $t_3$ to $\omega_3$ Fourier transform.) In the equatorial Ehrenfest method, we have 4 branched pure state densities of the form $|\phi_j\rangle\langle \phi_j|$ at the end of the $t_2$ evolution. The pure states are the same for the $\Phi_1$ and $\Phi_4$ response functions. The only difference is that the weight of each pure state in $\Phi_4$ is the complex conjugate of that in $\Phi_1$. Since $|\phi_j\rangle\langle\phi_j|\hat{\mu}_+$ is the same in both cases, the equatorial Ehrenfest evolutions through $t_3$ are the same for both response functions. So only one of them need be computed and the other can be obtained by complex conjugating its weights. Similar arguments apply to the other two pairs of response functions ($\Phi_2$ and $\Phi_5$, and $\Phi_3$ and $\Phi_6$.)}

\newpage
\section{Doubly focused sampling of spin-mapping variables}

The goal of the sampling of spin-mapping variables mentioned in Sec.~III.C of the text is to construct a set of $M$ normalized wavefunctions $\ket{c_m}$ such that
\begin{equation}
    {1\over M}\sum_{m=1}^M \hat{\rho}_{{\rm S},m} = \ket{\tilde{\phi}_j}\bra{\tilde{\phi}_j},
\end{equation}
where
\begin{equation}
    \hat{\rho}_{{\rm S},m} = \sqrt{M+1}\ket{c_m}\bra{c_m} - {{\sqrt{M+1}-1}\over M}\hat{I}_M. 
    \label{eq:rhoSm}
\end{equation}
To simplify this task, we first construct an orthonormal basis of $M$ states, $\{\ket{k}\}_1^M$, in which $\ket{\tilde{\phi}_j}$ is the first state with $k=1$. This can be done by Gram-Schmidt orthonormalization of $M-1$ randomly generated SU$(M)$ coherent states with respect to $\ket{\tilde{\phi}_j}$. The expansion coefficients of $\ket{\tilde{\phi}_j}$ in the resulting basis are $\langle k|\tilde{\phi}_j\rangle=\delta_{k,1}$, and those of $\ket{c_m}$ are $c_{k,m}=\langle k|c_m\rangle$. The aim is thus to choose these coefficients such that
\begin{equation}
    {1\over M} \sum_{m=1}^M \left[\sqrt{M+1}\,c_{k,m}c_{k',m}^*-{\sqrt{M+1}-1\over M}\delta_{k,k'}\right] = \delta_{k,1}\delta_{k',1}.
    \label{eq:constraint}
\end{equation}

The diagonal terms in this equation (those with $k=k'$) constrain the square moduli of the coefficients:
\begin{equation}
    {1\over M} \sum_{m=1}^M \left[\sqrt{M+1}\,|c_{k,m}|^2-{\sqrt{M+1}-1\over M}\right] = \delta_{k,1}.
    \label{eq:moduli}
\end{equation}
We are free to choose $|c_{k,m}|^2$ to be independent of $m$, as it is in the standard `focused sampling' of a pure state density in the generalised spin mapping method.\cite{Rune2020} Solving Eq.~\eqref{eq:moduli} then gives $|c_{k=1,m}|^2 = \alpha$ and $|c_{k>1,m}|^2 = \beta$, where
\begin{align}
\alpha &= {1\over M}+{1\over \sqrt{M+1}}-{1\over M\sqrt{M+1}},\\
\beta &= {1\over M}-{1\over M\sqrt{M+1}}.
\end{align}
Since $\alpha+(M-1)\beta=1$, this ensures that the states $\ket{c_m}$ are correctly normalized:
\begin{equation}
\langle c_m|c_m\rangle = \sum_{k=1}^M |c_{k,m}|^2 = 1.
\end{equation}

We can now write $c_{k=1,m}=\sqrt{\alpha}e^{+i\phi_{k,m}}$ and
$c_{k>1,m}=\sqrt{\beta}e^{+i\phi_{k,m}}$. Substituting these expressions into Eq.~\eqref{eq:constraint} gives the following constraints on the phases $\phi_{k,m}$,
\begin{equation}
{1\over M}\sum_{m=1}^M e^{{+i}(\phi_{k,m}-\phi_{k',m})} = 0 \ \hbox{for all}\  k\not=k'.
\label{eq:phase-constraints}
\end{equation}
\rev{
The most general way to satisfy these constraints is to set
\begin{equation}
\phi_{k,m} = \mu_k+\nu_m+2\pi(k-1)(m-1)/M,
\label{eq:phikm1}
\end{equation}
which gives
\begin{equation}
{1\over M}\sum_{m=1}^{M} e^{+i(\phi_{k,m}-\phi_{k',m})}
= e^{+i(\mu_k-\mu_{k'})}{1\over M}\sum_{m=1}^M e^{+2\pi i(k-k')(m-1)/M} = e^{+i(\mu_k-\mu_{k'})}\delta_{k,k'} = \delta_{k,k'}. 
\end{equation}
However, the overall phases $\nu_m$ of the states $\ket{c_m}$ are irrelevant because they do not affect $\hat{\rho}_{{\rm S},m}$, and therefore have no impact on either the dynamics or the target response function. We therefore set these to zero, sample each $\mu_k$ uniformly in the interval $[0,2\pi)$, and calculate $\langle k|c_m\rangle$ as $\sqrt{\alpha}e^{+i\phi_{k,m}}$ for $k=1$ and $\sqrt{\beta}e^{+i\phi_{k,m}}$ for $k>1$.} Finally, we dispense with the Gram-Schmidt basis by transforming to the excitonic basis, which in the case of the SE and ESA response functions is the basis of singly excited electronic states $\ket{n}$:
\begin{equation}
\langle n|c_m\rangle = \sum_{k=1}^M \langle n|k\rangle \langle k|c_m\rangle.
\end{equation}

\rev{
Because this sampling algorithm involves a stochastic element, we should in principle repeat it multiple times for each initial pure state density $\ket{\tilde{\phi}_j}\bra{\tilde{\phi}_j}$ and average over the resulting response functions to reduce the statistical error in the calculation, analogous to what is done in spin mapping with focused sampling.\cite{Rune2020}} However, there is less need to do this here than when using ordinary focused sampling because the error in that is dominated by a lack of phase cancellation, whereas we have ensured that the phases exactly cancel in the present `doubly focused' approach by enforcing the phase constraints in Eq.~\eqref{eq:phase-constraints}. Because of this, and because we have to sample the bath variables at the beginning of each trajectory in any case, we have found it preferable to stick with just one Gram-Schmidt basis per $\ket{\tilde{\phi}_j}\bra{\tilde{\phi}_j}$ pure state and reduce the statistical errors in our response functions by increasing the total number of trajectories. This effectively combines the Monte Carlo sampling of the initial bath variables with that of the Gram-Schmidt basis states.

\newpage
\section{Simulation details}
\label{sisec:simulation-details}
\subsection{Biexciton Model}
\label{sisec:biex-simulation-details}
The biexciton model consists of $N=2$ sites which yield two singly excited states and one doubly excited state. The site energies are $\varepsilon_1 = -50$\;cm$^{-1}$ and $\varepsilon_2 = 50$\;cm$^{-1}$, and there is just one site-site coupling matrix element, $J_{12}=100$\;cm$^{-1}$. Each site is coupled to a bath defined by a Debye spectral density with a reorganization energy $\lambda = 50~\cm$ and a characteristic frequency $\omega_c = 300~\cm$. We use a standard procedure\cite{Wang2001, Crai2007} to discretize this bath with $K=300$ oscillators. The bath frequencies, $\omega_k$, and couplings, $c_k$, are
\begin{subequations}
    \begin{equation}
        \omega_k = \omega_c \tan\left[ \frac{\pi}{2K}\left(k-\frac{1}{2}\right)\right], 
    \end{equation}
    \begin{equation}
        c_k = \omega_k\sqrt{\frac{2\lambda}{K}}.
    \end{equation}
\end{subequations}
We have chosen these bath parameters to be in a regime that is known to cause problems for the polar Ehrenfest method.\cite{Atsa2023}
Since the thermal energy $k_\mathrm{B}T = 208$\;cm$^{-1}$ is less than $\omega_c=300$ cm$^{-1}$, we used the Wigner distribution in Eq.~(40) in the main text to sample the bath modes.
The transition dipole operator has the elements $\mu_1=1$ and $\mu_2 = -0.2$ (Eqs.~(36)-(38) in the main text).
Simulations were run at 300 K with a 10 fs time step. 2DES spectra were obtained from correlation functions calculated for $t_2 \in \{ 0,~ 50,~ 100,~ 150,~ 200~\mathrm{fs}\}$ and for $t_1$ and $t_3$ out to 500~fs.
The HEOM results were obtained with \texttt{pyrho},\cite{Berk2020} using the convergence parameters $\texttt{L=15}$ and $\texttt{K=0}$.\cite{Atsa2023} 

\subsection{FMO Model}
\label{sisec:fmo-simulation-details}
We use a previously defined parameterization of the FMO model\cite{Kram2018} which consists of $N=7$ sites yielding 7 singly excited states and 21 doubly excited states. The singly excited block of the Hamiltonian in wavenumbers is
\begin{equation}
\begin{aligned}
    \hat H^s_\mathrm{S} \!=\!& \begin{pmatrix}
        \redspace 200 &\redspace -87.7 &\redspace 5.5 &\redspace -5.9 &\redspace 6.7 &\redspace \redspace -13.7 &\redspace -9.9 \redspace\\
        \redspace -87.7 &\redspace 320 &\redspace 30.8 &\redspace 8.2 &\redspace 0.7 &\redspace 11.8 &\redspace 4.3 \redspace\\
        \redspace 5.5 &\redspace 30.8 &\redspace 0 &\redspace -53.5 &\redspace -2.2 &\redspace -9.6 &\redspace 6.0 \redspace\\
        \redspace -5.9 &\redspace 8.2 &\redspace -53.5 &\redspace 110 &\redspace -70.7 &\redspace -17.0 &\redspace -63.3 \redspace\\
        \redspace 6.7 &\redspace 0.7 &\redspace -2.2 &\redspace -70.7 &\redspace 270 &\redspace 81.1 &\redspace -1.3 \redspace\\
        \redspace -13.7 &\redspace 11.8 &\redspace -9.6 &\redspace -17.0 &\redspace 81.1 &\redspace 420 &\redspace 39.7 \redspace\\
        \redspace -9.9 &\redspace 4.3 &\redspace 6.0 &\redspace -63.3 &\redspace -1.3 &\redspace 39.7 &\redspace 230 \redspace
    \end{pmatrix} \\
     &+ 12210\; \hat{I},
\end{aligned}
\end{equation}
where $\hat{I}$ is the $7\times 7$ identity matrix.
Each site is coupled to a bath defined by a Debye spectral density with a reorganization energy $\lambda = 35~\cm$ and a characteristic frequency $\omega_c = 106~\cm$. We discretize this bath with $K=60$ oscillators.
Simulations were run at 300 K with a 10 fs time step. 2DES spectra were obtained from correlation functions calculated for $t_2 \in \{ 0,~ 200,~ 400,~ 600~\mathrm{fs}\}$ and for $t_1$ and $t_3$ out to 500~fs.

The dipole moments of the seven bacteriochlorophylls are in the directions 
\begin{align}
    \bs d_1 &= \begin{pmatrix} -0.741, & -0.561, & -0.370 \end{pmatrix}, \\
    \bs d_2 &= \begin{pmatrix} -0.857, &  0.504, & -0.107 \end{pmatrix}, \\
    \bs d_3 &= \begin{pmatrix} -0.197, &  0.957, & -0.211 \end{pmatrix}, \\
    \bs d_4 &= \begin{pmatrix} -0.799, & -0.534, & -0.277 \end{pmatrix}, \\
    \bs d_5 &= \begin{pmatrix} -0.737, &  0.656, &  0.164 \end{pmatrix}, \\
    \bs d_6 &= \begin{pmatrix} -0.135, & -0.879, &  0.457 \end{pmatrix}, \\
    \bs d_7 &= \begin{pmatrix} -0.495, & -0.708, & -0.503 \end{pmatrix},
\end{align}
from which the full dipole moment operator $\hat{\bs \mu} = (\hat \mu_x, \hat \mu_y, \hat \mu_z)$ is constructed as
\begin{equation}
    \hat{\bs \mu} = \sum_{n=1}^7 (\ket{0}\bra{n} + \ket{n}\bra{0}) {\bs d}_{n}.
\end{equation}
Note that the magnitude of the dipole moment is immaterial because we are only interested in the relative intensities of the peaks in the spectrum.

To model an experiment performed on an ensemble of randomly oriented FMO complexes, we need to rotationally average the response function, or equivalently average over light polarizations.\cite{Hein2012} We begin by assuming that the light pulses all have the same polarization, so that all of the dipole moment operators in Eq.~(1) of the main text are the same.

For a given polarization vector $\bs e$, the dipole moment operator coupling the ground state to the singly excited manifold is
\begin{equation}
    \hat{\mu}^{sg} = \hat{\bs \mu} \cdot \bs e.
\end{equation}
Since the linear response function contains two dipole interactions, the relevant rotational average is
\begin{equation}
    \langle (\hat{\bs \mu} \cdot \bs e)^2\rangle = \frac{1}{4\pi}\int_{0}^{2\pi} \int_0^\pi (\hat{\bs \mu} \cdot \bs e)^2 \sin\theta \dd \theta \dd \varphi,
\end{equation}
where $\theta$ and $\varphi$ are the spherical polar angles of $\bs e$. This can be evaluated as
\begin{equation}
    \langle (\hat{\bs \mu} \cdot \bs e)^2\rangle = \frac{1}{3} \sum_{\alpha=x,y,z} (\hat{\bs \mu} \cdot \bs e_\alpha)^2,
    \label{eq:rotaveraging-linear}
\end{equation}
where $\bs e_x = (1, 0, 0)$, $\bs e_y = (0, 1, 0)$ and $\bs e_z = (0, 0, 1)$. 

The third order response function contains four dipole interactions. The relevant rotational average in this case is more complicated, but it can be evaluated in much the same way, by averaging over the ten polarization directions\cite{Hein2012} 
\begin{equation}
\label{eq:pol-2des}
    \begin{aligned}
        \bs e_1 &= (1,1,1), & \bs e_6 &= (0, -\frac{1}{\phi}, \phi), \\
        \bs e_2 &= (-1,1,1), & \bs e_7 &= (\frac{1}{\phi}, \phi, 0), \\
        \bs e_3 &= (1,-1,1), & \bs e_8 &= (-\frac{1}{\phi},\phi,0), \\
        \bs e_4 &= (-1,-1,1), & \bs e_9 &= (\phi, 0, \frac{1}{\phi}), \\
        \bs e_5 &= (0,\frac{1}{\phi},\phi), & \bs e_{10} &= (-\phi, 0, \frac{1}{\phi}),
    \end{aligned}
\end{equation}
where $\phi = \frac{1 + \sqrt{5}}{2}$ is the golden ratio. 

In practice, the rotational averaging therefore requires us to repeat the calculation for 3 different polarizations to calculate the linear response function, and for 10 polarizations to calculate the third order response function. For each polarization, $\bs e$, the dipole matrix elements, $\mu_n$, in Eqs.~(36) and (37) in the main text are given by $\mu_n = {\bs d}_n\cdot{\bs e}$. Note that the polarization vectors $\bs e_1$--$\bs e_{10}$ in Eq.~\eqref{eq:pol-2des} are not normalized: they each have a length of $\sqrt{3}$. However, this simply results in an overall scaling of the spectra and does not affect the relative intensities of the peaks.

\newpage
\section{Choice of bath distribution}
\label{sisec:bath-sampling}
\subsection{Biexciton Model}
\label{sisec:biex-bath-sampling}
\begin{figure}[t]
    \centering
    \includegraphics[width=0.75\textwidth]{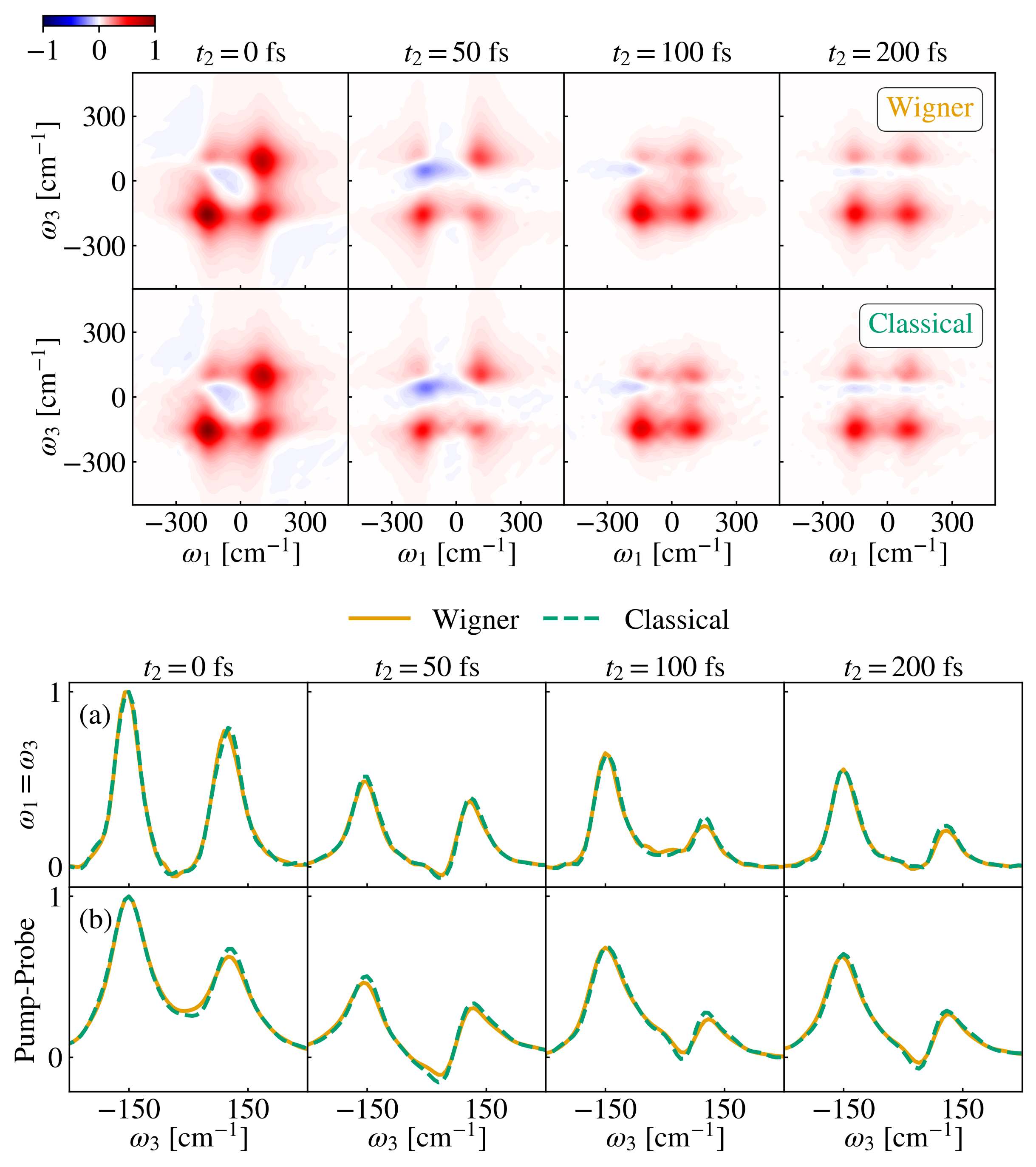}
    \caption{Comparison of Wigner and classical sampling of the bath modes for the biexciton model in an equatorial Ehrenfest calculation. Top: 2DES spectra. Bottom: diagonal cuts and pump-probe spectra.}
    \label{fig:biexc-bath-sampling}
\end{figure}

The baths in the biexciton model should be sampled using a Wigner distribution because $\omega_c=300~\cm > k_{\rm B}T=208~\cm$. However, this introduces an additional possible source of error: zero-point energy leakage. To investigate the significance of this effect, we have also simulated the biexciton model with a classically sampled bath. Figure~\ref{fig:biexc-bath-sampling} compares the 2DES contour plots, diagonal $(\omega_1=\omega_3)$ slices through the 2DES spectra, and pump-probe spectra obtained using Wigner and classical initial conditions. It is clear that the choice of bath distribution does not significantly change any of the predicted spectra in this parameter regime even though it has $\omega_c > k_{\rm B}T$ (presumably because $\omega_c$ and $k_{\rm B}T$ are still sufficiently close for thermal quantum effects to have a negligible impact). 

\subsection{FMO Model}
\label{sisec:fmo-bath-sampling}
\begin{figure}[t]
    \centering
    \includegraphics[width=0.6\textwidth]{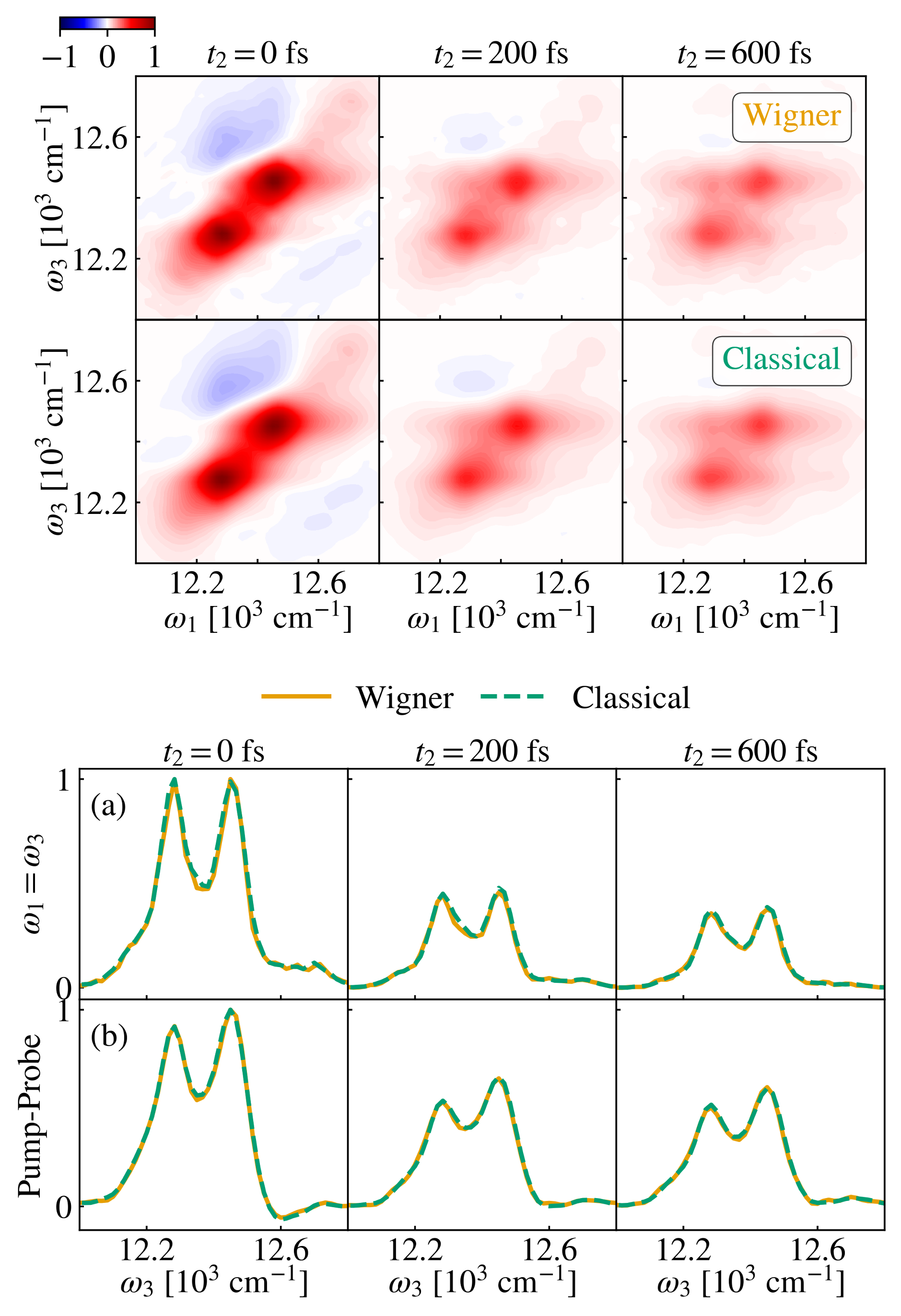}
    \caption{Comparison of Wigner and classical sampling of the bath modes for the FMO model in an equatorial Ehrenfest calculation. Top: 2DES spectra. Bottom: diagonal cuts and pump-probe spectra.}
    \label{fig:fmo-bath-sampling}
\end{figure}

We have used a classical Boltzmann distribution to sample the bath degrees of freedom for the FMO model because it has $\omega_c=106~\cm < k_{\rm B}T=208~\cm$. To verify that this is appropriate, we compare the results obtained from the equatorial approach using classical and Wigner distributions in Fig.~\ref{fig:fmo-bath-sampling}. The results obtained from the two distributions are graphically indistinguishable.

\newpage
\phantom{blah blah blah}

\newpage
\phantom{blah blah blah}

\newpage
\section{Numerical effort}
\label{sisec:effort}

Here we show how to obtain the number of force evaluations involved in simulating a single trajectory using the polar, equatorial, and spin-mapping approaches by progressing through each step of the simulation algorithm. 

In the polar approach, all 6 pathways must be calculated independently. We begin the trajectory for each pathway by decomposing the initial coherence into 4 pure states and evolving each through $N_{t_1}$ time steps, giving a total of $6\cdot4N_{t_1}$ force evaluations. Each of these four states is then decomposed into 4 new pure states which are evolved through $N_{t_2}$ time steps. Each step during the $t_2$ evolution either corresponds to a $t_2$ value for which we calculate the full spectrum (e.g. $t_2=100$~fs in the biexciton model), or an intermediate value (e.g. $t_2=30$~fs). Because we are interested in analyzing the spectra at a small number ($N_\mathrm{plot}$) of specific $t_2$'s, no $t_3$ evaluation is necessary for the intermediate $t_2$'s. This means that, for each trajectory, the number of force evaluations arising from intermediate $t_2$'s is
\begin{equation}
    6 \left(4N_{t_1}\right) \left[4(N_{t_2}-N_\mathrm{plot})\right].
\end{equation}
For the $t_2$'s we are interested in fully analyzing, we decompose each state into 4 further pure states after the next dipole interaction and evolve them through the remaining $N_{t_3}$ steps. This yields 
\begin{equation}
    6 \left(4N_{t_1}\right) \left( 4N_\mathrm{plot}\right) \left(4N_{t_3}\right)
\end{equation}
force evaluations for the target $t_2$'s. Ultimately, the total number of force evaluations for one polar trajectory is thus
\begin{equation}
    6 \left(4N_{t_1}\right) \left[4(N_{t_2}-N_\mathrm{plot})+\left( 4N_\mathrm{plot}\right) \left(4N_{t_3}\right)\right]. 
\end{equation}

In the equatorial approach, the same pattern applies but symmetries in the decomposition enable us to use the resummation trick in Eq.~(22) of the main text for the $t_1$ and $t_3$ evolutions. In addition, only three of the six pathways are independent in the equatorial decomposition, which yields an overall number of force evaluations for each trajectory of 
\begin{equation}
    \label{eq:equatorial-force-evaluations}
    3 \left( N_{t_1}\right) \left[ 4(N_{t_2}-N_\mathrm{plot})+\left(4N_\mathrm{plot}\right) \left( N_{t_3}\right)\right].
\end{equation}
 Here the first term requires a factor of 8 fewer force evaluations while the second requires a factor of 32 fewer force evaluations for the equatorial approach compared to the polar. Although we only consider a few target values of $t_2$, the second term invariably dominates as illustrated by the factor of 30.6 improvement from the polar to equatorial approach for the biexciton model. 

The spin-mapping approach is identical to the equatorial approach for the ground state bleaching (GSB) pathway. However, for the stimulated emission (SE) and excited state absorption (ESA) pathways, the spin-mapping approach further decomposes each state in a basis that spans the singly excited manifold at the start of the $t_2$ evolution, and then again in another basis of the same size at the start of the $t_3$ evolution, as described in Sec.~S1. This means for the SE and ESA pathways, the intermediate and plotted $t_2$'s require a factor of $N$ and $N^2$ additional force evaluations, respectively, where $N$ is the number of singly excited states in the model. In total, the GSB pathway involves one third of the total number of equatorial force evaluations in Eq.~\eqref{eq:equatorial-force-evaluations}, while the SE and ESA pathways together require
\begin{equation}
    2\left(N_{t_1}\right) \left[4N(N_{t_2}-N_\mathrm{plot})+4N^2\left( N_\mathrm{plot}\right) \left(N_{t_3}\right)\right].
\end{equation}
The plotting term is still dominant as illustrated by the factor of $3.9\simeq N^2$ increase in the number of force evaluations required by spin-mapping relative to the equatorial Ehrenfest method for the SE and ESA pathways of the biexciton model.

\newpage
\section{Convergence tests}
\label{sisec:convergence}

\subsection{Convergence with respect to the number of trajectories}
\label{sisec:convergence-trajectories}

\begin{figure}[b]
    \centering
    \includegraphics[width=0.9\linewidth]{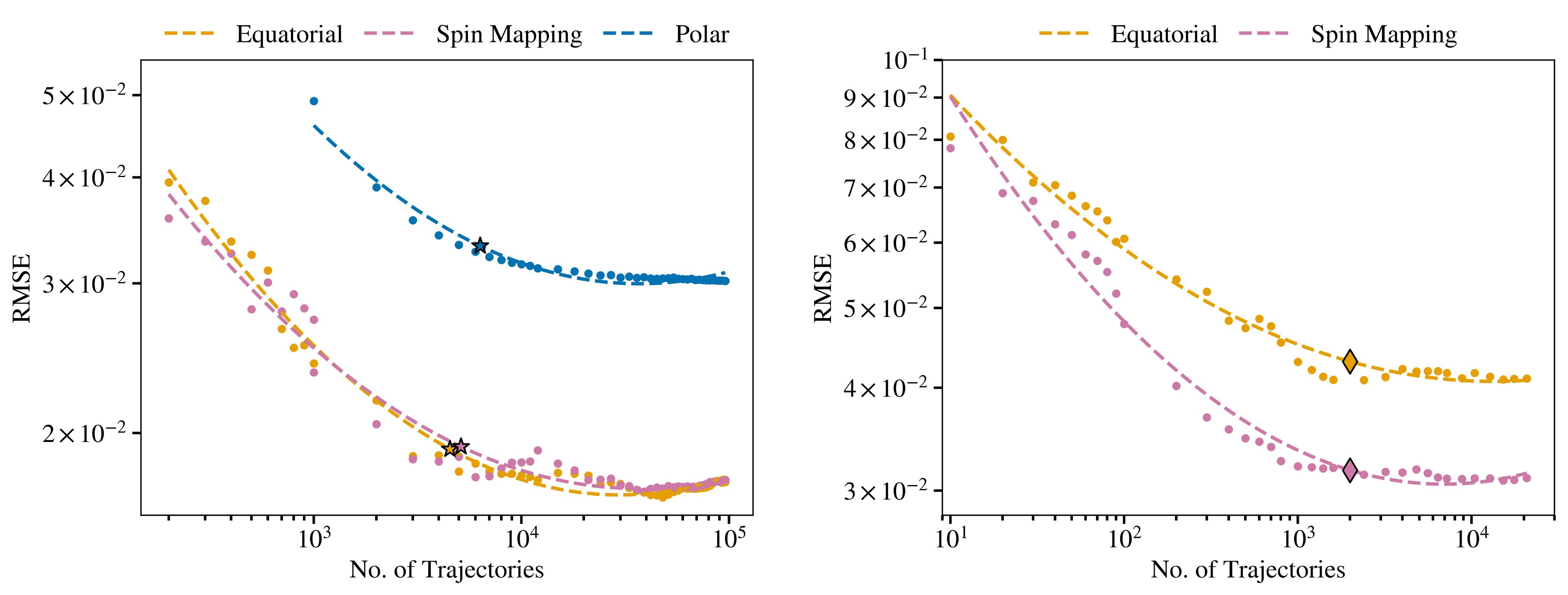}
    \caption{RMSEs of 2D spectra relative to HEOM. Left: Biexciton model. The stars mark the points where the RMSE is within 10\% of its final value. Right: FMO model. The diamonds mark the numbers of force evaluations needed to run 2,000 trajectories.}
    \label{fig:trajectory-error-convergence}
\end{figure}

While the equatorial, spin-mapping, and polar approaches have very different convergence behavior relative to the number of force evaluations (Figs.~6 and 10 in the main text), the convergence as a function of number of trajectories is more consistent (Fig.~\ref{fig:trajectory-error-convergence}). For the biexciton model (Fig.~\ref{fig:trajectory-error-convergence}, left), the equatorial and spin-mapping approaches converge at nearly the same rate and to the same value, while the polar approach also converges at a similar rate but to a less accurate result. For the FMO model (Fig.~\ref{fig:trajectory-error-convergence}, right), the spin-mapping approach is more accurate than equatorial approach but both converge at a similar rate. (The fact that the spin mapping and equatorial methods do converge at the same rate for both models as a function of the number of trajectories validates the statement we have made at the end of Sec.~S3 about the benefit of combining the Monte Carlo sampling of the initial bath variables with that of the spin mapping basis states. Sampling more than one spin mapping basis for each equatorial pure state $\ket{\tilde{\phi}_j}\bra{\tilde{\phi}_j}$ would be overkill because the statistical error is dominated by the sampling of the bath variables. If this were not the case, the equatorial Ehrenfest results  would converge more rapidly with respect to the number of trajectories than the spin mapping results in Fig.~\ref{fig:trajectory-error-convergence}.) 

\newpage
\subsection{Spectra within 10\% of the converged result for the biexciton model}
\label{sisec:biex-percenterror-spectra}

\begin{figure}[h]
    \centering
    \includegraphics[width=1\linewidth]{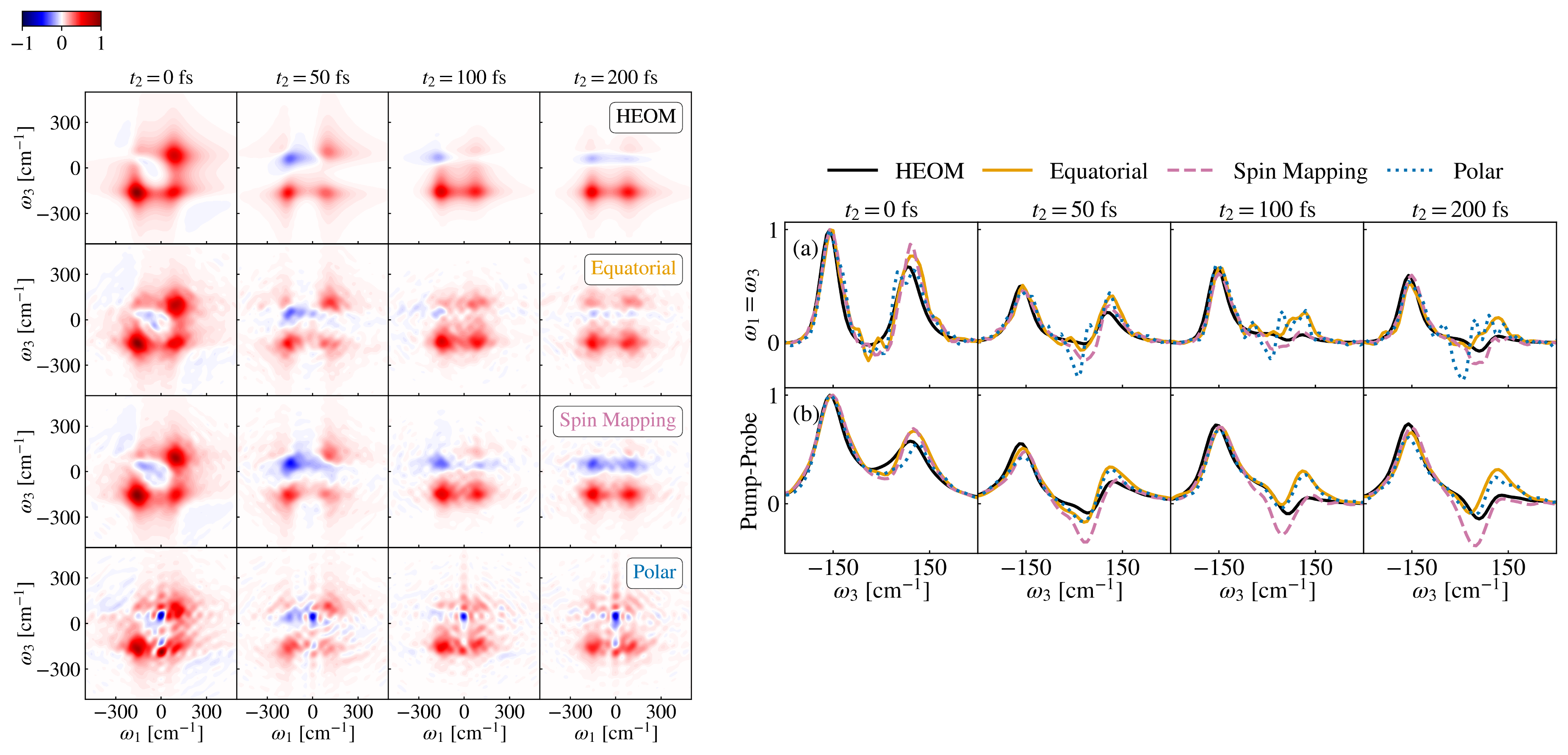}
    \caption{Less than fully converged spectral results for the biexciton model, obtained from sufficiently many trajectories to give a RMSE within 10\% of its final value.}    
    \label{fig:biex-10percent-results}
\end{figure}

Figure~\ref{fig:biex-10percent-results} shows the 2DES spectra, diagonal cuts, and pump-probe spectra obtained using data taken from the points where the stars are shown in Fig.~\ref{fig:trajectory-error-convergence} (and in Fig.~6 of the main text). While less than fully converged, these results from 4500, 5100, and 6300 trajectories of the equatorial, spin-mapping, and polar approaches, respectively, clearly reveal the same features as the fully converged spectra in Figs.~3 and 4 of the main text. 

\vspace{1cm}

\newpage
\subsection{Spectra from 2,000 trajectories for the biexciton model}

\begin{figure}[h]
    \centering
    \includegraphics[width=1\linewidth]{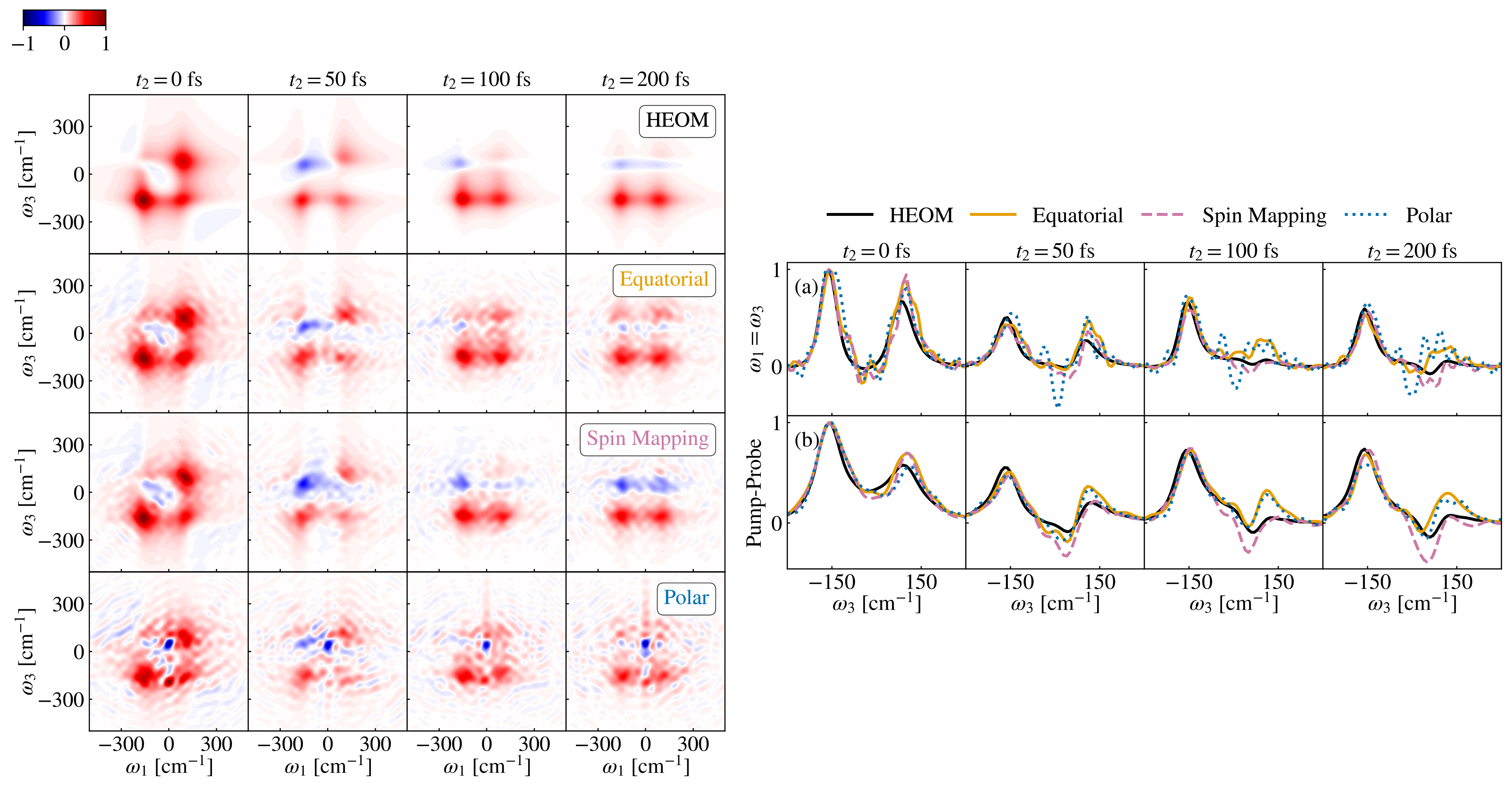}
    \caption{Unconverged spectral results for the biexciton model, obtained from just 2,000 trajectories of each method.}
    \label{fig:biex-2d-2000trajectories}
\end{figure}

As a more stringent test of the convergence of the various trajectory-based methods for the biexciton model, we show in Fig.~\ref{fig:biex-2d-2000trajectories} the spectra obtained from just 2,000 trajectories of each method. This is the same as the number of trajectories we used to illustrate what unconverged FMO spectra look like in Fig.~9 of the main text. Now the spectra generated by the polar approach are harder to interpret than they were in Fig.~\ref{fig:biex-10percent-results}, especially along the diagonal slices. However, the equatorial and spin-mapping approaches still yield perfectly interpretable spectra, highlighting just how little data is needed from these methods for a qualitative analysis.

\label{sisec:biex-2000traj-spectra}

\newpage
\subsection{Spectra from 2,000 trajectories for the FMO model}
\label{sisec:fmo-2000traj-spectra}

\begin{figure}[h]
    \centering
    \includegraphics[width=0.6\linewidth]{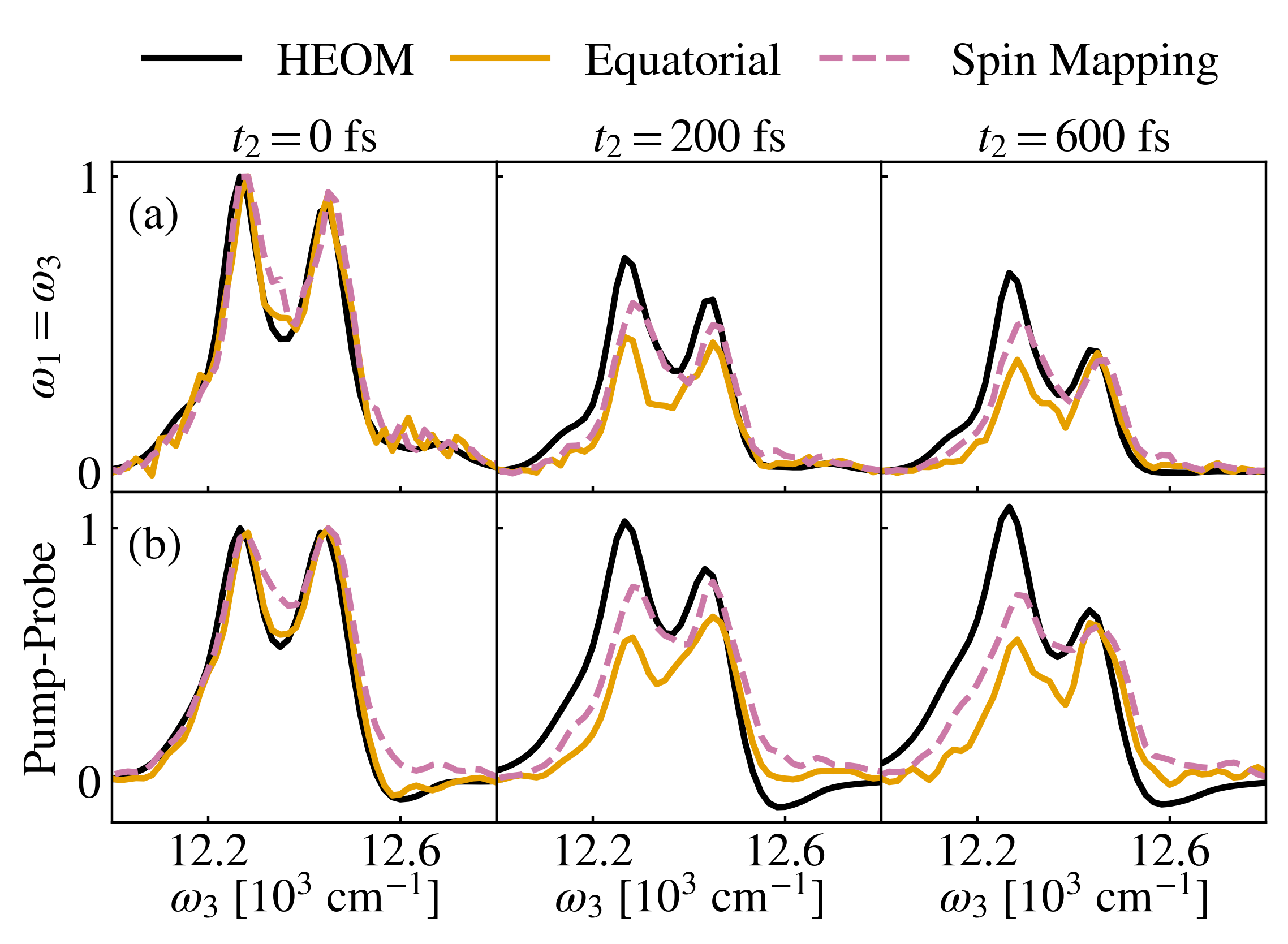}
    \caption{Diagonal slices and pump-probe spectra for the FMO model obtained from just 2,000 equatorial and spin mapping trajectories.}
    \label{sifig:fmo-1d-conv}
\end{figure}

Figure~9 of the main text shows the 2DES spectra of the FMO model obtained from just 2,000 trajectories of the equatorial and spin mapping methods. Here we show the corresponding diagonal slices and pump-probe spectra in Fig.~\ref{sifig:fmo-1d-conv}. While 2,000 trajectories are not enough for full convergence, they are enough to  reveal the key features of the fully converged diagonal slices and pump-probe spectra, which are shown in Fig.~8 of the main text. 

\vspace{1cm}

\newpage
\rev{
\section{Comparison with the mean classical path approach}}
\label{sisec:mcp}

\rev{
\subsection{Spectra from 2,000 trajectories for the FMO model}}

\rev{
Figs.~11 and 12 of the main text compare fully converged equatorial Ehrenfest and mean classical path spectra for the FMO model. Fig.~S8 shows how this comparison changes when we use just 2,000 trajectories for each method. The qualitative features of the spectra are already captured with this many trajectories in both cases.}

\vspace{1cm}

\begin{figure}[h]
\begin{subfigure}{0.49\textwidth}
    \centering
    \includegraphics[width=0.95\linewidth]{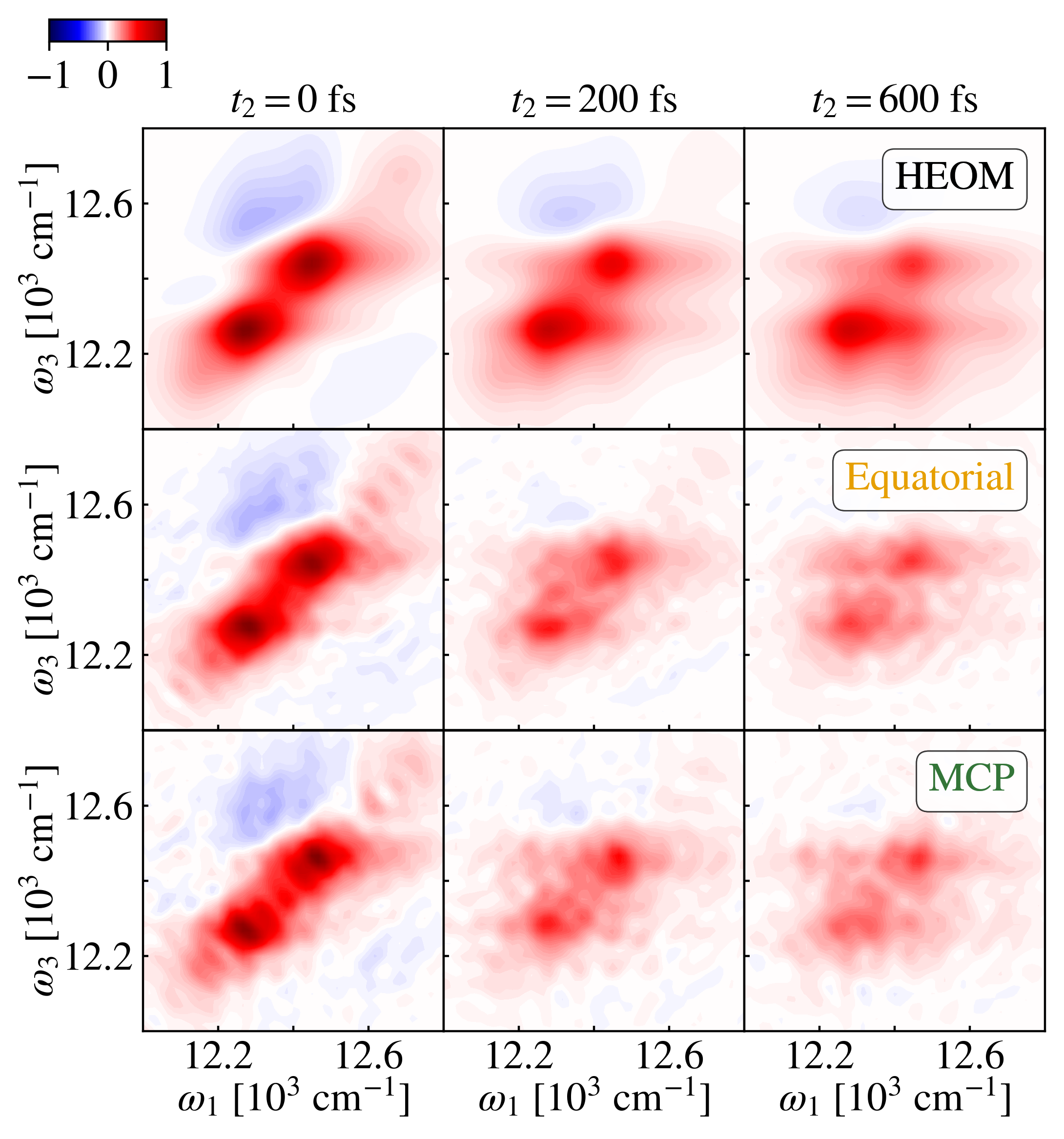}
    \label{sifig:fmo-eq-vs-mcp-2des-2000traj}
\end{subfigure}
\begin{subfigure}{0.49\textwidth}
    \centering
     \includegraphics[width=0.95\linewidth]{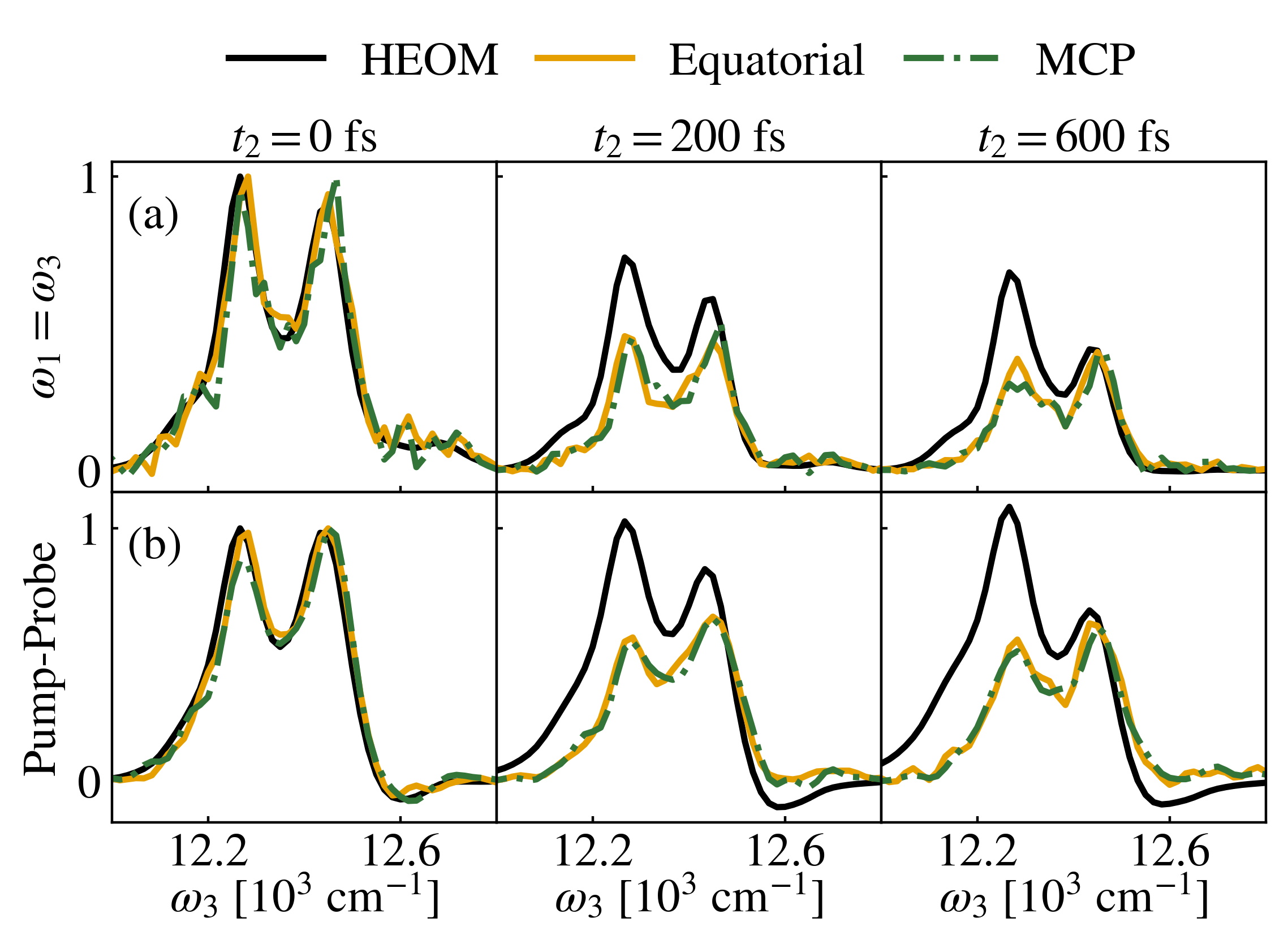}
    \label{sifig:biex-eq-vs-mcp-diag-pp-2000traj}
\end{subfigure}
\caption{\rev{Comparison of 2,000 trajectory equatorial Ehrenfest and mean classical path spectra for the FMO model. Left: 2DES spectra. Right: Diagonal slices and pump-probe spectra.}}
\end{figure}

\newpage
\rev{
\subsection{Fully converged spectra for the biexciton model}}

\rev{Here we present a comparison between the equatorial Ehrenfest and mean classical path spectra for the biexciton model. As for the FMO model, there are only very minor differences between the results of the two methods.}

\vspace{1cm}

\begin{figure}[h]
\begin{subfigure}{0.8\textwidth}
    \centering
    \includegraphics[width=0.8\linewidth]{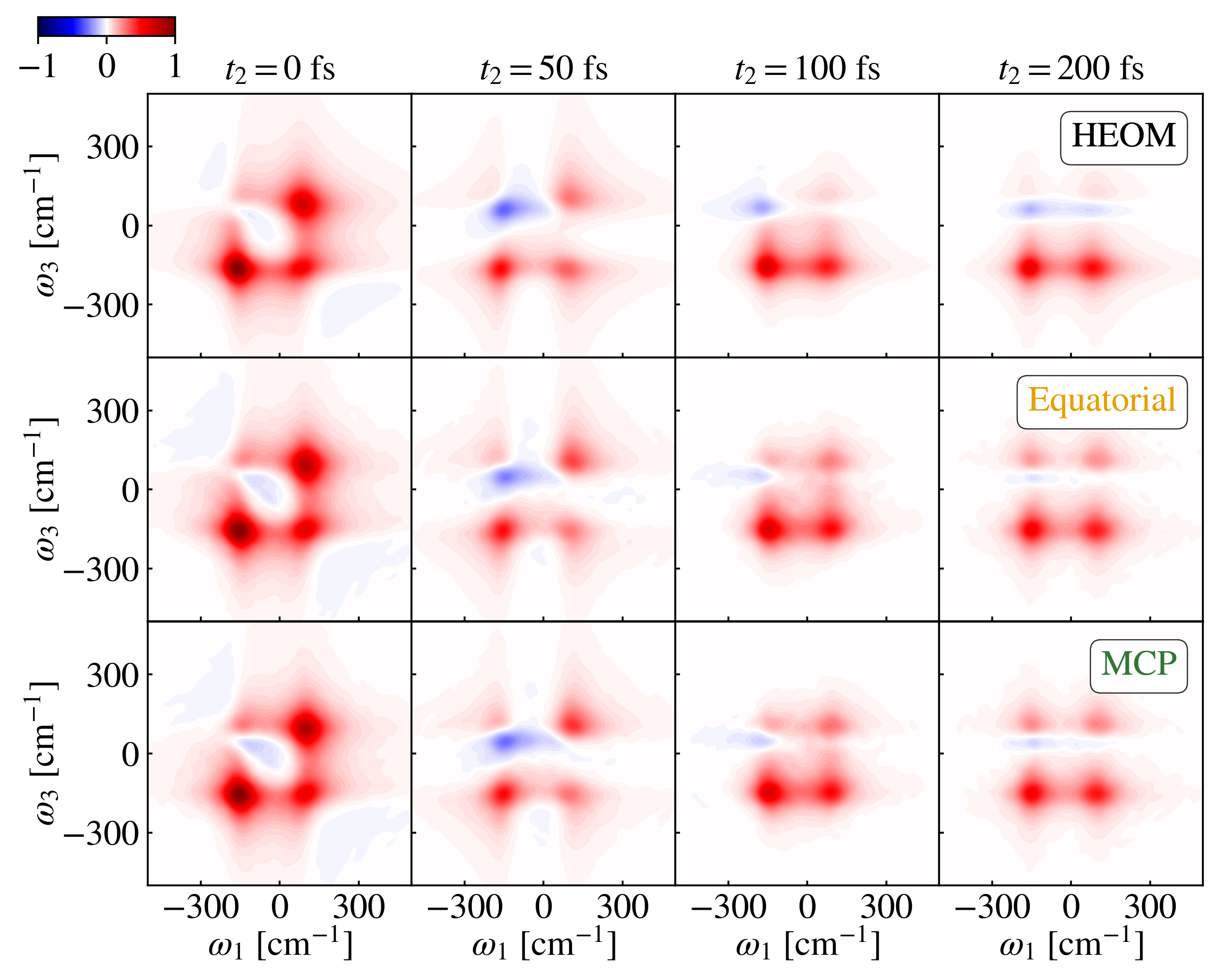}
\end{subfigure}
\vspace{0.5cm}
\begin{subfigure}{0.8\textwidth}
    \centering
    \includegraphics[width=0.8\linewidth]{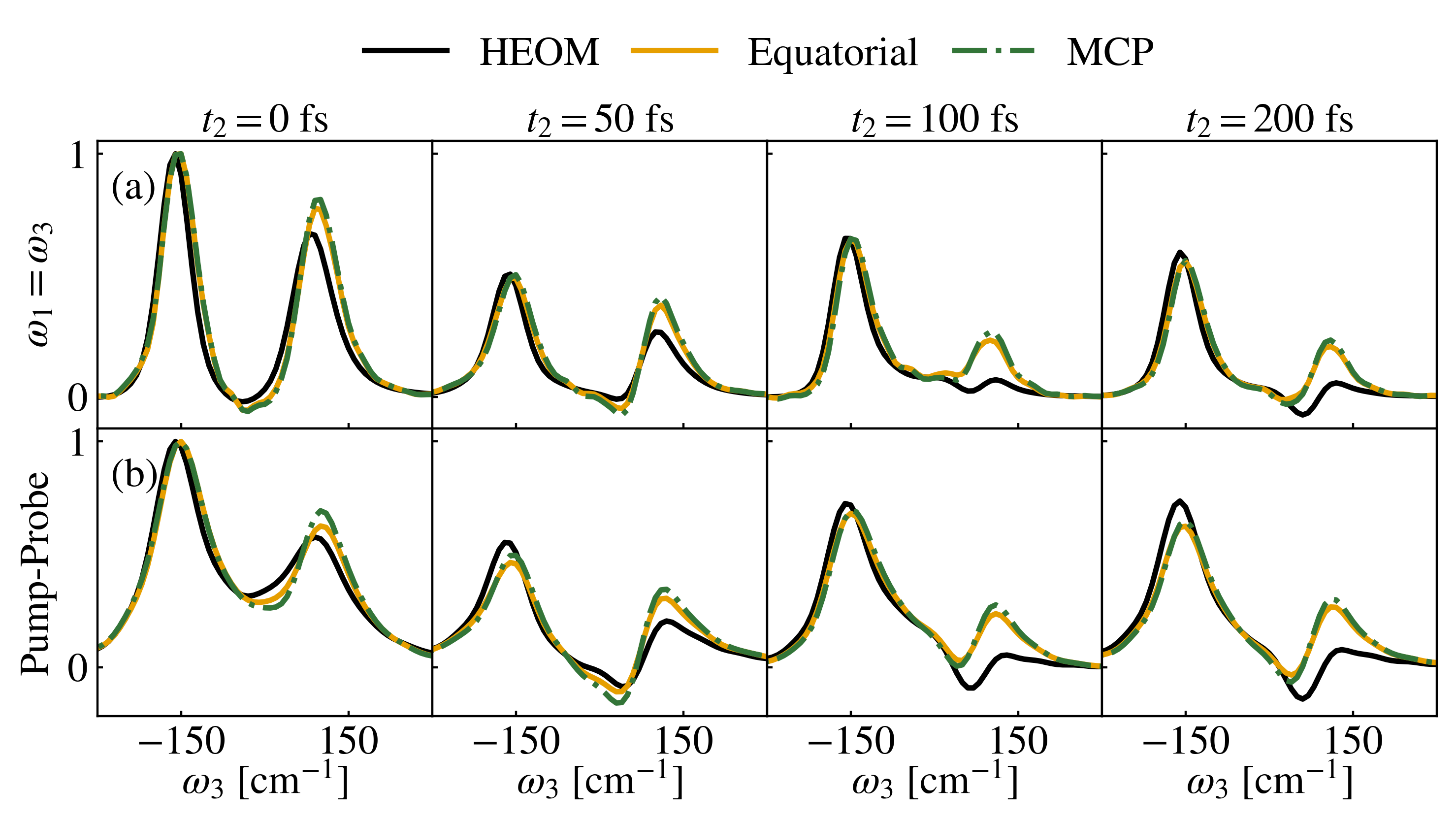}
    \label{sifig:biex-eq-vs-mcp-diag-pp}
\end{subfigure}
\caption{\rev{Comparison of 96,000 trajectory equatorial Ehrenfest and mean classical path spectra for the biexciton model. Top: 2DES spectra. Bottom: Diagonal slices and pump-probe spectra.}}
\end{figure}

\newpage
\small
\bibliography{nonlinear}